\documentclass[11pt, aps,superscriptaddress, onecolumn,prd, showpacs,preprintnumbers,nofootinbib, amsmath,amssymb,
aps, superscriptaddress]{revtex4-1}

\usepackage{graphicx}
\usepackage{amsmath,amssymb}
\usepackage{amsfonts}
\usepackage{xspace} 
\usepackage[usenames]{color}
\usepackage{dcolumn}
\usepackage{bm}
\usepackage{mathrsfs}
\usepackage{multirow}
\usepackage[colorlinks=true]{hyperref}
\usepackage[all]{hypcap} 
\usepackage[utf8]{inputenc} 
\usepackage{lipsum}
\usepackage{etoolbox}
\usepackage{tikz}
\usepackage{url}
\usepackage{subfigure}
\usepackage{adjustbox}
\usepackage{caption}
\usepackage{slashed}
\usepackage{multirow}
\usepackage{array}
\usepackage{booktabs}
\usepackage{siunitx}
\usepackage{comment}
\usepackage{lineno}
\usepackage{orcidlink}
\usepackage{soul}
\usepackage{mathtools}

\def\be{\begin{equation}}
\def\ee{\end{equation}}
\def\bea{\begin{eqnarray}}
\def\eea{\end{eqnarray}}

\newcommand{\beq}{\begin{eqnarray}}
\newcommand{\eeq}{\end{eqnarray}}
\usepackage{calrsfs}
\DeclareMathAlphabet{\pazocal}{OMS}{zplm}{m}{n}

\newcommand*\Bell{\ensuremath{\boldsymbol\ell}}
\newcommand*\Bn{\ensuremath{\boldsymbol{n}}}
\newcommand*\Bm{\ensuremath{\boldsymbol{m}}}

\usepackage{amsthm}

\theoremstyle{definition}
\newtheorem{definition}{Definition}[section]
\newtheorem{note}{Note}[section]
\newtheorem{theorem}{Theorem}[section]
\newtheorem{corollary}{Corollary}[section]
\newtheorem{lemma}{Lemma}[section]
\newtheorem{proposition}{Proposition}[section]
\newtheorem{example}{Example}[section]

\newcommand{\bef}{\begin{definition}}
\newcommand{\eef}{\end{definition}}
\newcommand{\ben}{\begin{note}}
\newcommand{\een}{\end{note}}
\newcommand{\bep}{\begin{proposition}}
\newcommand{\eep}{\end{proposition}}
\newcommand{\bet}{\begin{theorem}}
\newcommand{\eet}{\end{theorem}}
\newcommand{\bee}{\begin{example}}
\newcommand{\eee}{\end{example}}
\newcommand{\bec}{\begin{corollary}}
\newcommand{\eec}{\end{corollary}}
\newcommand{\bel}{\begin{lemma}}
\newcommand{\eel}{\end{lemma}}

\begin{document}

\title{Asymmetric  excitation of left- vs right-handed photons\\ in accelerating waveguides}

\author{Adri\'an del R\'{\i}o\,\orcidlink{0000-0002-9978-2211}}\email{adrdelri@math.uc3m.es}
 \affiliation{Universidad Carlos III de Madrid, Departamento de Matem\'aticas.\\ Avenida de la Universidad 30 (edificio Sabatini), 28911 Legan\'es (Madrid), Spain.}

\begin{abstract}

The electromagnetic duality symmetry of Maxwell's equations in vacuum implies that the  circular polarization $Q$ of classical electromagnetic waves is conserved. In  quantum field theory, the normal-ordered operator $\hat Q$ represents the difference between the number operators  of right- and left-handed photons. Previous studies have shown that its expectation value is not conserved  for observers propagating in a gravitational field. Here, we show that this Noether symmetry can also be realized in empty waveguides with duality-preserving boundary conditions, and we quantize the source-free Maxwell theory inside a long, cylindrical waveguide undergoing both linear and rotational acceleration from rest. In the  vacuum $|0\rangle$ associated to inertial observers, we find that the expectation value $\langle 0| \hat Q |0\rangle $  fails to be conserved for  observers co-moving with the waveguide. In particular, frame-dragging effects induce a spectral asymmetry between the right- and left-handed field modes at late times. As a  consequence, accelerated detectors co-moving with the rotating waveguide  can detect   photon-pair excitations from the quantum vacuum, exhibiting an imbalance between opposite helicity modes. This is a  relativistic quantum effect, which shows that the classical conservation law associated with duality symmetry is broken in the quantum theory even in flat spacetime, provided we work with non-inertial systems. Our analysis provides a concrete proof of concept for testing this  effect in analogue gravity platforms.

{\let\clearpage\relax  \let\sectionname\relax \tableofcontents}
\end{abstract}
\maketitle

\section{Introduction}

Symmetry principles have long stood at the core of theoretical physics, providing both aesthetic guidance and powerful constraints in the formulation of fundamental laws \cite{2010syph.book.....B}. Through Noether's theorem, continuous symmetries acquire dynamical significance, linking invariance with conservation laws.  In many cases, symmetry considerations not only constrain the form of admissible field equations but also point toward new physical entities and deeper layers of theoretical consistency.  The deep interplay between symmetry and dynamics has guided major developments---from general relativity, where dipheomorphism invariance encodes the universality of gravitation, to quantum field theory, where local gauge invariance dictates the structure of the Standard Model.

Among all known symmetries in Physics, the electric-magnetic duality of Maxwell's equations holds a particularly interesting status. The original discrete formulation, corresponding to the interchange of the electric and magnetic fields in the source-free theory, inspired Dirac to postulate the existence of magnetic monopoles as a means to explain the quantization of electric charge \cite{Dirac1931kp}, and decades later played a central role in gauge and string theories by relating strong and weak coupling regimes \cite{SEIBERG199419}. The continuous version of this duality, corresponding to a global $U(1)$ rotation in the space of  electromagnetic wave solutions, was first analyzed in flat spacetime by Calkin \cite{10.1119/1.1971089} and subsequently extended to curved backgrounds by Deser and Teitelboim \cite{PhysRevD.13.1592}, who further identified the associated Noether charge in terms of the electromagnetic fields. Despite its conceptual simplicity, this continuous symmetry received comparatively little attention for several decades, resurfacing  recently in some studies which explore its  implications in the quantum theory  \cite{PhysRevLett.118.111301, PhysRevD.98.125001}. 

A remarkable prediction of quantum field theory is the occurrence of quantum anomalies, which have played a central role in modern particle physics \cite{10.1093/acprof:oso/9780198507628.001.0001}. Anomalies arise when classical Noether symmetries fail to survive quantization, leading to the non-conservation of the corresponding quantum local current or global charge in any physical state of the theory. The paradigmatic example is the Adler-Bell-Jackiw (ABJ) axial anomaly for spin-$1/2$ fermions in quantum electrodynamics \cite{PhysRev.177.2426, 1969NCimA..60...47B}, which resolved the long-standing puzzle of the neutral pion decay and, in its non-Abelian form, contributed to explaining the $U(1)$ problem in quantum chromodynamics through the use of instantons \cite{PhysRevLett.37.8, THOOFT1986357}.

Remarkably, the electric-magnetic duality symmetry of source-free Maxwell theory also exhibits an anomaly when the field is quantized in a general curved spacetime \cite{PhysRevLett.118.111301, PhysRevD.98.125001}, providing the spin-1 analogue of the ABJ anomaly. In this case, the associated chiral charge $Q$ counts the difference between right- and left-handed photon modes. After renormalizing the Noether current, the vacuum expectation value of this charge ceases to be conserved, its evolution being dictated by the underlying spacetime geometry:
\bea 
\langle 0|\hat Q(t_2) |0\rangle -\langle 0|\hat Q(t_1) |0\rangle =\frac{-1}{96\pi^2} \int_{(t_1,t_2)\times \Sigma} d^4x \sqrt{-g} R_{abcd}{^*R}^{abcd}, \label{gravcp}
\eea
where $R^a_{\, \, \, bcd}$ denotes the Riemann tensor, $^*$ the Hodge dual, and $\Sigma$ is a spacelike Cauchy hypersurface of the spacetime manifold\footnote{A similar anomaly has been recently reported for linearized gravity in vacuum \cite{th76-hy1r}.}. Subsequent studies have shown that, for asymptotically flat spacetimes with well-defined notions of future $(\mathcal J^+)$ and past $(\mathcal J^-)$ null infinities, this duality anomaly is triggered whenever the gravitational background carries a flux of circularly polarized gravitational waves \cite{PhysRevLett.124.211301, PhysRevD.104.065012}, such as those emitted by binary black-hole mergers with net helicity \cite{PhysRevD.108.044052, 1nnp-w5w4}. If $h_+^{\ell m}(\omega)$, $h_\times^{\ell m}(\omega)$ refer to the multipoles of  frequency $\omega$ of the two gravitational-wave strain polarizations emitted by some astrophysical source, the non-conservation reads
\bea
\langle 0|\hat Q(\mathcal J^+) |0\rangle -\langle 0|\hat Q(\mathcal J^-) |0\rangle  = \int_0^{\infty} \frac{d\omega \omega^3}{24\pi^3} \sum_{\ell m} \left[| h_+^{\ell m}(\omega)-ih^{\ell m}_{\times}(\omega)|^2-|h_+^{\ell m}(\omega) + ih^{\ell m}_{\times}(\omega)|^2 \right] \, ,
\eea
Physically, this anomaly results in the spontaneous excitation of photon pairs from the quantum vacuum with an imbalance between opposite helicities \cite{doi:10.1142/S0218271817420019, sym10120763}, demonstrating how gravitational dynamics, and particularly the development of spacetime mirror asymmetries, can imprint chiral asymmetries on the electromagnetic vacuum itself.\footnote{While not  related to conservation laws, in the context of gauge theories discrete $\mathbb Z_2$ duality transformations have also been found to be ``anomalous'', in the sense that partition functions fail to be modular-invariant.  This was first studied in electromagnetism \cite{Witten_1995_06, PhysRevD.96.045008} and recently in linearized gravity \cite{3nkh-t15m}.} The physical picture is similar to the fermionic ABJ anomaly \cite{PhysRevD.21.1591}.

Although the theoretical implications of the electromagnetic duality anomaly are profound, its direct observational signatures in realistic astrophysical environments are expected to be exceedingly small. This last point is analogous to the limitation of observing Hawking's radiation from black holes \cite{cmp/1103899181} directly, whose  luminosity was estimated very low to be detected with astrophysical techniques \cite{PhysRevD.13.198, PhysRevD.14.3260, PhysRevD.16.2402}. This limitation naturally motivates the search for analogue realizations of the anomaly in controllable laboratory settings. The rapidly developing field of analogue gravity provides precisely such a framework, enabling the simulation of quantum field phenomena in curved spacetimes through condensed-matter systems, fluid analogues, or photonic platforms (see e.g. \cite{Barcel__2005, Jacquet:2020bar} and references therein). In this context, the Equivalence Principle suggests that the essential features of a gravitational field, and particularly  gravitational helicity, should also emerge for observers undergoing acceleration or rotational motion. It is therefore natural to ask whether non-inertial motion alone can induce an electromagnetic duality anomaly analogous to the gravitational case. The aim of this work is to explore this possibility by quantizing the source-free Maxwell field inside a long, cylindrical waveguide undergoing both linear and rotational acceleration, and by analyzing how non-inertial effects modify the conservation of the associated duality charge.

In this  work, we introduce a simplified theoretical model that captures the essence of this quantum effect within a controllable, non-inertial setting, thereby opening a pathway toward laboratory investigations of the electromagnetic duality anomaly under more realistic conditions. In contrast to the gravitational analyses of Refs. \cite{PhysRevLett.118.111301, PhysRevD.98.125001}, the present approach yields a direct evaluation of the vacuum expectation value of the duality charge, without resorting to formal regularization techniques in generic curved spacetimes, and makes explicit the physical link between the anomaly and the asymmetric excitation of right- and left-handed photon pairs from the quantum vacuum. This work is a follow up of a previous article \cite{delrio2025electromagneticdualityanomalyaccelerating}, and includes all technical details and further developments not previously communicated.\footnote{See also Ref.~\cite{PhysRevD.98.065016} for a  study related to the generation of chiral electromagnetic currents in a thermal photon gas confined in a cavity and subject to duality-violating boundary conditions.}

The manuscript is organized as follows. First, in Sec. II we establish the boundary conditions under which the classical duality symmetry for the source-free Maxwell theory holds when constrained within a generic waveguide. Then, in Sec. III we will describe the waveguide model employed in this work to analyze the emergence of the quantum anomaly. In Sec. IV we describe the classical covariant phase space for right-handed potentials, and perform quantization at early times. In Sec. V we explain the quantization at late times when the waveguide is rotating. In Sec. VI we evaluate the vacuum expectation value of the relevant operator and assess its time evolution via Bogoliubov transformtations. Finally, in Sec VII we summarize our main findings, and discuss future lines of research.

Regarding notation, we adopt geometric units with $G = c = 1$, while keeping Planck's constant explicit ($\hbar \neq 1$) in order to emphasize quantum effects in later sections. In this system of units, the electromagnetic potential and field tensor have dimensions $[A_a] = \sqrt{\hbar}/L$ and $[F_{ab}] = \sqrt{\hbar}/L^{2}$, respectively, where $L$ denotes a characteristic length scale and $\hbar$ sets the scale of angular momentum. Throughout the paper, $\eta_{ab}$ and $\nabla_a$  denote the flat Minkowski metric  and its associated Levi-Civita connection, respectively.

\section{Electric-magnetic duality  in empty waveguides}\label{secII}

Let $(M, \eta_{ab})$ denote the portion of Minkowski spacetime contained inside an empty cylindrical waveguide. The manifold can be decomposed as $M\simeq \mathbb{R}\times \Sigma$, where $\Sigma\simeq D\times \mathbb{R}$ is a spacelike Cauchy hypersurface, with $D\subset\mathbb{R}^2$ a closed disk. The boundary of this spacetime region is $\partial M\simeq \mathbb{R}\times \partial\Sigma$, where $\partial\Sigma\simeq \mathbb{S}^1\times \mathbb{R}$ represents the cylindrical surface of the waveguide.

The dynamics of electromagnetic waves in vacuum is governed by the source-free Maxwell equations $\nabla_a F^{ab}=\nabla_a {^*}F^{ab}=0$, where ${^*}F^{ab}=\frac{1}{2}\epsilon^{abcd}F_{cd}$ is the Hodge dual of the  two-form $F_{ab}$. These equations remain invariant under  the continuous transformations $F_{ab}\to \cos \theta F_{ab}+\sin \theta \, {^*F}_{ab}$, ${^*F}_{ab}\to \cos \theta \, {^*F}_{ab}-\sin \theta \, {F}_{ab}$, for any $\theta\in\mathbb R$, known as {\it electric-magnetic duality rotations}. Previous works have shown that such duality transformations correspond to a Noether symmetry in free space, even in curved spacetimes \cite{PhysRevD.13.1592, PhysRevD.98.125001}. More precisely, invariance of the free Maxwell action leads to the  conserved Noether charge
\bea
Q_0(A)=\frac{1}{2}\int_\Sigma d\Sigma\,  n_b\left[ A_a {^*F}^{ba}  -  Z_a F^{ba} \right]\, , \label{freeq}
\eea
where $n^a$ denotes the unit normal to $\Sigma$, and $Z_a$ is the potential dual to $A_a$. However, if the electromagnetic field is confined to a finite region of spacetime, such as inside the type of waveguides considered here, this conclusion is no longer obvious, since the field couples to the waveguide's surface. In this case, the field does not propagate freely but interacts effectively with the microscopic constituents through boundary conditions. In this section we revisit this question and analyze under which boundary conditions electromagnetic duality rotations can still generate a Noether symmetry.

There exist two complementary frameworks to describe the dynamical evolution of fields: the \emph{canonical} and the \emph{covariant} approaches. In the more traditional canonical formalism, one considers a Lagrangian of second order in field derivatives and defines a configuration space $\mathcal{Q}$ over a Cauchy surface $\Sigma_0\subset M$, where each element of $\mathcal{Q}$ corresponds to a pair of fields $(\phi_0,\dot\phi_0)$ providing initial data. If the theory is well-posed, such data determine a unique solution on the full spacetime $M$, and a natural symplectic structure can be constructed on the cotangent space of $\mathcal{Q}$ to identify symmetries and conservation laws. The more modern covariant approach, by contrast, deals directly with field solutions over the full spacetime $M$, forming a vector space known as the \emph{covariant phase space}~\cite{1987thyg.book..676C, 10.1063/1.528801,Harlow_2020,PhysRevD.103.025011}. Once a local action is specified, this method provides a systematic procedure to obtain a symplectic structure without resorting to a Hamiltonian formulation, and the result can be shown to coincide with the canonical one~\cite{PhysRevD.105.L101701}. In recent years, this formalism has gained popularity due to its versatility and manifest covariance, and it has been extended to field theories with boundaries~\cite{Harlow_2020,PhysRevD.103.025011}. We shall employ this method here to analyze the electromagnetic duality symmetry in waveguides, following closely the approach described in~\cite{PhysRevD.103.025011}.

\subsection{Review of some standard results}

To illustrate the power of this framework and the subtleties introduced by boundaries, we first review the standard derivation of  Maxwell's theory in terms of a magnetic potential. 

The first step in constructing the symplectic structure is to specify an action functional on the space $\mathcal F$ of ``kinematically allowed'' field configurations, $S:\mathcal F \to \mathbb R$. In our problem, $\mathcal F$  will be the  space of smooth 1-forms  over $M$, $\mathcal F=\Omega^1(M)$, representing all admissible magnetic potentials $A_a(x)$. This is an infinite-dimensional manifold on which familiar operations such as the exterior derivative $\delta$, Lie derivative $\mathcal L$, and interior product $i$ are naturally defined. In particular, given a field configuration $A\in\mathcal{F}$, variations of the field are represented by 1-forms $\delta A$ in the cotangent space of $\mathcal F$. The action of the  theory is determined then by a pair of Lagrangians as $S[A]=\int_M L[A]- \int_{\partial M}\bar\ell(A)$. The variation of the  ``bulk'' Lagrangian $L[A]$ yields the Euler-Lagrange field equations in $M$, while the ``boundary'' Lagrangian $\bar\ell(A)$ encodes a fixed set of boundary conditions  for the fields on $\partial M$ \cite{PhysRevD.103.025011}. 

In the source-free Maxwell theory one typically takes $L[A]=\frac{1}{2} F\wedge {^*F}$  and $\bar \ell(A)=0$, where $F=dA$ is the covariant  field strength derived from the magnetic potential. The choice $\bar \ell(A)=0$  is always implicitly assumed. When $\partial M=\emptyset$, as in free Minkowski space, $M=\mathbb R^4$, this choice effectively enforces suitable falloff conditions at  infinity. However, when $\partial M\neq \emptyset$, the particular choice of $\bar \ell(A)$ specifies a definite set of boundary conditions on $\partial M$, which may  modify the usual symplectic structure in free space by adding boundary contributions. Here we focus on the simplest case, $\bar \ell(A)=0$,  to illustrate the essential features. 

Once the action  $S[A]$ has been prescribed, the next step is to study variations of this functional on the configuration space $\mathcal F$. Varying the bulk Lagrangian $L[A]=\frac{1}{2} F\wedge {^*F}$  gives
\bea
\delta L=-d\, {^*F}\wedge \delta A+d(\delta A\wedge {^*F})\equiv E(A)\wedge \delta A + d \Theta^L(A)\, .
\eea
From this  we immediately  read off the field  equations $E(A)=0$, and the symplectic potential:
\bea
E(A)=-d\, {^*F}\, ,\quad  \Theta^L(A) = \delta A \wedge {^*F}\, .
\eea
Since $\bar \ell(A)=0$, on $\partial M$ one  further obtains 
\bea
\delta \bar \ell-j^*\Theta^L= - \delta \overline{A} \wedge\overline{^*F} \equiv \bar b (A) \wedge \delta \overline A- d\bar\theta^{(L,\bar \ell )}\, ,
\eea
where $j: \partial M \xhookrightarrow{} M$ is the inclusion map and barred quantities like $\bar A=j^*A$ represent pullbacks to $\partial M$. 
Following  \cite{PhysRevD.103.025011}, this result implies that the boundary conditions  $\bar b(A)=0$ are determined by:
\bea
\bar b(A)= \overline{^*F}\, ,\quad \bar\theta^{(L,\bar \ell )}= 0\, .
\eea
The space of solutions is given then by ${\rm Sol(S)}=\{A\in \Omega^1(M)\, /\, E(A)=0, \, \bar b(A)=0\}\subset \mathcal F$. Locally, the boundary condition can be expressed as
\bea
  \overline{^*F} = {^*F}_{ab}(j(x)) d(x\circ j)^a\wedge  d(x\circ j)^b \propto {^*F}_{ab}(j(x)) \epsilon^{abc}_{\quad d}\rho_c  d(x\circ j)^d \propto F_{cd}(j(x))\rho^c  d(x\circ j)^d\, ,
\eea
with $\rho^a$  the unit normal to the boundary. Therefore,  the solution space  associated with    $S[A]$ is
\bea
{\rm Sol(S)}=\{A_a\in \Omega^1(M)\quad/ \quad \nabla_a F^{ab}=0,\, \left. F^{ab} \rho_b \right |_{\partial M}=0, \quad {\rm for}\quad  F=dA\}\, ,
\eea
which forms a vector space owing to the linearity of the equations. 

It is important to note that these boundary conditions do not respect the invariance of the field equations under electric-magnetic duality rotations:  if $F\in {\rm Sol(S)}$, then generally  $^*F\notin {\rm Sol(S)}$, precisely because $\left. {^*F}^{ab} \rho_b \right |_{\partial M}\neq 0$. Hence, electromagnetic duality rotations do not constitute a Noether symmetry in this waveguide: the boundary conditions break the Noether symmetry present in free space.

The symplectic structure canonically associated with the action $S[A]$ follows from the symplectic potentials $\Theta^L$, $\bar \theta^{(L,\bar\ell)}$ as 
\bea
\Omega_S=\delta \left(\int_\Sigma i^*\Theta^L(A) -\int_{\partial \Sigma} \bar i^*\bar \theta^{(L,\bar \ell )}\right)\, ,
\eea
 where $i^*$ and $\bar i^*$ represent the pullback of the inclusion maps  $i: \Sigma \xhookrightarrow{} M$ and $\bar i: \partial\Sigma \xhookrightarrow{} \partial M$, respectively. In our case, this expression becomes
\bea
\Omega_S=\delta \int_\Sigma \delta A_\Sigma\wedge {^*F}_{\Sigma}=\int_\Sigma \delta A_\Sigma \wedge  \,{^*\delta F}_{\Sigma}\, ,
\eea
where $A_\Sigma=i^* A$ and $F_{\Sigma}=i^*{F}$ denote the pullbacks on the Cauchy slice $\Sigma$.
This $\Omega_S$ is a 2-form on the configuration space  $\mathcal F$. In general, it depends on the choice of Cauchy slice $\Sigma$. However, it can be shown that its pullback $i_S^* \Omega_S$ to the space of solutions with the natural embedding $i_S: {\rm Sol(S)}  \xhookrightarrow{} \mathcal F$ is independent of  $\Sigma$, and therefore conserved in time \cite{PhysRevD.103.025011}.  For notational clarity, in what follows we do not write the explicit pullbacks, as no ambiguity will arise.

Given two vectors $X_{A_1}$, $X_{A_2}$ tangent to ${\rm Sol(S)}$ at a point $A\in {\rm Sol(S)}$,  characterized by $\mathcal L_{X_{A_i}}A=i_{X_{A_i}}\delta A=A_i$\footnote{In a finite-dimensional manifold this would correspond, in coordinates,  to the identification $X_{A_i}=A_i\frac{\delta}{\delta A}$.},    the symplectic product is defined as
\bea
\Omega_S(A_1,A_2)\equiv \Omega_S(X_{A_1},X_{A_2})= \frac{1}{2}\int_{\Sigma} (A_1 \wedge {^*F_2} - A_2 \wedge {^*F_1})\, , 
\eea
which coincides with the familiar expression in free space.
This symplectic product provides a direct method to identify conserved charges associated with continuous symmetries. Specifically, if a symmetry transformation is infinitesimally generated by the vector field $\delta_Q A$ on the solution space, the function $Q: {\rm Sol(S)}\to \mathbb R$, defined by $Q(A):=\frac{1}{2}\Omega_S(A,\delta_Q A)$, is the  Noether charge generating that symmetry in phase space \cite{Chernoff1974PropertiesOI}. Conservation follows from the time-independence of the symplectic product $\Omega_S$. 

Electric-magnetic duality rotations are generated by the infinitesimal transformation $\delta_QA = Z$, where $Z_a$ is the dual ``electric''  potential satisfying the Gauss constraint $D_a E^a=0$ in the source-free theory \cite{PhysRevLett.118.111301, PhysRevD.98.125001}, in exact analogy with the familiar equation $D_aB^a = 0$ for the magnetic potential $A_a$. If $\delta_QA$ generated a genuine Noether symmetry on Sol(S), then  the associated Noether charge would be
\bea
Q(A)=\frac{1}{2}\Omega_S(A,Z)=\frac{1}{2}\int_\Sigma d\Sigma\,  n_b\left[ A_a {^*F}^{ba}  -  Z_a F^{ba} \right]\, ,
\eea
which agrees with the expression in free space given in (\ref{freeq}). However,  this quantity need not be conserved inside the waveguide. Although $A\in{\rm Sol(S)}$, and  $Z$ satisfies the vacuum Maxwell equations, its field strength $dZ$  typically fails to obey the required boundary conditions,  as discussed above. Hence,  $Z\notin {\rm Sol(S)}$ in general. As emphasized above, $\Omega_S$ is  independent of $\Sigma$ only when evaluated on any pair $A_1,A_2\in  {\rm Sol(S)}$; if one argument lies outside the solution space, $\Omega_S$ will generally depend on $\Sigma$. Consequently, the putative charge $Q(A)$ above will generically evolve in time in this waveguide. Notice how this breakdown occurs even when working with Maxwell equations in pure vacuum, once boundaries are present.

\subsection{A self-dual formulation of Maxwell equations inside a waveguide}

The example discussed in the previous subsection illustrates that boundary conditions must be treated carefully if one wishes to preserve the electric-magnetic duality symmetry of free-space Maxwell theory inside a bounded region. Since our goal is to explore potential quantum anomalies, we wish to identify situations in which the corresponding Noether charge $Q$ is conserved at the classical level.

The main difficulty arises from the boundary Lagrangian $\bar \ell(A)$, which can enforce boundary conditions that break duality invariance. A natural strategy is to begin by prescribing boundary conditions that respect duality and then determine the appropriate $\bar \ell$. However, not every set of boundary conditions can be derived from a variational principle, and working within a variational framework is desirable because it automatically provides a natural symplectic structure from which conserved charges can be identified easily. To guarantee that the boundary conditions derived from a given $\bar \ell$ remain duality invariant, we shall work directly within a new formulation where the canonical variables in the configuration space are manifestly self-dual or anti-self-dual. In this subsection, we construct such a formulation for the source-free Maxwell theory inside a generic cylindrical waveguide.

We consider the Lagrangian
\bea
L[A_R,A_L,\lambda_1, \lambda_2]= {F^+}\wedge {^*F^-} + (F^+-i{^*F}^+)\wedge \lambda_1 + (F^-+i{^*F}^-)\wedge \lambda_2\, ,
\eea
where $F^+=dA_R$ and $F^-=dA_L$, with $A_L=\overline A_R$. In general,   $F^+$ and $F^-$ are not self- or antiself- dual, respectively. The role of the 2-forms $\lambda_1$ and $\lambda_2$, which act as Lagrange multipliers, is precisely to enforce these duality conditions on-shell, for which the bulk term ${F^+}\wedge {^*F^-}$ vanishes and the Lagrangian itself $L$ reduces to zero. 

Taking variations of this Lagrangian produces
\bea
\delta L[A_R,A_L,\lambda_1, \lambda_2]&=&E(A_R)\wedge \delta A_R + E(A_-)\wedge \delta A_- + E(\lambda_1)\wedge \delta \lambda_1 + E(\lambda_2)\wedge \delta \lambda_2\nonumber\\
&&+d\Theta^L(A_R)+d\Theta^L(A_L)+d\Theta^L(\lambda_1)+d\Theta^L(\lambda_2)\, ,
\eea
with the Euler-Lagrange equations 
\bea
E(A_R)=-d{^*F}^- - d(\lambda_1-i{^*\lambda_1})=0\,,  &\quad & E(\lambda_1)=F^+-i{^*F}^+=0 \, ,\\
E(A_L)=-d{^*F}^+ - d(\lambda_2+i{^*\lambda_2})=0 ,&\quad & E(\lambda_2)=F^-+i{^*F}^-=0\, .
\eea
The  two right-handed equations imply that $F^\pm\propto F\pm i {^*F}$ for some real 2-form $F$. They also lead to $d\, {^*F}^\pm=\mp i dF^{\pm}=0$ by virtue of $F^+=dA_R$, $F^-=dA_L$. Together, these relations imply $dF=d\, {^*F}=0$, so $F^+$ ($F^-$) can be physically interpreted as a self(antiself) dual solution of the source-free Maxwell equations. Substituting these conditions back into the two left-handed equations yields $d(\lambda_1-i{^*\lambda_1})=d(\lambda_2+i{^*\lambda_2})=0$, which in a simply connected manifold like $M$ it implies $\lambda_1-i{^*\lambda_1}=dV_1$, $\lambda_2+i{^*\lambda_2}=dV_2$ for some 1-forms $V_1$, $V_2$.

On the other hand, the symplectic potentials have the form
\bea
\Theta^L(A_R)=\delta A_R\wedge \left[ {^*F}^-+\lambda_1-i{^*\lambda_1} \right] \, , &\quad& \Theta^L(\lambda_1)= 0\, , \\
\Theta^L(A_L)=\delta A_L\wedge \left[ {^*F}^++\lambda_2+i{^*\lambda_2} \right]\ \, ,&\quad& \Theta^L(\lambda_2)= 0\, .
\eea
Note that $\lambda_1-i{^*\lambda_1} $ is antiself-dual whereas $\lambda_2+i{^*\lambda_2}$ is self-dual. Besides $\lambda_1-i{^*\lambda_1}=dV_1$, $\lambda_2+i{^*\lambda_2}=dV_2$, the Lagrange multipliers $\lambda_1$ and $\lambda_2$ remain undetermined, reflecting a gauge symmetry of the form $A_R\to A_R+V_1$, $A_L\to A_L+ V_2$. We are free to choose the values of $V_1$ and $V_2$ equal to zero, without loss of generality. 

Let us introduce now some   2-forms  $F^\pm_B$, with fixed boundary values on $\partial M$, related to each other by complex-conjugation. We define the boundary lagrangian  $\bar\ell(A_R,A_L)=\overline{A_R}\wedge \overline{{^*F}_B^-} + \overline{A_L}\wedge \overline{{^*F}_B^+}$. Varying this term gives
\bea
\delta \bar \ell-j^*\Theta^L = \bar b (A_R) \wedge \delta \overline{A_R}+\bar b (A_L) \wedge \delta \overline{A_L} - d\bar \theta^{(L,\bar \ell )}\, ,
\eea
where $\bar\theta^{(L,\bar \ell )}=0$ and the boundary conditions read
\bea
\bar b (A_R) = \overline{^*F^-}-\overline{{^*F}_B^-}\, , &\quad&  \bar b (A_L) = \overline{^*F^+}-\overline{{^*F}_B^+}\, . \label{bvalues}
\eea
Thus,
\bea
{\rm Sol(S)}=\{A_R\in \Omega^1(M)\, /\,   d \,{^*F}^{+}=0,\, i^*F^+=F^+\, , \left. F^{+,ab} \rho_b \right |_{\partial M}= \left. F_B^{+,ab} \rho_b \right |_{\partial M}, \quad {\rm for}\quad  F^+=dA_R\}\, .\nonumber
\eea

If either $F_B^+$ or $F_B^-$ fail to be self- or anti-self-dual, respectively, the boundary conditions (\ref{bvalues}) are incompatible and the solution space is empty, ${\rm Sol}(S)=\emptyset$. Conversely, when the boundary terms are self- and anti-self-dual, the space of solutions is nontrivial. In the rest of this work we will fix $F^+_B$ subject to the conditions $\rho_d\epsilon^{dabc}\nabla_a F_{B,bc}^+=\phi_d\epsilon^{dabc}\nabla_a {F}_{B,bc}^+=0$, $t^bz^cF_{B,bc}^+=0$  on $\partial M$, where $\{t^a,\phi^a,z^a\}$ are vectors tangent  to $\partial M$. The first two represent continuity conditions for $F^\pm$, while the last one imposes a nontrivial constraint. All three are invariant under duality rotations, hence  electromagnetic duality rotations remain an exact Noether symmetry on ${\rm Sol}(S)$.

Finally, the symplectic structure canonically associated with this action  is then 
\bea
\Omega_S&=&\delta \left(\int_\Sigma \Theta^L(A_R)+\Theta^L(A_L) -\int_{\partial \Sigma} \bar\theta^{(L,\bar \ell )}\right)=\int_\Sigma (\delta A_R \wedge \delta {^*F}^-+\delta A_L \wedge \delta {^*F}^+)\, , \nonumber\\
&=&-\int_\Sigma d\Sigma \,n_b(\delta A_{R,a}\wedge  \delta {F}^{-,ab}+\delta A_{L,a} \wedge \delta {F}^{+,ab}) = \int_\Sigma d\Sigma \,(\delta A_{R,a}\wedge  \delta {H}^{L,a}+\delta A_{L,a} \wedge \delta {H}^{R,a}) \nonumber\, ,
\eea
where we have defined $H_a^R:=F^{+}_{ab}n^a$,  $H_a^L:=F^{-}_{ab}n^a$. The symplectic product can be expressed as
\bea
\Omega_S(A_R^1,A_R^2)= \int_\Sigma d\Sigma \,{\rm Re}( \overline{A_{R,a}^1}   {H}^{R,2a}- A_{R,a}^2   \overline{{H}^{R,a,1}})\, .\label{symplecticfundamental}
\eea
 This expression will be used in later sections for doing the quantization.

\section{Description of the waveguide model:   boundary conditions and dynamics}

As usual in quantum field theory in curved spacetimes, one first fixes some background and then quantizes the field of interest on that classical background.

In this work, the background system consists of an infinitely long, empty cylindrical waveguide of radius $R$ in Minkowski spacetime $(\mathbb{R}^4, \eta_{ab})$, where $\eta_{ab}$ denotes the flat spacetime metric. To mimic the effect of gravitational helicity on the electromagnetic field modes propagating inside the waveguide, the latter is assumed to start from rest and then undergo a period of acceleration until it reaches a constant angular velocity $\Omega$ at late times. For technical reasons to be discussed below, we will also let the waveguide experience a longitudinal acceleration along its symmetry axis, reaching a constant linear velocity $v$ asymptotically. In this final stationary configuration, the background carries nonzero helicity, thereby breaking the mirror symmetries of the initial setup and inducing a spectral asymmetry between right- and left-handed photon modes.

To quantize the electromagnetic field inside the waveguide, we must first determine the complete set of allowed solutions to the source-free Maxwell equations compatible with the boundary conditions---that is, to characterize the corresponding covariant phase space of the theory. We will start by discussing the static case in detail,  introducing the duality-preserving model, and  incorporate later the background dynamics to derive the mode basis in the rotating frame at late times.

\subsection{Preliminary example:  electric and magnetic  modes allowed in a perfectly conducting  waveguide}

In this subsection we review the standard procedure for solving Maxwell's equations inside a waveguide, focusing on the well-known case of perfectly conducting boundaries. This is standard material that can be found in many textbooks (see, e.g., \cite{2006iii}). The purpose of this review is twofold: first, to build intuition for the model to be introduced in the next subsection, which will be our main focus; and second, to illustrate explicitly with a simple example how  boundary conditions can lead to the violation of the classical electric--magnetic duality symmetry.

The source-free Maxwell equations for the electromagnetic  field $F_{ab}$ are  $\nabla_a F^{ab}=\nabla_a {^*}F^{ab}=0$, where ${^*}F^{ab}=\frac{1}{2}\epsilon^{abcd}F_{cd}$ is the Hodge dual of the two-form $F_{ab}$. The cylindrical symmetry of the background motivates the use of cylindrical coordinates $\{t,\rho,\phi,z\}$ to solve these equations, with $z$ denoting the axis of the cylinder. In these coordinates, the flat Minkowski metric takes the form
\bea
ds^2= \eta_{ab}( x) d x^a d x^b=-dt^2+d\rho^2+\rho^2d\phi^2+dz^2 \label{metric}\, ,
\eea
with $t,z\in \mathbb R$, $\rho\in (0,R)$, $\phi\in (0,2\pi)$, and each spacetime point carries a natural  orthonormal basis of  1-forms $\{ t_a, \rho_a,  \phi_a,  z_a\}=\{-\nabla_a  t, \nabla_a  \rho,  \rho\nabla_a  \phi, \nabla_a  z\}$, where $ t^a$ physically represents  the 4-velocity of  inertial observers, i.e.  worldlines in Minkowski with vanishing acceleration, $a^b=t^c\nabla_c t^b=0$.

  Modes of the  electromagnetic field inside the waveguide are solutions of Maxwell equations subject to some  boundary conditions. Let $E_a  =  t^b F_{ba}$ and $B_a  =  t^b {^*F}_{ba}$ denote the electric and magnetic fields measured by the inertial observers. For perfectly conducting boundaries, the tangential components of the electric field must vanish on the surface $E_a z^a=E_a \phi^a=0$, while the normal component of the magnetic field must also vanish $\rho^aB_a=0$. These conditions ensure that no surface currents appear on the conductor walls.

Because the background is stationary and axisymmetric, we can work in a basis of solutions with definite  frequency $\omega$, angular momentum $m$ and linear momentum $k$ along the $z$-axis.  Therefore, we can expand the most general solution in normal modes as
\bea
F_{ab}(t,\rho,\phi,z)&=&\int_{-\infty}^{\infty} d\omega \int_{-\infty}^{\infty} dk  \sum_{m=-\infty}^{\infty} e^{-i(\omega t+m\phi+kz)}F^{\omega k m}_{ab}(\rho) \label{F}\, ,
\eea
where $m\in \mathbb Z$ due to the periodicity  $\phi\in\mathbb S^1$ and  $\omega, k\in \mathbb R$. Substituting the ansatz  (\ref{F})  into Maxwell's equations $\nabla_a F^{ab}=\nabla_a {^*F}^{ab}=0$ produces the following 4 equations:
\bea
E_{\rho} & = & \frac{-i}{\omega^2-k^2}\left[k\partial_{\rho} E_z+i\frac{m\omega}{\rho}B_z \right]\, , \label{erho}\\
B_{\phi} & = & \frac{i}{\omega^2-k^2}\left[\omega\partial_{\rho} E_z+i\frac{mk}{\rho}B_z \right] \, , \label{bphi}\\
E_{\phi} & = & \frac{-i}{\omega^2-k^2}\left[ -i\frac{k m}{\rho}E_z+\omega \partial_{\rho} B_z \right]\, , \label{ephi}\\
B_{\rho} & = & \frac{i}{\omega^2-k^2} \left[ i\frac{ m\omega}{\rho}E_z-k \partial_{\rho} B_z \right] \, .\label{brho}
\eea
 On the other hand, Maxwell's equations $t_az_d\epsilon^{abcd}\nabla_c \nabla_e F^{e}_{\hspace{0.15cm}b}=0$, and $t_az_d\epsilon^{abcd}\nabla_c \nabla_e {^*F}^{e}_{\hspace{0.15cm}b}=0$, produce two independent second-order differential equations for $E_z$ and $B_z$:
\bea
\left[\frac{d^2}{d\rho^2}+\frac{1}{\rho}\frac{d}{d\rho}+\left(\omega^2-k^2-\frac{m^2}{\rho^2}\right) \right]E_z=0 \label{eqE}\, ,\\
\left[\frac{d^2}{d\rho^2}+\frac{1}{\rho}\frac{d}{d\rho}+\left(\omega^2-k^2-\frac{m^2}{\rho^2}\right) \right]B_z=0 \label{eqB}\, ,
\eea
which are simply the Bessel equation. Note that, once $E_z$ and $B_z$ are known, the rest of the field components follow algebraically from (\ref{erho})-(\ref{brho}). 

For perfect conductors, the boundary conditions at $\rho=R$ reduce to  the two conditions $E_z|_{\rho=R}=0$ and $\partial_\rho B_z|_{\rho=R}=0$. Because these two boundary conditions cannot be satisfied simultaneously for nontrivial solutions, we need to consider two separate cases, yielding two different families of  allowed modes: 

\noindent {\bf Transverse electric (TE) modes:} $E_z=0$, $B_z\neq 0$. The general solution  to (\ref{eqE})  is a linear combination of $J_m(\sqrt{\omega^2-k^2}\rho)$ and $Y_m(\sqrt{\omega^2-k^2}\rho)$. Demanding regularity at the axis $\rho=0$ selects 
\bea 
E_z=0, \hspace{1cm }  B_z=A_{\omega km}J_m(\sqrt{\omega^2-k^2}\rho)\, , \label{te} 
\eea
for some complex constant $A_{\omega km}$.
The electric and magnetic mode solutions are then $E_a=E_{\rho}\nabla_a \rho +  E_{\phi}\, \rho \nabla_a \phi + E_z \nabla_a z$, and $B_a=B_{\rho}\nabla_a \rho +  B_{\phi} \,\rho \nabla_a \phi + B_z \nabla_a z$, where\footnote{Note that primes denote derivatives of the function with respect to its argument, hence adimensional.}
\bea
E_\rho&=&A_{\omega km}\frac{m\omega}{\rho(\omega^2-k^2)}J_m(\sqrt{\omega^2-k^2}\rho) \label{erhoTE}\, ,\\
E_\phi&=&A_{\omega km} \frac{-i  \omega}{\sqrt{\omega^2-k^2}}J_m'(\sqrt{\omega^2-k^2}\rho)\, ,\\
E_z &=& 0\, ,\\
B_\rho&=&A_{\omega km}\frac{-ik}{\sqrt{\omega^2-k^2}}J'_m(\sqrt{\omega^2-k^2}\rho)\, ,\\
B_\phi&=&A_{\omega km} \frac{- m k}{\rho(\omega^2-k^2)}J_m(\sqrt{\omega^2-k^2}\rho)\, ,\\
B_z &=&  A_{\omega km} J_m(\sqrt{\omega^2-k^2}\rho) \label{ezTE}\, .
\eea
Applying the boundary conditions gives $\partial_\rho J_m(\sqrt{\omega^2-k^2}R)=0$, which is satisfied when $\sqrt{\omega^2-k^2}R=j'_{mn}$, for $\{j'_{mn}\}_{n\in \mathbb N}$  the zeros of the derivative of the Bessel $J$ function, $\partial_\rho J_m(j'_{mn})=0$. The effect of imposing these boundary conditions is to discretize the allowed frequencies inside the waveguide: $\omega_{kmn}^{TE}=\sqrt{k^2+\frac{j'_{mn}}{R}}$. Note that, because $J_{-m}(z)=(-1)^m J_m(z)$, the zeros satisfy $j'_{-mn}=j'_{mn}$\footnote{For $m>-1$, $J_m(z)$ has infinitely many, non-repeating  zeros, and all of them are real (see  8.541 in \cite{gradshteyn2007}).  If $j_{m1}>0$ is the smallest positive zero of $J_m(z)$ for $m>0$ (for which $j_{m0}=0$), then $j_{m1}>m$, so $j_{mn}>m$ for $n>0$ (see 8.544 in \cite{gradshteyn2007}). For $m\geq 0$, the number of  zeros of $J_m(x)$ contained in the interval $(0, (N+\frac{m}{2}+\frac{1}{4})\pi)$ is exactly $N$ (see  8.545 in \cite{gradshteyn2007}). For large $n$, $j_{m n}\sim (n+\frac{m}{2}-\frac{1}{4})\pi$ (see  8.547 in \cite{gradshteyn2007}).  }.

\noindent {\bf Transverse magnetic (TM) modes:} $E_z\neq 0$, $B_z= 0$. A similar reasoning yields
\bea
E_z=B_{\omega km} J_m(\sqrt{\omega^2-k^2}\rho)\, , \hspace{1cm } B_z=0\, ,  \label{tm}
\eea 
for some complex constant $B_{\omega km}$. The electric and magnetic fields are 
$E_a=E_{\rho}\nabla_a \rho +  E_{\phi}\, \rho \nabla_a \phi + E_z \nabla_a z$, and $B_a=B_{\rho}\nabla_a \rho +  B_{\phi} \,\rho \nabla_a \phi + B_z \nabla_a z$, where now
\bea
E_\rho&=&B_{\omega km}\frac{-ik}{\sqrt{\omega^2-k^2}}J'_m(\sqrt{\omega^2-k^2}\rho)\label{erhoTM}\, ,\\
E_\phi&=&B_{\omega km} \frac{- m k}{\rho(\omega^2-k^2)}J_m(\sqrt{\omega^2-k^2}\rho)\, ,\\
E_z &=& B_{\omega km} J_m(\sqrt{\omega^2-k^2}\rho)\, ,\\
B_\rho&=&B_{\omega km}\frac{-m\omega}{\rho(\omega^2-k^2)}J_m(\sqrt{\omega^2-k^2}\rho)\, ,\\
B_\phi&=&B_{\omega km} \frac{i  \omega}{\sqrt{\omega^2-k^2}}J_m'(\sqrt{\omega^2-k^2}\rho)\, ,\\
B_z &=& 0 \label{bzTM}\, .
\eea
Boundary conditions imposed by the perfect conductor at $\rho=R$ require $J_m(\sqrt{\omega^2-k^2}R)=0$, which is satisfied now for $\sqrt{\omega^2-k^2}R=j_{mn}$, where $\{j_{mn}\}_{n\in \mathbb N}$ are the zeros of the  Bessel $J$ function, $J_m(j_{mn})=0$. The result of imposing these boundary conditions is to discretize again the allowed frequencies inside the waveguide, according to $\omega^{TM}_{kmn}=\sqrt{k^2+\frac{j_{mn}}{R}}$. Again, we have $j_{-mn}=j_{mn}$.

Notice that, prior to imposing boundary conditions, the TM modes (\ref{erhoTM})-(\ref{bzTM}) are related to the TE modes (\ref{erhoTE})-(\ref{ezTE}) by a duality transformation (provided we take $B_{\omega km}=-A_{\omega km}$). That is, 
\bea
E_{a,\omega k m}^{TE}=-B_{a,\omega k m}^{TM}\, ,\quad B_{a,\omega k m}^{TE}=E_{a,\omega k m}^{TM}\, .
 \eea
Now, after imposing boundary conditions and discretizing the allowed frequencies, one finds
\bea
E_{a,\omega^{TE}_{kmn} k m}^{TE}\neq -B_{a,\omega^{TM}_{kmn} k m}^{TM}\, ,\quad B_{a,\omega^{TE}_{kmn} k m}^{TE}\neq E_{a,\omega^{TM}_{kmn} k m}^{TM}\, ,
\eea
precisely because $\omega_{kmn}^{TE}\neq \omega_{kmn}^{TM}$. As a result, performing an electric-magnetic duality rotation on a given solution takes you out of the physical solution space defined by the boundary conditions. This illustrates how boundary conditions can effectively break the classical electric-magnetic duality symmetry of free space.

\noindent {\bf Other modes}. There are no nontrivial transverse electromagnetic (TEM) modes: setting $E_z=B_z=0$ in  (\ref{erho})-(\ref{brho}) produce $ E_a= B_a=0$. Hybrid modes with both $E_z\neq 0$, $B_z\neq 0$ are also excluded, since  they would require  $\sqrt{\omega^2-k^2}R=j_{mn}=j'_{mn}$ for the same set of mode parameters  $\{\omega,m,k\}$, which never occurs (the Bessel zeros satisfy $j_{mn}<j'_{mn}<j_{mn+1}<j'_{mn+1}<\dots$ \cite{P_lmai_2011}).

\subsection{Electric/magnetic  modes allowed in our waveguide model}

As we have seen above, for (\ref{freeq}) to represent a classically conserved Noether charge, it is not enough that Maxwell's equations inside the waveguide be invariant under electric-magnetic rotations (which is always true in the absence of sources). It is also necessary that the boundary conditions remain invariant under duality transformations; otherwise, applying a duality rotation to a given solution takes it out of the space of admissible solutions.

In order to study the quantum anomaly, it will be convenient to consider a system in which the charge $Q$ is classically conserved, even if this requires boundary conditions that are not entirely realistic from an experimental standpoint. This simplification is harmless for our purposes, since the anomaly will always manifest itself through an explicitly time-dependent contribution to $\langle \hat Q\rangle$, regardless of the boundary. Proceeding this way allows us to provide a concrete proof of concept for the existence of a purely quantum violation of duality in non-gravitational settings. In realistic scenarios, both the classical and quantum mechanisms of symmetry breaking would coexist\footnote{A closely related situation occurs for Dirac fermions: the chiral symmetry may be broken explicitly at the classical level by a mass term, while at the same time the axial current remains subject to a quantum (Adler-Bell-Jackiw) anomaly.}.  The experimental challenge would then be to isolate the genuinely quantum component, a question that lies beyond the scope of the present work.

In this work we will therefore impose boundary conditions that are manifestly duality-invariant. More precisely, as advanced in Sec. II, we consider the following covariant equations
\bea
F_{ab}t^a z^b|_{\rho=R}=0\, ,\quad {^*F}_{ab}t^a z^b|_{\rho=R}=0\, , \label{boundaryconditions0}
\eea
at the waveguide surface $\rho=R$. These conditions eliminate the longitudinal electric and magnetic components simultaneously, and hence remain invariant under the transformation $F_{ab}\to {^*F_{ab}}$, ${^*F_{ab}}\to -F_{ab}$.

With such duality-preserving boundary conditions, it is more convenient to work with the complex field $H^R_a=E_a+iB_a=t^b\, {^+F}_{ba}$,
which unifies Maxwell's equations into a single complex form and compactly encodes the duality invariance. Namely, the boundary conditions reduce simply to $H_z^R|_{\rho=R}=0$,  and the 6 Maxwell equations (\ref{erho})-(\ref{eqB}) diagonalize into:
\bea
H^R_{\rho} = \frac{-i}{\omega^2-k^2}\left[k\partial_{\rho} +\frac{m\omega}{\rho}\right] H^R_z \, ,\label{h1} \\
H^R_{\phi}  =  \frac{-1}{\omega^2-k^2}\left[ \frac{k m}{\rho}+\omega \partial_{\rho}  \right] H^R_z\, , \label{h2}\\
\left[\frac{d^2}{d\rho^2}+\frac{1}{\rho}\frac{d}{d\rho}+\left(\omega^2-k^2-\frac{m^2}{\rho^2}\right) \right] H^R_z=0 \label{h3}\, .
\eea
Equation (\ref{h3}) is decoupled and can be solved independently for  $H^R_z$, while (\ref{h1})-(\ref{h2}) provide the remaining field components algebraically. The {\it most general solution} for $H^R_z$ that is bounded on the cylinder axis is $H^R_z\propto J_m(\sqrt{\omega^2-k^2}\rho)$. The boundary condition then selects a single family of allowed modes, with discretized frequencies $\omega_{kmn}=\sqrt{k^2+\frac{j_{mn}^2}{R^2}}$. Because $J_{-m}(z)=(-1)^m J_m(z)$ for integer $m$, the zeros satisfy $j_{-mn}=j_{mn}$, thus $\omega_{-k-mn}=\omega_{kmn}$.

\subsection{Right- and left-handed potentials inside the waveguide model}

Although most treatments of waveguide electrodynamics focus solely on the allowed electric and magnetic field modes, a formulation in terms of electromagnetic potentials becomes essential when extending the analysis beyond classical wave propagation into the quantum regime. This is because, in the quantum theory, the canonical variable is the potential, not the field strength. To the best of my knowledge, for waveguides this formulation is lacking in the literature. Because of this, we describe here in some detail the  procedure to obtain the (self-dual) potentials from the allowed electric and magnetic modes inside the waveguide. 

As introduced in Sec. II, the self-dual field $H^R_a=t^b{^+}F_{ba}$ in vacuum satisfies the Gauss constraint $D_a H^{R,a}=0$, where $D_a$ is the spatial covariant derivative induced on  spacelike Cauchy hypersurfaces $\Sigma=\{t=const\}$. This constraint  can be integrated  in terms of a complex ``right-handed'' potential $A_a^R$, yielding 
\bea
H^{R,a}=i \epsilon^{abc}D_b A_c^R \, , \quad H^{L,a}=-i \epsilon^{abc}D_b A_c^L=\overline{H^{R,a}}\, ,
\eea 
whose real part gives  the usual magnetic potential, $A_a=\sqrt{2}\, {\rm Re} A^R_a$, while the imaginary one yields the dual potential $Z_a=\sqrt{2}\, {\rm Im} A^R_a$. The dynamics of $A_a^R$ follows from $\nabla^a (\nabla_a A_b^R-\nabla_b A_a^R)=0$. To solve this equation we will (partially) fix the gauge with the Lorentz condition $\nabla_a A^{R,a}=0$, which significantly reduces the dynamical equations to $\nabla^a\nabla_a A_b^R=0$. 
Using the ansatz for the normal modes
\bea
A^R_a(t,\rho,\phi,z)&=& \int_{-\infty}^{\infty} d\omega \int_{-\infty}^{\infty} dk  \sum_{m=-\infty}^{\infty}  e^{-i(\omega t+m\phi+kz)} A^R_a(\rho) \label{A}\, ,
\eea
with $A^R_a=A^R_t\nabla_a t+A^R_{\rho}\nabla_a \rho +  A^R_{\phi} \, \rho\nabla_a \phi + A^R_z \nabla_a z$, a straightforward but somewhat tedious computation yields
\bea
\left[\frac{d^2}{d\rho^2}+\frac{1}{\rho}\frac{d}{d\rho}+\left(\omega^2-k^2-\frac{m^2}{\rho^2}\right) \right]A^R_t&=&0 \label{eqAt}\, ,\\
\left[\frac{d^2}{d\rho^2}+\frac{1}{\rho}\frac{d}{d\rho}+\left(\omega^2-k^2-\frac{(m\mp 1)^2}{\rho^2}\right) \right]A^R_{\pm}& =&0 \label{eqA}\, ,\\
\left[\frac{d^2}{d\rho^2}+\frac{1}{\rho}\frac{d}{d\rho}+\left(\omega^2-k^2-\frac{m^2}{\rho^2}\right) \right]A^R_z&=&0 \label{eqAz}\, ,
\eea
where we introduced the variables $A^R_{\pm}=\frac{1}{\sqrt{2}}(A^R_\rho\pm iA^R_\phi)$. These are standard Bessel equations. Regularity conditions at the axis $\rho=0$ yield
\bea
A^R_t & = & C^R_t J_m(\sqrt{\omega^2-k^2}\rho) \, ,\label{At}\\
A^R_{\pm} & = & C^R_{\pm} J_{m\mp 1}(\sqrt{\omega^2-k^2}\rho) \, ,\label{apm}\\
A^R_z & = & C^R_z J_m(\sqrt{\omega^2-k^2}\rho) \label{Az}\, .
\eea
This is the general solution of the source-free Maxwell equations for the right-handed potential in cylindrical coordinates. The left-handed potential satisfies identical equations, 
\bea
A^L_t & = & C^L_t J_m(\sqrt{\omega^2-k^2}\rho) \, ,\label{At}\\
A^L_{\pm} & = & C^L_{\pm} J_{m\mp 1}(\sqrt{\omega^2-k^2}\rho) \, ,\label{apm}\\
A^L_z & = & C^L_z J_m(\sqrt{\omega^2-k^2}\rho) \label{Az}\, .
\eea

The remaining coefficients $C^R_t,  C^R_{\pm}, C^R_z, C^L_t,  C^L_{\pm}, C^L_z$ are determined by imposing the boundary conditions on the self-dual fields, $H_z^R|_{\rho=R}=H_z^L|_{\rho=R}=0$, and by fixing the residual gauge freedom. Let us focus first on the right-handed potential. By using the identities  $J_{m-1}-J_{m+1}=\frac{2\partial_\rho J_m(\sqrt{\omega^2-k^2}\rho) }{\sqrt{\omega^2-k^2}}$, $J_{m-1}+J_{m+1}=\frac{2m J_m(\sqrt{\omega^2-k^2}\rho) }{\rho\sqrt{\omega^2-k^2}}$ (see 8.471.1, 8.471.2 in \cite{gradshteyn2007}) we can rewrite 
\bea
A^R_{\pm} & = & \frac{C^R_{\pm}}{\sqrt{\omega^2-k^2}} \left[\partial_\rho\pm \frac{m}{\rho} \right]J_m(\rho \sqrt{\omega^2-k^2})\, .
\eea
Now, the defining equation for the right-handed potential, $H^{R,a}=i \epsilon^{abc}D_b A_c^R$, yields 3 independent equations:
\bea
H^R_\rho\pm i H^R_\phi&=&\left[\frac{-i \sqrt{2} k C^R_+}{ \sqrt{\omega^2-k^2}}\mp C^R_z\right]\left[\partial_\rho\pm \frac{m}{\rho} \right]J_m(\rho \sqrt{\omega^2-k^2})\, ,\\
H^R_z&=&\frac{(C^R_++C^R_-)\sqrt{\omega^2-k^2}}{\sqrt{2}}J_m(\rho \sqrt{\omega^2-k^2})\, ,
\eea
while  equations (\ref{h1})-(\ref{h3}) yield
\bea
H^R_\rho\pm i H^R_\phi&=&\mp \frac{i}{\omega\mp k}\left[\partial_\rho\pm \frac{m}{\rho} \right]H^R_z(\rho)\, ,\\
H_z^R&=& J_m(\rho \sqrt{\omega^2-k^2})\, ,
\eea
where the normalization factor of $H_z^R$ has been set to 1 without loss of generality (we will discuss in later sections how to suitably obtain the adequate value). Comparing both sets of equations we obtain
\bea
C^R_-&=&\frac{\sqrt{2}}{\sqrt{\omega^2-k^2}} - C_+^R\, ,\\
C^R_z &=& -i \left[\frac{1}{k-\omega}+\frac{\sqrt{2}k C_+^R}{\sqrt{\omega^2-k^2}} \right]\, .
\eea
On the other hand, the Lorentz condition $\nabla_a A^{R,a}=0$ produces
\bea
 A^R_t(\rho)=\frac{2i C_z^R k-\sqrt{2}(C_-^R-C_+^R)\sqrt{\omega^2-k^2})}{2i \omega}J_m(\rho \sqrt{\omega^2-k^2})\, ,
\eea
thus
\bea
C_t^R= -i \left[\frac{1}{k-\omega}+\frac{\sqrt{2}\omega C_+^R}{\sqrt{\omega^2-k^2}} \right]\, .
\eea
The freedom to choose $C_+^R$ corresponds to the remaining freedom of performing gauge transformations $A^R_a\to A^R_a+\nabla_a \phi^R$ subject to $\Box \phi^R=0$. To fully fix the gauge we  impose the condition $\bar m^aA_a^R=0$, with $\bar m_a=\frac{1}{\sqrt{2}} (\nabla_a\rho+i \rho \nabla_a\phi)$. This sets $C^R_+=0$ and therefore $C_z^R=C_t^R=\frac{i}{\omega-k}$, and $C_-^R=\frac{\sqrt{2}}{\sqrt{\omega^2-k^2}}$.

Repeating exactly the same steps as before, but now using  $H^{L,a}=-i \epsilon^{abc}D_b A_c^L$ (notice the relative minus sign) we obtain
 \bea
C^L_-&=&-\frac{\sqrt{2}}{\sqrt{\omega^2-k^2}} - C_+^L\, ,\\
C^L_z &=& i \left[\frac{1}{k+\omega}-\frac{\sqrt{2}k C_+^L}{\sqrt{\omega^2-k^2}} \right]\, ,\\
C_t^L&=& -i \left[\frac{1}{k+\omega}+\frac{\sqrt{2}\omega C_+^L}{\sqrt{\omega^2-k^2}} \right]\, .
\eea
The gauge condition imposed before is equivalent to $m^aA_a^L=0$ (after complex conjugating), which implies now $C^L_-=0$, so $C_z^L=C_t^R=\frac{i}{\omega-k}$, and $C_+^L=-\frac{\sqrt{2}}{\sqrt{\omega^2-k^2}}$.

Collecting all results, the right- and left-handed potentials allowed inside our waveguide take the compact forms:
\bea
A_{a,hkmn}^{R}\propto e^{-i(h\omega_{kmn} t+k z+m \phi)} \label{ar}   \left[  J_{m+1}\left(  \frac{j_{mn}}{R} \rho\right) {\overline{  \Bm_a}}-\frac{iR}{j_{mn}} (h\omega_{kmn}+k) {\bf \Bell_a} J_{m}\left(  \frac{j_{mn}}{R} \rho\right) \right]  \, , \\
A_{a,hkmn}^{L} \propto  e^{-i(h\omega_{kmn} t+k z+m \phi)}  \label{al} \left[    J_{m-1}\left(  \frac{j_{mn}}{R} \rho\right)  { \Bm_a}+\frac{iR}{j_{mn}} (h\omega_{kmn}+k){\bf \Bell_a} J_{m}\left(  \frac{j_{mn}}{R} \rho\right) \right] \, ,
\eea
 where, as mentioned in the previous subsection, the boundary conditions give  $\omega_{kmn}=+\sqrt{k^2+\frac{j_{mn}^2}{R^2}}$ and $h\in \{+1,-1\}$ tracks the overall frequency sign. These expressions have been written in terms of the cylindrical Newman-Penrose null tetrad  \cite{Torres2003}
 \bea
 {\{  \Bn_a,\Bell_a,  \Bm_a, \overline{ \Bm_a}\}}=\frac{1}{\sqrt{2}}\{ t_a + z_a,   t_a -   z_a,\rho_a- i  \phi_a, \rho_a+ i  \phi_a\}\nonumber\, ,
 \eea
satisfying  $\Bell^a  \Bn_a=-1$, $\Bm^a\bar \Bm_a=1$, and zero for any other contractions. In particular, notice that (\ref{ar}) and (\ref{al}) are null vectors: $A_{a,hkmn}^RA_{h'k'm'n'}^{a,R}=0$, $A_{a,hkmn}^LA_{h'k'm'n'}^{a,L}=0$. Together with
\bea
\overline{A_{a,-h-k-mn}^L}&=&(-1)^m A_{a,hkmn}^R\, , \label{relationRL}
\eea 
we conclude that, $\{A_{a,hkmn}^{R}\}$ form a complete set of linearly independent vector solutions of Maxwell equations.  These are eigenfunctions of the Hodge duality operator, in the  sense:
\bea
\frac{1}{2}\epsilon_{ab}^{\quad cd}\nabla_{[c}A^{R/L}_{d],h k m n} = \pm i \nabla_{[a}A^{R/L}_{b],h k m n}\, , \quad \frac{1}{2}\epsilon_{ab}^{\quad cd}\nabla_{[c}\overline{A^{L/R}_{d],h k m n}} &=& \pm i \nabla_{[a}\overline{A^{L/R}_{b],h k m n}}\, .
\eea
corresponding to definite handedness/self-duality equal to $\pm 1$.

Besides, these vector fields  have well-definite spin weight, and therefore helicity. To see this, notice that  the vector $\Bell_a$ denotes the null ``outgoing'' direction of propagation along the $z$ axis (namely, $A_{a,hkmn}^R$ propagates towards increasingly positive values of $z$ as time evolves), while $\bar \Bm_a$ indicates the direction of rotation in the plane orthogonal to it.  For positive-frequency modes, $A_{a,hkmn}^R$ rotates right-handedly along its propagation direction (thus, positive helicity), while for negative-frequency modes it rotates left-handedly as time evolves along its propagation direction (thus, negative helicity). Similarly, $A_{a,hkmn}^L$ rotates left-handedly along its propagation direction for positive-frequency modes (thus, negative helicity), while for negative-frequency modes it rotates right-handedly along its propagation direction (thus, positive helicity). This is, our results satisfy the expected relation between duality and helicity \cite{AA86}.

This last observation motivates us to decompose (\ref{ar})-(\ref{al}) as the sum of a ``transverse'' contribution plus a ``longitudinal'' term. In particular, it is not difficult to see from (\ref{ar})-(\ref{al}) that the most general solution (\ref{A}) has  the form
\bea
A_{a}^{R}=A_{a}^{R, T}+\Bell_a \mathcal L_{\Bn} \phi \label{longitudinaldecomposition}\, ,
\eea
where $\mathcal L_{\Bn}$ denotes the Lie derivative with respect to the vector $\Bn^a$, and $\phi\propto e^{-i(h\omega_{kmn} t+k z+m \phi)} J_{m}(  \frac{j_{mn}}{R} \rho)$ satisfies the Klein-Gordon equation. The two physical degrees of freedom are captured in the transverse contribution $A_{a}^{R, T}$, and can be obtained by contracting (\ref{ar})-(\ref{al}) with $\{\Bm_a, \bar \Bm_a\}$, which carry the two transverse photon polarizations. In contrast, the longitudinal degree of freedom, represented by this $\phi$, is not physical, but rather the result of imposing the Lorentz gauge fixing. It does not contribute to any observable, as can be checked from the fact that the symplectic structure (\ref{symplecticfundamental}) is independent of it. Taking into account that $\nabla_a\Bell_b=0$, we have $F^+=dA^{R,T}-\Bell \wedge  \mathcal L_{\Bn} d\phi$, thus
\bea
\Omega_S&=&i\int_\Sigma (\delta A_R \wedge \delta {F}^--\delta A_L \wedge \delta {F}^+)\nonumber\\
&=& i \int_\Sigma \left[\delta A_R^T\wedge \delta F^{-,T}-\delta A^T_R\wedge \ell\wedge \mathcal L_n d\bar \phi + \ell\wedge \mathcal L_n \phi\,  d \delta \overline{A_R^{T}}    \right.\nonumber\\
&& \quad \left. -\delta \overline{A_R^T}\wedge \delta F^{+,T}+\delta \overline{A^T_R}\wedge \ell\wedge \mathcal L_n d \phi - \ell\wedge \mathcal L_n \bar\phi\,  d \delta {A_R^{T}} \right]\nonumber\, .
\eea
Since $\Bell\wedge \mathcal L_{\Bn} \phi\,  d \delta \overline{A_R^{T}}=\Bell\wedge d(\mathcal L_{\Bn} \phi\,   \delta \overline{A_R^{T}})-\Bell\wedge d \mathcal L_{\Bn} \phi \wedge  \delta \overline{A_R^{T}}=d(\Bell\wedge \mathcal L_{\Bn} \phi\,   \delta \overline{A_R^{T}})-  \delta \overline{A_R^{T}}\wedge\Bell \wedge d \mathcal L_{\Bn} \phi$,  similarly with its complex conjugate, and $\mathcal L_{\Bn} \phi$ vanishes identically at the boundary $\rho=R$, we conclude
\bea
\Omega_S&=&i\int_\Sigma (\delta A^T_R \wedge \delta {F}^{-,T}-\delta A^T_L \wedge \delta {F}^{+,T})\, .
\eea

For completeness, the  field strengths associated to the vector modes (\ref{ar})-(\ref{al}), measured by inertial observers $t^a$, are:
\bea
H_{hkmn}^{R, a}&=&i\, t_d\epsilon^{dabc}\nabla_b A_{c,hkmn}^{R}\label{hr}\\
& \propto& e^{-i(h\omega_{kmn} t+k z+m \phi)}   \left[ \frac{i(h\omega_{kmn}-k)}{2} J_{m+1}\, { \overline{  \Bm_a}}+ \frac{i(h\omega_{kmn}+k) }{2}J_{m-1} { {\Bm_a}}-z_a\frac{j_{mn}}{\sqrt{2}R} J_{m} \right]  \, , \nonumber\\
H_{hkmn}^{L, a}&=&-i\, t_d\epsilon^{dabc}\nabla_b A_{c,hkmn}^{L}\label{hl}\\
& \propto& e^{-i(h\omega_{kmn} t+k z+m \phi)}   \left[ \frac{i(h\omega_{kmn}+k)}{2} J_{m+1}\, { \overline{  \Bm_a}}+ \frac{i(h\omega_{kmn}-k)}{2}J_{m-1}\, { {\Bm_a}}+z_a\frac{j_{mn}}{\sqrt{2}R} J_{m} \right]  \, , \nonumber
\eea
with all Bessel functions  evaluated at $\frac{j_{mn}}{R} \rho$. These modes also satisfy $\overline{H_{a,-h-k-mn}^L}=(-1)^m H_{a,hkmn}^R$.

\subsection{Acceleration of the waveguide and frame-dragging of basis vectors}

As indicated above, our waveguide is a hollow cylinder with duality-invariant boundary conditions (\ref{boundaryconditions0}) for the electromagnetic field modes. Initially static, the cylinder accelerates both tangentially and longitudinally until it reaches a constant angular velocity $\Omega_0$ and a constant linear velocity $v_0$ along its symmetry axis. In previous subsections we  obtained the basis of electromagnetic modes allowed inside the waveguide at early times, when it is static. Here we will derive the corresponding basis at late times in the co-rotating frame. In this late-time stationary configuration, the waveguide exhibits a handedness and thus breaks the mirror symmetries of the initial state. This feature will be crucial for inducing an asymmetry between the two chiral sectors of the allowed set of electromagnetic modes inside the waveguide in the co-rotating frame.

 To describe the dynamics of the accelerated waveguide at later times, we introduce a fiducial reference frame $\tilde O$ attached to inertial observers and decoupled from the waveguide, where the motion becomes explicit. In cylindrical coordinates $\{\tilde t,\tilde \rho,\tilde\phi,\tilde z\}$ adapted to this frame, the flat Minkowski metric has its canonical diagonal form everywhere, $ds^2= \tilde \eta_{ab}( \tilde x) d \tilde x^a d \tilde x^b=-d\tilde t^2+d\tilde \rho^2+\tilde\rho^2d\tilde\phi^2+d\tilde z^2$, and each spacetime point carries a natural orthonormal basis of 1-forms, written in these coordinates as $\{\tilde t_a, \tilde\rho_a, \tilde \phi_a, \tilde z_a\}=\{-\tilde\nabla_a \tilde t, \tilde\nabla_a \tilde \rho, \tilde \rho\,\tilde\nabla_a \tilde \phi, \tilde\nabla_a \tilde z\}$, with $\tilde t^a$ the 4-velocity of inertial observers. The self-dual solutions of Maxwell equations for $H^R_a$ and the right-handed potential $A^R_a$ are then given by the expressions obtained in the previous subsections (with tildes). More precisely,  self-dual solutions of Maxwell's equations that are eigenfunctions of the Killing Vector Fields (KVFs) $\partial_{\tilde t}$, $\partial_{\tilde \phi}$ and $\partial_{\tilde z}$ have the form (\ref{ar})-(\ref{al}).

We now wish to describe the quantum field from the viewpoint of a non-inertial reference frame $O$ associated with observers co-rotating and co-propagating with the waveguide. In this accelerating frame the constituents of the cylinder remain at rest at all times. Let $\{t,\rho,\phi,z\}$ denote cylindrical coordinates that remain constant along the worldlines of such observers at rest.  Specifically, if $\gamma_{\rho,\phi,z}(t)$ denotes one such curve from the congruence, labelled by $(\rho,\phi,z)$ and with proper time $t$, then its coordinates in this frame are $x^a(\gamma_{\rho,\phi,z}(t))=(t,\rho,\phi, z)$. Observers that follow these curves propagate with 4-velocity $u=\partial/\partial t$ and non-trivial 4-acceleration $a_a=u^b\nabla_b u_a\neq 0$. 

At early times, both frames coincide ($u^a=t^a$, hence $a^a=0$), and the metric in the non-inertial frame $O$ is also $ds^2=\eta_{ab}(x)\, dx^a dx^b \sim -dt^2 + d\rho^2 + \rho^2 d\phi^2 + dz^2$.  At later times, however, the non-inertial frame differs from the inertial one, as the co-rotating observers experience a centripetal acceleration, $a^a a_a = \rho^2\gamma^4 \Omega_0^4 \neq 0$. The specific dynamics of the waveguide is encoded in the coordinate functions $(\tilde t(x^a), \tilde \rho(x^a), \tilde \phi(x^a), \tilde z(x^a))$, which depend on the particular interpolating functions $\Omega(t)$ and $v(t)$ between $0$ and the final values $\Omega_0$, $v_0$. At sufficiently late times, once the waveguide reaches a stationary configuration, the two coordinate systems are simply related by $\rho=\tilde \rho$, $\phi=\tilde\phi-\Omega_0 \tilde t$ (time-dependent rotation),  $z=\gamma(\tilde z+v_0 \tilde t)$, $t=\gamma(\tilde t+v_0 \tilde z)$ (Lorentz transformation with boost parameter $\gamma^{-2}=1-v_0^2$). Under this transformation, the line element naturally acquires off-diagonal terms encoding non-inertial effects (such as frame dragging):
\bea
ds^2\sim -(1-\rho^2\gamma^2\Omega_0^2)dt^2+d\rho^2+\rho^2 d\phi^2+(1+v_0^2\rho^2\gamma^2\Omega_0^2)dz^2\nonumber\\
+2\gamma\Omega_0 \rho^2 { dtd\phi} -2v_0\gamma^2 \Omega_0^2 \rho^2 {  dtd z} -2v_0\gamma \Omega_0 \rho^2  { d\phi d z } \label{metricX}\, . 
\eea

To obtain the basis of self-dual solutions to Maxwell equations  inside the rotating waveguide at late times, we need to  explicitly incorporate the effects of frame-dragging  of the fiducial tetrad on the  electromagnetic basis modes (\ref{ar})--(\ref{al}). The fiducial tetrad, written in these coordinates as $\{e_a^{\,I}(x)\}_{I=0}^3=\{t_a,\rho_a,\phi_a,z_a\}$, is adapted to the motion of inertial observers, in the sense that each $e_a^{\,I}(x)$ is parallel transported along the integral curves of the inertial four-velocity: $t^a\nabla_a e_b^{\,I}=0$. In contrast, the accelerated observers at the waveguide's surface follow integral curves of $u=\partial_t$, which at late times relate to the inertial basis as $u^a|_{t\to\infty}\sim \gamma( t^a+ v_0 z^a)+\Omega_0 \gamma\, \rho\, \phi^a $. Hence, in this accelerated frame the inertial basis is expected to rotate as a function of time $t$. Indeed, using (\ref{metricX}) one finds
\bea
\frac{D \rho_a}{dt}\equiv u^b\nabla_b \rho_a= \Omega_0 \gamma \phi_a\, , \quad  \frac{D \phi_a}{dt}\equiv u^b\nabla_b \phi_a= -\Omega_0 \gamma \rho_a\, . \nonumber
\eea
i.e. $\{\rho_a,\phi_a\}$ fail to be parallel transported along the integral curves of $u^a$ and undergo  rotation with angular frequency $\Omega_0\gamma$ along the curves  $\gamma_{\rho,\phi,z}(t)$.

These relations   can  be diagonalized by introducing the Newman-Penrose cylindrical vectors $\sqrt{2} \Bm_a := \rho_a - i \phi_a$ and $\sqrt{2} \bar \Bm_a := \rho_a + i \phi_a$,  which satisfy $u^b \nabla_b \Bm_a = i \Omega_0 \gamma \Bm_a$, and $u^b \nabla_b \bar \Bm_a =- i \Omega_0 \gamma \bar \Bm_a$. Although $\Bm_a$ and $\bar \Bm_a$ are not parallel transported along $u^a$, their time-dependent phases can be factored out explicitly: 
\bea\label{framedragging}
\Bm_a=e^{i\Omega_0 \gamma t} \Bm_a^0\, , \quad \overline{\Bm_a}=e^{-i\Omega_0 \gamma t} \overline{\Bm_a^0}\, , \label{framedragging}
\eea
for some  $\Bm_a^0$, $\bar \Bm_a^0$ that are now parallel transported along $u^a$ ($u^b\nabla_b \Bm_a^0=0$).  This  time-dependent rotation of the inertial vectors $\{\Bm_a, \bar \Bm_a\}$, as seen from the  non-inertial observers propagating with 4-velocity $u^a$, captures the  frame-dragging effect of the rotating waveguide. Using this result, the electromagnetic circularly polarization basis  takes the following form in the non-inertial frame at late times:
\bea
 A_{a,hkmn}^{R}\sim e^{-i\left[(h\omega_{kmn}+m\Omega_0-kv_0)\gamma t+(k-h\omega_{kmn} v-m v_0 \Omega_0)\gamma z+m\phi\right]}    \left[  J_{m+1} {\overline{\Bm^0_a}}{\bf e^{-i \Omega_0 \gamma t}}- \frac{i R}{j_{mn}} (h\omega_{kmn}+k){\bf \Bell_a} J_{m} \right]  \label{arlate}\, ,\\
 A_{a,hkmn}^{L} \sim e^{-i\left[(h\omega_{kmn}+m\Omega_0-kv_0)\gamma t+(k-h\omega_{kmn} v-m v_0 \Omega_0)\gamma z+m\phi\right]}   \left[   J_{m-1}  {\Bm^0_a}{\bf e^{+i \Omega_0 \gamma t}} +  \frac{i R}{j_{mn}} (h\omega_{kmn}+k){\bf \Bell_a} J_{m} \right] \label{allate}\, ,
\eea 
with all Bessel functions  evaluated at $\frac{j_{mn}}{R} \rho$. Notice how the accelerating frame ``rotates'' the  the right-handed modes with global factor $e^{-i\Omega_0 \gamma t}$ and left-handed modes with the opposite factor $e^{+i\Omega_0 \gamma t}$. This is because the field $A_a^R$ has a well-definite (positive) helicity, rotating as a spin-1 field,  while $A_a^L=\overline{A_a^R}$  has the opposite one\footnote{As indicated below Eq. (\ref{longitudinaldecomposition}), the longitudinal contribution does not play any physical role.}. This frame-dragging effect will be crucial for introducing a spectral asymmetry in the quantum theory. 

The effects of frame-dragging also manifest on the form of the boundary conditions for the electric and magnetic fields, which become dynamical in the non-inertial frame. More precisely, the electric and magnetic fields measured by the accelerating observers at the waveguide's surface are given by $E_b=u^aF_{ab}$ and $B_b=u^a\, {^*F}_{ab}$. Using the orthogonal frame $\{u^a,\rho^a, \Phi^a, Z^a\}:=\{ \gamma( t^a+ v_0 z^a)+\Omega_0 \gamma\, \rho\, \phi^a,\rho^a, \phi^a+\gamma^2 \Omega_0\rho (t^a+v_0 z^a) , \gamma(z^a+v_0t^a)\}$ associated with the family of accelerating observers at late times, the  boundary conditions (\ref{boundaryconditions0}) imply
\bea
E_Z|_{\rho=R}= \gamma \Omega_0 B_\rho|_{\rho=R} \label{eframe}\, , \quad B_Z|_{\rho=R}=-\gamma \Omega_0 E_\rho|_{\rho=R}\, . \label{bframe}
\eea
Thus, the  form of the boundary conditions on the electric and magnetic fields changes in the accelerating frame, while respecting duality invariance. In particular, the physical modes $(E_Z,B_Z)$, which vanish  at early times, evolve continuously to the expressions above at late times due to the rotational background. By contrast, the form of the boundary conditions in the inertial orthonormal  frame $\{t^a,\rho^a,\phi^a, z^a\}$ read $\tilde E_z|_{\rho=R}=\tilde B_z|_{\rho=R}=0$ all the time. 

Finally, we remark that it is not necessary to model the transient phase of acceleration in detail. The final value of the vacuum expectation value of the Noether charge will depend only on the final state of motion of the waveguide (namely, its constant angular and linear velocities), and not on the precise acceleration history. This is precisely the expected behavior in the presence of a chiral anomaly \cite{PhysRevD.100.085014, PhysRevD.108.105025}. Anomalies are topological in nature and therefore insensitive to the detailed time profile of the background, provided the initial and final configurations are well defined. In this sense, only the asymptotic states of motion of the waveguide contribute to the observable effect, while the intermediate evolution acts merely as an interpolation between them.

\section{Quantization at early times} \label{seciv}

While quantization of the free electromagnetic field is a well-developed topic in globally hyperbolic spacetimes, introducing boundaries---especially time-dependent or accelerating ones---requires careful treatment and specific computations on a case-by-case basis\footnote{For some universal algebraic properties see e.g. \cite{Benini_2018}}. In particular, one cannot simply assume \emph{a priori} that the quantum field admits an expansion in creation and annihilation operators. 
A mathematically rigorous treatment of the electromagnetic field quantization  in 3+1 waveguides appears to be scarce in the literature, to the best of our knowledge. The aim of this and the following section is therefore to construct, in a precise manner, the ``in'' and ``out'' Fock spaces used throughout the paper\footnote{Even in studies of the (dynamical) Casimir effect, where quantum fields interacting with boundaries are routinely considered, the mode decomposition is typically introduced heuristically and the focus lies primarily on computing observable quantities. On the other hand, in  quantum optics  one usually quantizes only a finite set of selected modes. The anomaly analysed here depends crucially on the full infinite-dimensional structure of the quantum field, and hence requires a more robust functional-analytic framework.}.

To obtain a quantum description of the electromagnetic field inside the waveguide we first need to specify its underlying classical theory. For that, we adopt the covariant phase-space approach~\cite{1987thyg.book..676C,10.1063/1.528801}, in particular incorporating the recent developments that account for the presence of boundaries~\cite{PhysRevD.103.025011}. As reviewed in Sec.~II, in this formalism one works directly with field solutions defined over the entire spacetime, and a natural symplectic structure~$\Omega$ can be constructed once a local action functional is specified, including possible boundary contributions. Here we will work with Eq.~(\ref{symplecticfundamental}). Quantization can then be carried out using standard techniques in quantum field theory~\cite{Wald:1995yp,doi:10.1142/S0217751X13300238}. In particular, the construction of a  Fock  space  requires specification of a well-defined vacuum state. This can be achieved by e.g.  constructing a  complex structure $J$  that  is {\it compatible} with ~$\Omega$~\cite{Ashtekar1975zn}.  Physically, this allows to distinguish between positive and negative frequency modes. This will be the strategy adopted in this work.

In this section we will first perform the quantization of the theory at early times, when the waveguide remains static, leaving the corresponding construction at late times to the next section. The two resulting Fock spaces, $\mathcal{H}_{\text{in}}$ and $\mathcal{H}_{\text{out}}$, will enable us to carry out a Bogoliubov transformation analysis, necessary to evaluate the \emph{in}-vacuum expectation value of the Noether charge operator at late times. As we will see, the detailed profile of the intermediate acceleration phase of the waveguide will not be required to obtain the final result in closed form. This is because the anomalous time dependence of~$\langle \hat Q\rangle$ originates from a ``topological'' contribution that depends only on the \emph{boundaries} of our spacetime~\cite{PhysRevD.100.085014,PhysRevD.108.105025}, namely, on the initial and final spacelike hypersurfaces.

\subsection{Classical covariant phase space (standard variables)}

As a preliminary exercise, it is instructive to perform the  quantization using the usual real-valued magnetic potential~$A_a$ as the canonical variable. This will provide a familiar reference point before moving on to the self-dual variables in the next subsection.

The Fourier modes $A_{a,hkmn}$ of the real-valued  potential  must satisfy the usual reality condition $A_{a,-h-k-mn}=\overline{A_{a,hkmn}}$. From the right- and left-handed mode basis obtained in the previous section, (\ref{ar})-(\ref{al}), which obey $\overline{A_{a,-h-k-mn}^{L}}=(-1)^m A_{a,hkmn}^{R}$, we can build two linearly independent vector combinations satisfying the same reality property: $A_{a,hkmn}^1:=\frac{1}{\sqrt{2}}(A^R_{a,hkmn}+ A^L_{a,hkmn})$ and $A_{a,hkmn}^2:=-\frac{i}{\sqrt{2}}(A^R_{a,hkmn}- A^L_{a,hkmn})$, which are orthogonal to each other for fixed $h,k,m,n$: $A_{a,hkmn}^1A_{hkmn}^{a,2}=0$. Physically, $\{A^R_{a,hkmn}, A^L_{a,hkmn}\}$ correspond to circularly polarized photon modes, while $\{A^1_{a,hkmn}, A^2_{a,hkmn}\}$ represent the two linear polarization vectors. With this basis,  any real-valued vector solution of the source-free Maxwell equations (with the prescribed boundary conditions) can be written as the linear combination 
\bea
A_a(x)&=&\sum_{n=1}^{\infty}\sum_{m=-\infty}^{\infty}\int_{-\infty}^{\infty} dk \sum_{\lambda=1,2}  \sum_{h=\pm 1} z_{hkmn}^\lambda A^{\lambda}_{a,hkmn}(x)\nonumber\\
&=&\sum_{n=1}^{\infty}\sum_{m=-\infty}^{\infty}\int_{-\infty}^{\infty} dk \sum_{\lambda=1,2 } \left[ z_{1kmn}^\lambda A^{\lambda}_{a,1kmn}(x)+ \overline{z_{1kmn}^\lambda }\overline{ A^{\lambda}_{a,1kmn}}(x) \right] \label{basicA}\, ,
\eea
where $x$ is shorthand for $(t,\rho,\phi,z)$, and $z_{hkmn}^\lambda$ are some complex scalars  satisfying the reality condition $\overline{z_{hkmn}^\lambda}=(-1)^mz_{-h-k-mn}^\lambda$. We remark that, in our duality-invariant waveguide model, both linear polarization modes share the same frequency spectrum. In contrast, for boundary conditions that break electric-magnetic duality (e.g. perfectly conducting walls), the two polarizations acquire distinct spectra.

The covariant phase space  is defined as the real vector space $\Gamma$ spanned by this (complex-valued) basis of solutions, $\left\{A^{\lambda}_{a,hkmn}(x)\right\}$. Each element of $\Gamma$ has the form (\ref{basicA}) and is labelled by the set of Bargmann-Segal coordinates  $\{z_{hkmn}^\lambda\}_{k\in \mathbb R,m\in \mathbb Z,n\in \mathbb N,\lambda\in \mathbb Z_2}$,  which are assumed to decay sufficiently fast for $|k|, |m|, n\to\infty$ to ensure convergence of the integral and sums. The canonical symplectic product $\Omega_0$ defined on this covariant phase space is 
\bea
\Omega_0(A_1,A_2)&=&\frac{1}{2}\int_\Sigma d\Sigma  \, n_b\left[A_{a,1} F^{ab}_2 - A_{a,2} F_1^{ab} \right] \in \mathbb R  \label{sp0}\, ,
\eea
where $\Sigma_{t_0}=\{t=t_0\}$ is a spacelike Cauchy hypersurface, and $n_b=-\frac{1}{\sqrt{\nabla_a t\nabla^a t}}\nabla_b t=-\nabla_b t$ its future-directed normal. In our conventions $A_a$  carries dimensions of $\sqrt{\hbar}/L$, so that $\Omega_0$ has dimensions of $\hbar$, as expected for the classical precursor of a quantum commutator.

This symplectic structure is manifestly anti-symmetric, $\Omega_0(A_1,A_2)=-\Omega_0(A_2,A_1)$. Its time independence can be explicitly verified by computing the Lie derivative along $n^a$ (see Appendix B). For practical calculations, the integral expression for $\Omega$ can be rewritten as
\bea
\Omega_0(A_1,A_2)& =& \frac{1}{2}\int_\Sigma d\rho  dz d\phi \, \rho \, \left[A_{a,1}  \dot A_2^a  - A_{a,2} \dot A_1^a - A_{1}^a \partial_a (n_bA_{2}^b)  + A_{2}^a \partial_a (n_bA_{1}^b)  \right] \\
& =& \frac{1}{2}\int_\Sigma d\rho  dz d\phi \, \rho \, \left[A_{a,1} \dot A_2^a  - A_{a,2} \dot A_1^a \right]  
\, , 
\eea
where $\dot A_b\equiv n^a \partial_a A_b$, and the second line can be obtained after some straightforward manipulations, and using the property $n^a A_a|_{\rho=R}=0$.

We now decompose the general solution (\ref{basicA}) as $A_a=A_a^++A_a^-$, where $A_a^+$ and $A_a^-$  are, respectively,  the positive- and negative-frequency parts  with respect to the timelike KVF $\partial/\partial t$ at early times. We can then define a linear map $J: \Gamma \to \Gamma$ by $JA_a^\pm =\pm i A_a^\pm$, so that $JA_a=i A_a^+-i A_a^-$ for any $A_a$. Since $J^2=-\mathbb I$, this defines   a complex structure on $\Gamma$.  The doublet $(\Gamma, J)$ becomes a complex vector space, on which the multiplication by complex numbers is defined by $(a +j b) A_a := a A_a+ b J A_a$  with $j$  the imaginary unit,  for any $a, b\in \mathbb R$ and any (real-valued)  $A_a\in \Gamma$.

This complex structure $J$  is compatible with the symplectic structure, in the sense that $\Omega_0(JA_1, JA_2)=\Omega_0(A_1,A_2)$. Indeed, $\Omega_0(JA_1,JA_2)=\Omega_0(i A_1^+,i A_2^+)+\Omega_0(i A_1^+,-i A_2^-)+\Omega_0(-i A_1^-,i A_2^+)+\Omega_0(-i A_1^-,-i A_2^-)$, where
\bea
\Omega_0(i A_1^+,i A_2^+)&=& -\Omega_0( A_1^+, A_2^+)\, ,\\ 
\Omega_0(i A_1^+,-i A_2^-)&=& \Omega_0( A_1^+, A_2^-)\, ,\\ 
\Omega_0(-i A_1^-,i A_2^+)&=& \Omega_0( A_1^-, A_2^+)\, ,\\ 
\Omega_0(-i A_1^-,-i A_2^-)&=& -\Omega_0( A_1^-, A_2^-)\, ,
\eea
by virtue of (\ref{sp0}), and 
\bea
\Omega_0( A_1^+, A_2^+) = \frac{1}{2}\sum_{kmn\lambda} \sum_{k'm'n'\lambda'} z^{\lambda}_{1,1kmn}z^{\lambda'}_{2,1k'm'n'}\int_\Sigma d\rho  dz d\phi \, \rho 
 \left[(-i\omega^{}_{k'm'n'})-(-i\omega^{}_{kmn}) \right]A^{\lambda}_{a,1kmn}A_{1k'm'n'}^{a,\lambda'} =0\, ,\nonumber
\eea
where in the last equality we used Proposition \ref{circbasis2}. Similarly, we obtain $\Omega_0( A_1^-, A_2^-)=0$  by taking the complex conjugate. Therefore, $\Omega_0(JA_1,JA_2)=\Omega_0(i A_1^+,-i A_2^-)+\Omega_0(-i A_1^-,i A_2^+)=\Omega_0( A_1^+,A_2^-)+\Omega_0(A_1^-,A_2^+)=\Omega_0(A_1,A_2)$. This property further implies the identity $\Omega_0(A_1, J A_2)=\Omega_0(JA_1, J^2 A_2)=-\Omega_0(J A_1,  A_2)=\Omega_0(  A_2, J A_1)$ for any  $A_1, A_2\in \Gamma$.

The pair ($\Omega_0$, $J$) endows our phase space with  the (dimensionless) hermitian inner product:
\bea
\langle A_1,A_2\rangle_0=\frac{1}{2\hbar}\left[\Omega_0(A_1,JA_2)+i\Omega_0(A_1,A_2) \right] \in \mathbb C.
\eea
More precisely,  $\langle A_1,A_2\rangle_0 =\overline{ \langle A_2,A_1\rangle_0}$, and $\langle  A_1,  (a+b J)A_2 \rangle_0  = a \langle A_1, A_2\rangle_0 + i b \langle A_1, A_2\rangle_0$, which follows directly from the properties mentioned above. Furthermore, since
\bea
\Omega_0( A_1^+,i A_2^+)&=& i\Omega_0( A_1^+, A_2^+)=0\, ,\\ 
\Omega_0( A_1^+,-i A_2^-)&=& -i\Omega_0( A_1^+, A_2^-)\, ,\\ 
\Omega_0( A_1^-,i A_2^+)&=& i\Omega_0( A_1^-, A_2^+)\, ,\\ 
\Omega_0( A_1^-,-i A_2^-)&=& -i\Omega_0( A_1^-, A_2^-)=0\, ,
\eea
we have $\langle A_1,A_2\rangle_0=\frac{i\Omega_0( A_1^-, A_2^+)}{\hbar}$. Choosing a convenient normalization of the mode functions (\ref{ar})-(\ref{al}), the inner product takes the simple form
\bea
\langle A_1,A_2\rangle_0=R\sum_{n=1}^{\infty}\sum_{m=-\infty}^{\infty}\int_{-\infty}^{\infty} dk \sum_{\lambda=1,2} \overline{z_{1kmn}^{\lambda, 1}} z_{1kmn}^{\lambda, 2} \, , \label{standardinnerp}
\eea
which  is positive definite, $\langle A,A\rangle_0\geq 0$ (and $\langle A,A\rangle_0=0$ only when $z^{\lambda}_{1kmn}=0$, i.e. $A=0$).

In the quantum theory, the complex vector space  $(\Gamma,J,\langle\cdot, \cdot \rangle_0)$ constitutes the one-particle Hilbert space, and forms the basis for  Fock quantization at early times.

\subsection{Classical covariant phase space (self-dual variables)}

As a next step, let us repeat the covariant phase-space construction using the self-dual variables introduced earlier. This formulation will later make the connection with the chiral anomaly more transparent.

Any self-dual  solution of the source-free Maxwell equations with the boundary conditions specified before can be written as a linear combination of the mode solutions (\ref{ar}) previously derived. Explicitly, 
\bea
A^R_a&=& \sum_{h=\pm 1}\sum_{n=1}^{\infty} \sum_{m=-\infty}^{\infty}\int_{-\infty}^{\infty} dk  \, z_{h k m n}^{R}A_{a,hkmn}^R \label{ARb2}\\
&=&\sum_{n=1}^{\infty} \sum_{m=-\infty}^{\infty}\int_{-\infty}^{\infty} dk\left[ z_{1 k m n}^{R}A_{a,1kmn}^R+ \overline{ z_{1 k m n}^{L}} \, \overline{ A_{a,1kmn}^L} \right]  \nonumber \, ,
\eea
where $z_{h k m n}^{R}$ are some complex coefficients.  The second line follows from the relation between the right- and left-handed modes  (\ref{relationRL}), and $z_{h k m n}^{L}$ is defined by $z_{h k m n}^{L}:=(-1)^m\overline{z_{-h -k -m n}^{R}}$. The notation ``L'' is conveniently chosen so that the antiself-dual potential, which is obtained by complex conjugating $A^R_a$, reads
\bea\label{ALb2}
A^L_a&=&\sum_{n=1}^{\infty} \sum_{m=-\infty}^{\infty}\int_{-\infty}^{\infty} dk\left[ z_{1 k m n}^{L}A_{a,1kmn}^L+ \overline{ z_{1 k m n}^{R}} \, \overline{ A_{a,1kmn}^R} \right]   \\
&=& \sum_{h=\pm 1}\sum_{n=1}^{\infty} \sum_{m=-\infty}^{\infty}\int_{-\infty}^{\infty} dk  \, z_{h k m n}^{L}A_{a,hkmn}^L \nonumber\, .
\eea
Hence, a self-dual configuration is entirely determined by the set of right-handed amplitudes $z_{h k m n}^{R}$, the left-handed ones being fixed by complex conjugation.

 The (right-handed) covariant phase space is defined as the real vector space $\Gamma_R$  spanned by the (complex-valued) basis solutions $\left\{A^{R}_{a,hkmn}(x)\right\}$. Each element has the form (\ref{ARb2}) and can be represented by the collection of Bargmann-Segal coordinates   $\{z^R_{1kmn},z^L_{1kmn}\}_{k\in \mathbb R,m\in \mathbb Z,n\in \mathbb N}$, assumed to decay sufficiently fast for large $|k|, |m|, n$ to ensure convergence. The canonical symplectic structure $\Omega$ on this covariant phase space $\Gamma_R$ is (see Eq.~(\ref{symplecticfundamental}) in Sec. II)
\bea \label{sympR}
\Omega(A_R^1,A_R^2)  :=\frac{1}{2} \int_\Sigma d\Sigma \, {\rm Re} \, (\overline{A^1_R}  H_R^2-A^2_R  \overline{H_R^1})\in \mathbb R\, , 
\eea
where $H_{R, a}=n^bF^+_{ba}$ with $n_b=-\frac{1}{\sqrt{\nabla_a t\nabla^a t}}\nabla_b t=-\nabla_b t$ the unit future-directed timelike normal to the Cauchy slice $\Sigma_t=\{t={\rm constant}\}$. In geometric units $G=c=1$, the potential $A_R$ has dimensions of $\sqrt{\hbar}/L$ and therefore the symplectic product  has dimensions of $\hbar$, as it should.

This bilinear form is antisymmetric, $\Omega(A_R^1,A_R^2)=-\Omega(A_R^2,A_R^1)$, and also real-valued.
Moreover, it coincides with the usual symplectic structure obtained for real potentials. Indeed, by writing $A=\sqrt{2}{\rm Re}\, A_R$ and $Z=\sqrt{2}{\rm Im}\, A_R$ one can express the integral as
\bea
\Omega(A_R^1,A_R^2)  &=&\frac{1}{2} \int_\Sigma d\Sigma \, {\rm Re} \,n_b (A^1_{a,L}  {^+F}^{ba, 2}-A^2_{a,R}  {^-F}^{ba, 1})\nonumber\\
&=& \frac{1}{2} \int_\Sigma d\Sigma \, {\rm Re} \, (\overline{A^1_R}\wedge  {^*d}A_R^2-{A^2_R}\wedge  {^*d}\overline A_R^1)\nonumber\\
&=& \frac{1}{2} \int_\Sigma d\Sigma \, {\rm Im} \, (\overline{A^1_R}\wedge  {d}A_R^2+{A^2_R}\wedge  {d}\overline A_R^1)\nonumber\\
&=& \frac{1}{4} \int_\Sigma d\Sigma \,  ({A^1}\wedge  {d}Z^2-{Z^1}\wedge  {d}A^2-{A^2}\wedge  {d}Z^1+{Z^2}\wedge  {d}A^1)\nonumber\\
&=& \frac{1}{2} \int_\Sigma d\Sigma \,  ({A^1}\wedge  {d}Z^2-{A^2}\wedge  {d}Z^1-\frac{1}{2}d({A^2}\wedge  Z^1+{Z^2}\wedge  A^1))\nonumber\\
&=& \frac{1}{2} \int_\Sigma d\Sigma \,  ({A^1}\wedge  {^*}F^2-{A^2}\wedge  {^*F}^1)- \frac{1}{4}\int_{\{\rho=R,  t=t_0\}}  ({A^2}\wedge  Z^1+{Z^2}\wedge  A^1)\nonumber\\
&=& \Omega_0(A^1,A^2)+\frac{1}{4}\int_{\{\rho=R, t=t_0\}}dS \, \phi^{[a} z^{b]}(Z_a^1  A_b^2-Z_a^2 A_b^1)\nonumber \\
&=&\Omega_0(A^1,A^2)\, ,
\eea
where  $S$ denotes the cylindrical boundary of the waveguide at $t=t_0$, with unit radial normal $\rho_a=\nabla_a \rho$. In the last line the surface term vanishes because $ z^a\, A_a(t,\rho=R,\theta,\phi)=0$, $z^a\, Z_a(t,\rho=R,\theta,\phi)=0$. Consequently, the symplectic product is conserved in time (\ref{sympR}), $\partial_t \Omega(A_R^1,A_R^2)=0$. 

In coordinates, one may equivalently write
\bea \label{sympR2}
\Omega(A_R^1,A_R^2)  &=&\frac{1}{2}\int_\Sigma d\rho  dz d\phi \, \rho \, {\rm Re} \, (\overline{A^1_R} \dot A_R^2-{A^2_R} \dot{\overline{A_R^1}})  \in \mathbb R\, , 
\eea
which follows from the Lorentz gauge, and using $n^aA_{R.a}(\rho=R)=0$ after integrating by parts.

As in the previous subsection, we can now decompose each solution as $A_R=A^{+}_R+A^{-}_R$, where the positive- $A^{+}_R$ and negative- $A^{-}_R$ frequency parts are defined with respect to the timelike KVF $u=\partial/\partial t$. Then,  the linear map $J: \Gamma_R \to \Gamma_R$ defined by $JA_R^\pm=\pm i A_R^\pm$, satisfies $JA_R=i A_R^+-i A_R^-$ and thus $J^2=-\mathbb I$, endowing the phase space $\Gamma_R$ with a natural complex structure. The pair $(\Gamma_R, J)$ becomes a complex vector space, with complex multiplication $(a +j b) A_R := a A_R+ b J A_R$ for any $A_R\in \Gamma_R$.

This complex structure $J$ is compatible with the symplectic structure, in the sense that $\Omega(JA_R^1, JA_R^2)=\Omega(A_R^1,A_R^2)$. More precisely, we have $\Omega(JA_R^1,JA_R^2)=\Omega(i A_R^{1,+},i A_R^{2,+})+\Omega(i A_R^{1,+},-i A_R^{2,-})+\Omega(-i A_R^{1,-},i A_R^{2,+})+\Omega(-i A_R^{1.-},-i A_R^{2,-})$, where now
\bea
\Omega(i A_R^{1,+},i A_R^{2,+})&=& \Omega( A_R^{1,+},A_R^{2,+})\label{c1}\, ,\\ 
\Omega(i A_R^{1,+},-i A_R^{2,-})&=&-\Omega(A_R^{1,+}, A_R^{2,-})\label{c2}\, ,\\ 
\Omega(-i A_R^{1,-},i A_R^{2,+})&=&-\Omega( A_R^{1,-}, A_R^{2,+})\label{c3}\, ,\\ 
\Omega(-i A_R^{1.-},-i A_R^{2,-})&=& \Omega( A_R^{1.-}, A_R^{2,-})\label{c4}\, .
\eea
which follows from (\ref{sympR}). Now
\bea 
\Omega( A_R^{1,+}, A_R^{2,-}) = \frac{1}{2}\int dk \int dk' \sum_{mn} \sum_{m'n'}{\rm Re}\, \overline{z_{1,1kmn}^R}\overline{z^L_{2,1k'm'n'}}\int_\Sigma d\rho  dz d\phi \, \rho \,  i (\omega_{k'm'n'}-\omega_{kmn})\overline{A_{a, 1kmn}^R}\overline{A^{a,L}_{1k'm'n'}}=0\nonumber \, ,
\eea 
which vanishes by virtue of  Proposition \ref{circbasis2} and the fact that $\omega_{-k-mn}=\omega_{kmn}$. As explained in the previous subsection, this condition is enough to make the complex and symplectic structures on the phase space  compatible.  As a consequence of this compatibility, we find the identity $\Omega(A_R^1, J A_R^2)=\Omega(JA_R^1, J^2 A_R^2)=-\Omega(J A_R^1,  A_R^2)=\Omega(  A_R^2, J A_R^1)$ for any field configurations $A_R^1, A_R^2\in \Gamma_R$.

{\it Remark:} we can also decompose the left-handed field as $A_L=A^{+}_L+A^{-}_L$, where $A^{+}_L$ is  the positive-frequency part and $A^{-}_L$  the negative-frequency part with respect to the timelike KVF $u=\partial/\partial t$.  In this case $\overline{A_R^{\pm}}=A_L^{\mp}$.
 
The pair ($\Omega, J$) induces now a natural (dimensionless) hermitian inner product on the phase space $\Gamma_R$,
\bea
\langle A_R^1,A_R^2\rangle=\frac{1}{2\hbar}\left[\Omega(A_R^1,JA_R^2)+i\Omega(A_R^1,A_R^2) \right]. \label{innerR}
\eea
Namely,  $\langle A_R^1,A_R^2\rangle =\overline{ \langle A_R^2,A_R^1\rangle}$ and 
\bea
\langle  A_R^1,  (a+b J)A_R^2 \rangle  = a \langle A_R^1, A_R^2\rangle + i b \langle A_R^1, A_R^2\rangle \, ,\label{antilinearity}
\eea
 for any pair of real numbers $a,b$.  To see the positive-definiteness, first note that $\Omega(A_R^1,A_R^2)=\Omega( A_R^{1,+},A_R^{2,+})+\Omega(A_R^{1,+}, A_R^{2,-})+\Omega(A_R^{1,-}, A_R^{2,+})+\Omega( A_R^{1.-}, A_R^{2,-})=\Omega( A_R^{1,+},A_R^{2,+})+\Omega( A_R^{1.-}, A_R^{2,-})$.
To obtain an explicit expression, we compute
\bea
\Omega( A_R^{1,+}, A_R^{2,+}) &=& \frac{1}{2} \int dk \int dk'\sum_{mn} \sum_{m'n'}{\rm Re}\, \overline{z^R_{1,1kmn}} z^R_{2,1k'm'n'} \int_\Sigma d\rho  dz d\phi \, \rho \,  (-i) (\omega_{k'm'n'}+\omega_{kmn})  \overline{A^R_{a,1kmn}} A^{R,a}_{1k'm'n'}   \nonumber \, ,\\
\Omega( A_R^{1,-}, A_R^{2,-}) &=& \frac{1}{2}  \int dk \int dk'\sum_{mn} \sum_{m'n'}{\rm Re}\, {z^L_{1,1kmn}} \overline{z^L_{2,1k'm'n'}} \int_\Sigma d\rho  dz d\phi \, \rho \,  (i) (\omega_{k'm'n'}+\omega_{kmn})  {A^L_{a,1kmn}} \overline{A^{L,a}_{1k'm'n'}}  \nonumber\, .
\eea
Using Proposition \ref{circbasis2} for the  integrals, the sum of these two terms gives 
\bea
\Omega( A_R^{1,+}, A_R^{2,+}) + \Omega( A_R^{1,-}, A_R^{2,-}) &=&\int dk  \sum_{mn}{\rm Re}\,(\overline{z^R_{1,1kmn}} z^R_{2,1kmn}+ \overline{z^L_{1,1kmn}} z^L_{2,1kmn})|A_{kmn}|^2 (-i)2\pi^2 \omega_{kmn} R^2   {J'}_m(j_{mn})^2 \nonumber\\
&=& \int dk \sum_{mn}{\rm Im}\,(\overline{z^R_{1,1kmn}} z^R_{2,1kmn}+ \overline{z^L_{1,1kmn}} z^L_{2,1kmn})|A_{kmn}|^2 2\pi^2 \omega_{kmn} R^2   {J'}_m(j_{mn})^2  \label{aux01}\, ,
\eea
where $A_{kmn}$ is an arbitrary normalization factor, to be fixed.
 On the other hand,  $\Omega(A_R^1,JA_R^2)=\Omega( A_R^{1,+},i A_R^{2,+})+\Omega( A_R^{1,+},-i A_R^{2,-})+\Omega(A_R^{1,-},i A_R^{2,+})+\Omega( A_R^{1.-},-i A_R^{2,-})$. Now, using Proposition \ref{circbasis2} we can get
\bea
\Omega( A_R^{1,+}, -i A_R^{2,-}) &=&\frac{1}{2} \int dk \int dk' \sum_{mn} \sum_{m'n'}{\rm Re}\, \overline{z_{1,1kmn}^R}\,\overline{z^L_{2,1k'm'n'}}\int_\Sigma d\rho  dz d\phi \, \rho \,   (\omega_{k'm'n'}-\omega_{kmn})  \overline{A_{a, 1kmn}^R}\overline{A^{a,L}_{1k'm'n'}}=0 \nonumber\, ,
\eea
and similarly for $\Omega( A_R^{1,-}, i A_R^{2,+})  = - \Omega( i A_R^{2,+}, A_R^{1,-})= - \Omega(  A_R^{2,+}, -i A_R^{1,-})=0$. Furthermore,
\bea
\Omega( A_R^{1,+}, iA_R^{2,+}) &=& \frac{1}{2} \int dk \int dk'\sum_{mn} \sum_{m'n'}{\rm Re}\, \overline{z^R_{1,1kmn}} z^R_{2,1k'm'n'} \int_\Sigma d\rho  dz d\phi \, \rho \, (\omega_{k'm'n'}+\omega_{kmn})  \overline{A^R_{a,1kmn}} A^{R,a}_{1k'm'n'}   \nonumber\, ,\\
\Omega( A_R^{1,-}, -i A_R^{2,-}) &=& \frac{1}{2} \int dk \int dk'\sum_{mn} \sum_{m'n'}{\rm Re}\, {z^L_{1,1kmn}} \overline{z^L_{2,1k'm'n'}} \int_\Sigma d\rho  dz d\phi \, \rho \,   (\omega_{k'm'n'}+\omega_{kmn})  {A^L_{a,1kmn}} \overline{A^{L,a}_{1k'm'n'}}  \nonumber\, ,
\eea
and, again, using Proposition \ref{circbasis2} for the  integrals, the sum of these two terms gives 
\bea
\Omega( A_R^{1,+}, i A_R^{2,+})+\Omega( A_R^{1,-}, -i A_R^{2,-}) & = &\int dk \sum_{mn}{\rm Re}\,(\overline{z^R_{1,1kmn}} z^R_{2,1kmn}+ \overline{z^L_{1,1kmn}} z^L_{2,1kmn})|A_{kmn}|^2 2\pi^2\omega_{kmn} R^2  {J'}_m(j_{mn})^2 \label{aux02}\, .
\eea
Combining these results, the inner product (\ref{innerR}) takes the form
\bea
\langle A_R^1,A_R^2\rangle& = & \frac{1}{2\hbar}\int dk\sum_{mn}\,(\overline{z^R_{1,1kmn}} z^R_{2,1kmn}+ \overline{z^L_{1,1kmn}} z^L_{2,1kmn})|A_{kmn}|^2 2\pi^2\omega_{kmn} R^2   {J'}_m(j_{mn})^2\, .
\eea
Choosing the normalization factor
\bea
A_{kmn} &=&\frac{\sqrt{\hbar}}{\pi\sqrt{ \omega_{kmn}R}} \frac{1}{{J'}_m(j_{mn})} \label{normalization} \, ,
\eea
we finally obtain
\bea \label{inner2}
\langle A_R^1,A_R^2\rangle&=&R\int_{-\infty}^{\infty} dk\sum_{m=-\infty}^{\infty}\sum_{n=1}^{\infty}\,(\overline{z^R_{1,1kmn}} z^R_{2,1kmn}+ \overline{z^L_{1,1kmn}} z^L_{2,1kmn})\\
&=& R\int_{-\infty}^{\infty} dk\sum_{m=-\infty}^{\infty}\sum_{n=1}^{\infty}\sum_{\lambda= 1,2} \overline{z_{1,1kmn}^\lambda} z_{2,1kmn}^\lambda=\langle A_1,A_2\rangle_0 \, .\nonumber
\eea
which coincides with the inner product (\ref{standardinnerp}) derived in the standard real-variable formulation. This inner product is positive definite and preserved in time for any pair $A_R^1, A_R^2\in \Gamma_R$, as expected from its definition (\ref{innerR}), where only $\Omega$ is involved. With the chosen normalization (\ref{normalization}), the Bargmann coefficients  $z_{kmn}^\lambda$ in (\ref{ARb2})  are dimensionless, ensuring dimensional consistency for the inner product. The properly normalized vector modes (\ref{ar})-(\ref{al}) yield
\bea
A_{a,hkmn}^R&=&\frac{ \sqrt{\hbar} \,e^{-i(h\omega_{kmn}t+kz+m\phi)}}{\pi\sqrt{ R\, \omega_{kmn}}J'_m(j_{mn})}\left[  J_{m+1}( j_{mn}\rho/R) \bar \Bm_a+i\frac{ R(h\omega_{kmn}+k)}{j_{mn} }\Bell_a J_m(  j_{mn}\rho/R) \right] \label{normalizedR}\, ,\\
A_{a,hkmn}^L &=&\frac{ \sqrt{\hbar} \,e^{-i(h\omega_{kmn}t+kz+m\phi)}}{\pi\sqrt{ R\, \omega_{kmn}}J'_m(j_{mn})}\left[ J_{m-1}(  j_{mn}\rho/R)  \Bm_a-i \frac{ R(h\omega_{kmn}+k)}{j_{mn} }\Bell_a J_m( j_{mn}\rho/R) \right] \, .\label{normalizedL}
\eea

In summary, the self-dual formulation yields the same phase-space structure and inner product as the standard quantization in terms of real potentials, but with a decomposition that isolates the chiral sectors. This will be crucial in the subsequent analysis of the anomaly and of the Bogoliubov transformations connecting the early- and late-time vacua.

\subsection{Quantization: construction of the in Fock space}

The completion of the vector space $(\Gamma_R, J)$ with respect to the inner product (\ref{innerR}) defines the 1-particle Hilbert space $\mathcal H$. From $\mathcal H$,  we can construct the bosonic Fock space $\mathcal F$ in the standard way. Elements of $\mathcal F$ are interpreted physically as quantum states with arbitrary photon occupation numbers in the allowed mode spectrum of the waveguide.

The quantized electromagnetic field is then represented by an operator-valued distribution acting on this $\mathcal F$. Using the orthonormal mode basis derived in the previous subsection, the quantum field  takes the form (compare with \eqref{ARb2})
\bea
\hat A^R_a=\sum_{n=1}^{\infty} \sum_{m=-\infty}^{\infty}\int_{-\infty}^{\infty} dk\left[ a_{1 k m n}^{R}A^R_{a, 1 k m n}+ a_{1 k m n}^{L, \dagger} \, \overline{ A^{L}_{a,  1k m n} } \right]  \label{inrep}\, ,
\eea 
where $a_{1 k m n}^{L, \dagger}$, $a_{1 k m n}^{R, \dagger}$ and $a_{1 k m n}^{L}$, $a_{1 k m n}^{R}$ are creation and annihilation operators, respectively, for left- and right-handed photons. They obey the canonical commutation relations, $[a_{1 k m n}^{\lambda},a_{1 k' m' n'}^{\lambda', \dagger}]= R^{-1}\delta(k-k')\delta_{m m'}\delta_{n n'}\delta_{\lambda, \lambda'}$, with all other commutators vanishing.  This normalization ensures that the equal-time canonical commutation relations between the field $\hat A^R_a$ and its conjugate momentum $\Pi_a^R=- H_a^{L}=-n^aF^-_{ab}$ (where we recall $n_a=-\nabla_a t$ is the unit normal to the spacelike hypersurfaces $\Sigma_t$) are satisfied. To verify this explicitly, one can compute the commutator  using the mode expansion above
\bea
[A^R_a(t, \vec x),\Pi^R_b(t, \vec x\, ')]&=&-  \sum_{n=1}^{\infty} \sum_{m=-\infty}^{\infty}\int_{-\infty}^{\infty} dk \,  R^{-1} \left[A^R_{a,1  k m n}(x) \overline{H^R_{b, 1 k m n}(x')}-H^L_{b,  1k m n}(x') \overline{A^L_{a, 1 k m n}(x)}\right] \nonumber\, .
\eea
To check the standard result, let us evaluate the trace of this equation only. Using (\ref{ar})-(\ref{al}) and (\ref{hr})-(\ref{hl}), and the normalization (\ref{normalization}), we get
\bea
[A^R_a(t,\vec x),\Pi^{R,a}(t, \vec x\, ')]&=& -\sum_{n=1}^{\infty} \sum_{m=-\infty}^{\infty}\int_{-\infty}^{\infty} dk \, R^{-1}\left[A^R_{a,1  k m n}(x) \overline{H^{R,a}_{1 k m n}(x')}-H^L_{a,  1k m n}(x') \overline{A^{L,a}_{1 k m n}(x)}\right]\nonumber\\
&=&\sum_{kmn} \, \frac{i\hbar e^{-i[k(z-z')+m(\phi-\phi')]}}{2\pi^2 \omega_{kmn}R^2 J'^2}\left[(J_{m+1}(x_{mn})J_{m+1}(x'_{mn})+J_{m}(x_{mn})J_{m}(x'_{mn}))\omega_{kmn} \right.\nonumber\\
&& \hspace{4.7cm}\left. -k(J_{m+1}(x_{mn})J_{m+1}(x'_{mn})-J_{m}(x_{mn})J_{m}(x'_{mn}))\right]\nonumber\\
&&+\sum_{kmn} \, \frac{i\hbar e^{i[k(z-z')+m(\phi-\phi')]}}{2\pi^2 \omega_{kmn}R^2 J'^2}\left[(J_{m-1}(x_{mn})J_{m-1}(x'_{mn})+J_{m}(x_{mn})J_{m}(x'_{mn}))\omega_{kmn} \right.\nonumber\\
&& \hspace{4.7cm}\left. -k(J_{m-1}(x_{mn})J_{m-1}(x'_{mn})-J_{m}(x_{mn})J_{m}(x'_{mn}))\right]\nonumber\\
&=& \sum_{m=-\infty}^{\infty}\int_{-\infty}^{\infty} dk\frac{i \hbar  R^2  \delta(\rho-\rho')  \left[  e^{-i[k(z-z')+m(\phi-\phi')]} +  e^{i[k(z-z')+m(\phi-\phi')]}\right]
}{2\rho (\pi^2 R^2 ) }  \nonumber\\
&=&\frac{i 4\hbar    \delta(\rho-\rho') \delta(\phi-\phi')\delta(z-z')   
}{\rho   } \nonumber\\
&=&4 \, i\hbar\,    \delta^3(\vec x-\vec x\, ') \nonumber\, ,
\eea
\begin{comment}
\bea
[A^R_a(t,\vec x),\Pi^{R,a}(t, \vec x\, ')]=i  \sum_{n=1}^{\infty} \sum_{m=-\infty}^{\infty}\int_{-\infty}^{\infty} dk \,\omega_{kmn} R^{-1}\left[A^R_{a,1  k m n}(x) \overline{A^{R,a}_{1 k m n}(x')}+A^L_{a,  1k m n}(x') \overline{A^{L,a}_{1 k m n}(x)}\right]\nonumber\\
= \frac{ \hbar i \sum_{n=0}^{\infty} \sum_{m=-\infty}^{\infty}\int_{-\infty}^{\infty} dk  \left[ J_{m+1}\left( x_{mn}\right)J_{m+1}\left(x_{mn}'\right) e^{-i[k(z-z')+m(\phi-\phi')]} +  J_{m-1}\left(x_{mn}\right)J_{m-1}\left(x_{mn}'\right)e^{i[k(z-z')+m(\phi-\phi')]}\right]
}{\pi^2 R^2\, {J'}^2_m(j_{mn})}  \nonumber\\
= \frac{ \hbar i R  \delta(\rho-\rho') \sum_{m=-\infty}^{\infty}\int_{-\infty}^{\infty} dk \left[  e^{-i[k(z-z')+m(\phi-\phi')]} +  e^{i[k(z-z')+m(\phi-\phi')]}\right]
}{2\rho\pi^2 R }  \nonumber\\
=\frac{2(2\pi) \hbar i   \delta(\rho-\rho')\delta(\phi-\phi')\delta(z-z') }{\rho\pi  }  \nonumber\\
=\frac{4 \hbar i   \delta(\rho-\rho')\delta(\phi-\phi')\delta(z-z') 
}{\sqrt{h}  }  \nonumber\\
=4 \, i\hbar\,    \delta^3(\vec x-\vec x\, ') \nonumber
\eea
\end{comment}
where  $x_{mn}\equiv j_{mn}\rho/R$,  and $\sum_{kmn}$ denotes $ \sum_{n=1}^{\infty} \sum_{m=-\infty}^{\infty}\int_{-\infty}^{\infty} dk$. In  the third equality we used the completeness identities (\ref{completeness}) and (\ref{completenessid2}); in the fifth equality we rewrote $\rho$ in terms of the determinant of the spatial metric $h_{ab}$. This result implies
\bea
[A^R_a(t, \vec x),\Pi^R_b(t,\vec x\, ')]&=& i\hbar\, (\eta_{ab}-L_{ab})   \delta^3(\vec x-\vec x\, ')\, ,
\eea
where $L_{ab}$  represents a longitudinal (pure-gauge) contribution, which has vanishing trace, $\eta^{ab}L_{ab}=0$, and carries no physical degrees of freedom. 

The physical degrees of freedom reside in the two transverse photon polarizations, which can be extracted by contracting the  field $A_a^R$ with the Newman-Penrose polarization  vectors $\Bm^a$ and $\bar \Bm^a$. Defining $A^R_2:=\Bm^aA_a^R$ and $\Pi^R_2:=\bar \Bm^a \Pi^R_a$, which are the two relevant (canonically conjugate) Newman-Penrose scalars,  one finds $[A^R_2(t, \vec x),\Pi^R_2(t,\vec x\, ')]=i\hbar\,\delta^3(\vec x-\vec x\, ')$. By taking the Hermitian conjugate, the same commutation relation follows for the left-handed field operator, $[A^L_2(t, \vec x),\Pi^L_2(t,\vec x\, ')]=i\hbar\,\delta^3(\vec x-\vec x\, ')$. $A_2^R$ and $A_2^L$ are two  scalar fields with spin weight 1 and $-1$, respectively, showing that these variables capture the two physical photon polarizations.

It is worth noting that the commutation relations for the creation and annihilation operators are dimensionally consistent. This follows from the fact that, with the normalization adopted in (\ref{normalization}), the coefficients $z_{1kmn}^\lambda$ ---and hence the operators $a_{1kmn}^\lambda$--- are dimensionless (see the comment below (\ref{inner2})).

\section{Quantization at late times}

We will refer to the expansion (\ref{inrep}) as the {\it in} representation of the quantum field, since it corresponds to the basis of mode functions defined in the static configuration of the waveguide at early times. The {\it in} vacuum state 
is defined by $a_{1 kmn}^R | {\rm in} \rangle=a_{1 kmn}^L | {\rm in}\rangle=0$ for all $k,m,n$. This state represents the absence of photons with respect to static observers. In this  section we  introduce an analogous {\it out} representation associated with the late-time configuration of the accelerating waveguide, and corresponding {\it out} vacuum $| {\rm out} \rangle$, and relate  both constructions later through a Bogoliubov transformation.

\subsection{A  late-time mode basis for the classical covariant phase space}

As indicated in Sec.~\ref{seciv}, particularly in  (\ref{ARb2}), any self-dual  solution of the source-free Maxwell equations that satisfies our boundary conditions can be expanded in the early-time (``{in}'') circular basis of right- and left-handed modes $A^{R,{\rm in}}_{a, 1 k m n}$, $\overline{A^{L,{\rm in}}_{a, 1 k m n}}$ given in (\ref{ar})-(\ref{al}). This basis  is natural at early times because  the mode functions oscillate with positive or negative frequency, respectively, with respect to the  timelike Killing field $t^a$ associated with inertial observers. Explicitly,
\bea
 i t^b\nabla_b \left.  A^{R,{\rm in}}_{\bar 2, 1 k m n}\right|_{t\to-\infty} &=& \Bm^b (i t^b \nabla_b \left. A^{R,{\rm in}}_{a, 1 k m n})\right|_{t\to-\infty} = \omega_{kmn}\Bm^b\left. A^{R,{\rm in}}_{a,1 k m n}\right|_{t\to-\infty}\, ,\\
 i t^b\nabla_b \left.  \overline{A^{L,{\rm in}}_{ 2, 1 k m n}}\right|_{t\to-\infty} &=& \bar \Bm^b (i t^b \nabla_b \left. \overline{A^{L,{\rm in}}_{a, 1 k m n}})\right|_{t\to-\infty} = -\omega_{kmn}\bar \Bm^b\left. \overline{A^{L,{\rm in}}_{a,1 k m n}}\right|_{t\to-\infty}\, .
\eea
However,  {\it not all} the basis modes $A^{R}_{a, 1k m n}$ in (\ref{ARb2}) oscillate with positive-frequency   when measured by the accelerated observers at late times, whose 4-velocity vector is  $u=\partial_t$ rather than $t^a$. This is because, although $u^a|_{t\to -\infty}\sim t^a$ at early times, at late times one has  $u^a|_{t\to \infty}\sim \gamma (t^a + v_0 z^a)+\rho \Omega_0 \gamma \phi^a$. Using (\ref{framedragging}), or equivalently (\ref{arlate})-(\ref{allate}), one  finds
\bea
 i u^b\nabla_b \left.  A^{R,{\rm in}}_{\bar 2, 1 k m n}\right|_{t\to\infty} &=& \Bm^{0,b} (i u^b \nabla_b \left. A^{R,{\rm in}}_{a, 1 k m n})\right|_{t\to\infty} = \gamma (\omega_{kmn}-m \Omega_0 -k v_0-\Omega_0)\Bm^{0,b}\left. A^{R,{\rm in}}_{a,1 k m n}\right|_{t\to\infty}\, ,\\
 i u^b\nabla_b \left.  \overline{A^{L,{\rm in}}_{ 2, 1 k m n}}\right|_{t\to\infty} &=&  \Bm^{0,b} (i u^b \nabla_b \left. \overline{A^{L,{\rm in}}_{a, 1 k m n}})\right|_{t\to\infty} = \gamma (-\omega_{kmn}+m \Omega_0 +k v_0-\Omega_0) \Bm^{0,b}\left. \overline{A^{L,{\rm in}}_{a,1 k m n}}\right|_{t\to\infty}\, .
\eea
Thus, it may happen that $(k,m,n)$ is such that $\omega_{kmn}-m \Omega_0 -k v_0- \Omega_0<0$ despite that  $\omega_{kmn}>0$. This motivates a second decomposition of the general solution $A^R_a\in \Gamma_R$ in a basis of field modes $A^{R/L,{\rm out}}_{a, h k m n}$ that are of positive-frequency  with respect to the timelike KVF $u=\frac{\partial}{\partial t}$ {\it at late times}:
\bea
A^R_a&=& \sum_{h=\pm 1}\sum_{n=1}^{\infty} \sum_{m=-\infty}^{\infty}\int_{-\infty}^{\infty} dk  \, z_{h k m n}^{R,{\rm out}}A_{a,hkmn}^{R,{\rm out}} \label{ARb3}\\
&=&\sum_{hkmn\in H_h^{+>}}  z_{h k m n}^{R,{\rm out}}A_{a,hkmn}^{R,{\rm out}}+\sum_{hkmn\in H_h^{+<}} { z_{h k m n}^{R,{\rm out}}} \, { A_{a,hkmn}^{R,{\rm out}}}   \nonumber \\
&=&\sum_{hkmn\in H_h^{+>}}  z_{h k m n}^{R,{\rm out}}A_{a,hkmn}^{R,{\rm out}}+\sum_{hkmn\in H_h^{->}} \overline{ z_{h k m n}^{L,{\rm out}}} \, \overline{ A_{a,hkmn}^{L,{\rm out}}}   \nonumber  \, .
\eea
In this expression the out basis modes are given by the late-time solutions (\ref{arlate})-(\ref{allate}),  the complex coefficients $(z_{h k m n}^{R,{\rm out}},z_{h k m n}^{L,{\rm out}})$ satisfy $z_{h k m n}^{L,{\rm out}}=(-1)^m\overline{z_{-h -k -m n}^{R,{\rm out}}}$, and 
\bea
H_{h}^{\pm>}:=\{(k,m,n)\in \mathbb R\times \mathbb Z \times \mathbb N   \, / \, h\sqrt{k^2+j_{mn}^2/R^2}-m\Omega_0-kv_0 \pm \Omega_0 > 0  \}\, ,\\
H_{h}^{\pm<}:=\{(k,m,n)\in \mathbb R\times \mathbb Z \times \mathbb N   \, / \,h \sqrt{k^2+j_{mn}^2/R^2}-m\Omega_0-kv_0 \pm \Omega_0 < 0  \}\, ,
\eea
define the sets $H_h^{+>}$ and $H_h^{->}$  of positive and negative-frequency with respect to $u^a$. They partition the full mode set, since $H_{ h}^{\pm >}\cup H_{ h}^{\pm <}=\mathbb R\times \mathbb Z \times \mathbb N$ and $H_{ h}^{\pm >}\cap H_{ h}^{\pm <}=\emptyset$. For $\Omega_0>0$, one finds $H_{ h}^{+ >}\cap H_{ h}^{- >}=H_{ h}^{- >}$ and  $H_{ h}^{+ <}\cap H_{ h}^{- <}=H_{ h}^{+<}$, while for $\Omega_0<0$ we get $H_{h}^{+ >}\cap H_{ h}^{- >}=H_{ h}^{+ >}$ and  $H_{ h}^{+ <}\cap H_{ h}^{- <}=H_{ h}^{-<}$. When both $\Omega_0=v_0=0$ then $H_{ h}^{\pm >}=\mathbb R\times \mathbb Z \times \mathbb N$ and $H_{ h}^{\pm <}=\emptyset$, as expected for the stationary waveguide.

The classical covariant phase space is  the same real vector space $\Gamma_R$ as  before, consisting of all  solutions $A^R_a$ of the form (\ref{ARb3}). Classically, the two decompositions (\ref{ARb3}) and (\ref{ARb2}) are equivalent, since they only differ by a change of basis in this vector space. In the quantum theory, however, the two splittings into positive and negative frequency parts lead to different complex structures and therefore to different representations of the canonical commutation relations. This difference in the quantum theory can have physical implications.

The vector space $\Gamma_R$ is  endowed with the same symplectic structure as before, specifically
\bea
\Omega(A_R^1,A_R^2)  &:=&\frac{1}{2}\int_{\Sigma_t} d\Sigma_t \, {\rm Re} \, (\overline{A^1_{R,a}}  H_{R}^{2,a}-A^2_{R,a}  \overline{H_R^{1,a}}) \label{symproduct2}\, ,
\eea
where  $H_{R, a}=n^bF^+_{ba}$ with $n_b=-\frac{1}{\sqrt{-\nabla_a t \nabla^at}}\nabla_b t=-\nabla_b t$ the unit future-directed timelike normal to $\Sigma_t=\{t={\rm constant}\}$\footnote{Note that $n^a\neq u^a$. In particular, $u^a$ is not even hypersurface orthogonal to any foliation $\Sigma$, since the Frobenius condition does not hold, $u\wedge du\neq 0$.}. Since the out basis modes are naturally defined at late times, it is convenient to evaluate this integral on a late-time hypersurface $\Sigma_t$, owing to the  time-independence of the symplectic structure (see Appendix B).  In practical calculations, it is further convenient to express (\ref{symproduct2}) in the fiducial coordinate system $\{\tilde t, \tilde \rho, \tilde \phi, \tilde z\}$. A somewhat tedious but straightforward calculation gives us a relatively simple expression:
\bea
\Omega(A_R^1,A_R^2)  &=&  \left.\frac{1}{2}\int d\tilde\rho  d\tilde z d\tilde\phi \, \tilde\rho \gamma\, {\rm Re} \, \left\{\overline{\tilde A^{1,b}_R}  (\partial_{\tilde t}+v_0 \partial_{\tilde z}) \tilde A_{R,b}^2-{\tilde A^{2,b}_R} (\partial_{\tilde t}+v_0 \partial_{\tilde z})\overline{\tilde A_{R,b}^1}  \right\}\right |_{(t \gamma^{-1}+v_0 \tilde z,\,\tilde \rho,\, \tilde \phi,\, \tilde z)}\, . \label{practic}
\eea

 In this phase space $\Gamma_R$ we have now two natural basis of positive-frequency field modes and, therefore, two natural choices of complex structures  to do the Fock quantization: $J_{\rm in}$, associated with the early-time decomposition; and  $J_{\rm out}$, associated with the late-time one.  As indicated in Sec.~\ref{seciv}, the decomposition   (\ref{ARb2}) motivates defining the complex structure $J_{\rm in}: \Gamma_R \to \Gamma_R$ by the operation $J_{\rm in} A_R=i A_R^+-i A_R^-$, where $ A_R^+$ and $ A_R^-$ are
\bea
A^+_R&=&\sum_{n=1}^{\infty} \sum_{m=-\infty}^{\infty}\int_{-\infty}^{\infty} dk \, z_{1 k m n}^{R,{\rm in} } A^{R,{\rm in}}_{a, 1k m n}\, ,\\
A^-_R&=&\sum_{n=1}^{\infty} \sum_{m=-\infty}^{\infty}\int_{-\infty}^{\infty} dk\, \overline{ z_{1 k m n}^{L, {\rm in}}} \,\overline{ A^{L,{\rm in}}_{a,1 k m n}}\, .
\eea
Because the symplectic structure (\ref{symproduct2}) is constant in time $t$, the calculations done in Sec.~\ref{seciv} at early times hold at any other time $t$. In particular, the inner product takes the constant form
\bea
\langle A_R^1,A_R^2\rangle_{\rm in} &=&R \sum_{n=1}^{\infty} \sum_{m=-\infty}^{\infty}\int_{-\infty}^{\infty} dk\, (\overline{z^{R, {\rm in}}_{1,1kmn}} z^{R, {\rm in}}_{2,1kmn}+ \overline{z^{L, {\rm in}}_{1,1kmn}} z^{L, {\rm in}}_{2,1kmn})\, , \label{inproductaux}
\eea
with normalization factor given in (\ref{normalization}).

Now, we can define the alternative complex structure $J_{\rm out}: \Gamma_R \to \Gamma_R$ by  $J_{\rm out} A_R=i A_R^+-i A_R^-$, where $ A_R^+$ and $ A_R^-$ are now, 
 \bea
 A_R^+&=&\sum_{hkmn\in H_h^{+>}}  z_{h k m n}^{R,{\rm out}}A_{a,hkmn}^{R,{\rm out}}=\sum_{(k,m,n) \in H_{1}^{+>}} z_{ 1kmn}^{R,{\rm out} } A^{R,{\rm out}}_{a, 1 kmn}+  \sum_{(k,m,n)\in H_{1}^{-<}} \overline{ z_{ 1k m n}^{L, {\rm out}}} \,\overline{ A^{L,{\rm out}}_{a,1 k m n} }\, ,\label{outpositive}\\
 A_R^-&=&\sum_{hkmn\in H_h^{->}} \overline{ z_{h k m n}^{L,{\rm out}}} \, \overline{ A_{a,hkmn}^{L,{\rm out}}}=\sum_{( k,m,n)\in H_{1}^{->}}\overline{ z_{  1kmn}^{L, {\rm out}}} \,\overline{ A^{L,{\rm out}}_{a, 1 kmn} } + \sum_{( k,m,n)\in H_{1}^{+<}} z_{  1kmn}^{R,{\rm out} } A^{R,{\rm out}}_{a, 1 kmn} \, . \label{outnegative}
 \eea
 Following the same steps as in Sec.~\ref{seciv}, we first check that this new complex structure is compatible with the symplectic product,  in the sense that $\Omega(J_{\rm out}A_R^1, J_{\rm out}A_R^2)=\Omega(A_R^1,A_R^2)$.  Taking into account the identities (\ref{c1})-(\ref{c4}), we only need to check that $\Omega( A_R^{1,+}, A_R^{2,-})=0$. To compute this, we evaluate the symplectic structure at late times in the coordiantes of the co-moving frame using equation (\ref{practic}) directly. Thus, 
 \bea
\Omega( A_R^{1,+}, A_R^{2,-}) &=& \frac{1}{2}{\rm Re}\, \sum_{i\in H_1^{+>}}\sum_{j\in H_1^{->}}\overline{z_{1,1i}^{R,{\rm out}}}\overline{z^{L,{\rm out}}_{2,1j}}\int_\Sigma d\tilde\rho  d\tilde z d\tilde\phi \, \tilde\rho\,  i (\tilde\omega_{j}-\tilde\omega_i)  \overline{\tilde A_{a, 1i}^{R,{\rm out}}}\overline{\tilde A^{L,{\rm out}}_{b,1j}}\tilde\eta^{ab}\nonumber\\
&+&\frac{1}{2}{\rm Re}\, \sum_{i\in H_1^{+>}}\sum_{j\in H_1^{+<}}\overline{z_{1,1i}^{R,{\rm out}}}{z^{R,{\rm out}}_{2,1j}}\int_\Sigma d\tilde\rho  d \tilde z d\tilde\phi \, \tilde\rho\,  i (-\tilde\omega_{j}-\tilde\omega_i)  \overline{\tilde A_{a, 1i}^{R,{\rm out}}}{\tilde A^{R,{\rm out}}_{b,1j}}\tilde\eta^{ab}\nonumber\\
&+& \frac{1}{2}{\rm Re}\, \sum_{i\in H_1^{-<}}\sum_{j\in H_1^{->}}{z_{1,1i}^{L,{\rm out}}}\overline{z^{L,{\rm out}}_{2,1j}}\int_\Sigma d\tilde\rho  d\tilde z d\tilde\phi \, \tilde\rho\,  i (\tilde\omega_{j} +\tilde\omega_i)  {\tilde A_{a, 1i}^{L,{\rm out}}}\overline{\tilde A^{L,{\rm out}}_{b,1j}}\tilde\eta^{ab}\nonumber\\
&+&\frac{1}{2}{\rm Re}\, \sum_{i\in H_1^{-<}}\sum_{j\in H_1^{+<}}{z_{1,1i}^{L,{\rm out}}}{z^{R,{\rm out}}_{2,1j}}\int_\Sigma d\tilde \rho  d\tilde z d\tilde \phi \, \tilde\rho\,  i (-\tilde\omega_{j} +\tilde\omega_i)  {\tilde A_{a, 1i}^{L,{\rm out}}}{\tilde A^{R,{\rm out}}_{b,1j}}\tilde\eta^{ab}\nonumber\\
&=&0\, , \label{consistencyJ}
\eea
where $i$ and $j$ are shorthands for $kmn$ and $k'm'n'$, respectively, and $\tilde \omega_{kmn}:=\gamma(\omega_{kmn}+v_0 k)$. The second and third lines above are each of them identically zero because, for $(kmn)\in H_1^{\pm >}$ and $(k'm'n')\in H_1^{\pm <}$, the integrals produce a term $\delta_{m,m'}\delta_{k,k'}\delta_{n,n'}$    by virtue of Proposition (\ref{circbasis2}) (setting $h=h'=1$), and  $H_1^{\pm >}\cap H_1^{\pm <}=\emptyset$. The first and fourth lines are identically zero each of them because the integrals produce a term $\delta_{m,-m'}\delta_{k,-k'}\delta_{n,n'}$ using again Proposition (\ref{circbasis2}), and $\omega_{-k,-m,n}=\omega_{k,m,n}$.\footnote{The integrals actually produce a $\delta(\tilde k-\tilde k')$ where $\tilde k=\gamma(k+v_0 \omega_{kmn})$ and $\tilde k'=\gamma(k'+v_0 \omega_{k'm'n'}')$, because the integral is over constant $t$, and $e^{-i(\omega \tilde t+k \tilde z)}=e^{-i\omega t \gamma^{-1}}e^{-i(k+v_0\omega)\tilde z}$.  However, we have $\tilde \omega^2-\tilde k^2=\omega^2-k^2=\frac{j_{mn}^2}{R^2}$. Therefore, $\tilde \omega=\sqrt{\tilde k^2+\frac{j_{mn}^2}{R^2}}$ and $\delta(\tilde k_1-\tilde k_2)$ is proportional to $\delta(\tilde\omega_1-\tilde \omega_2)$. Now, if $\tilde \omega_1=\tilde \omega_2$ and $\tilde k_1=\tilde k_2$ hold, then $0=\omega_1-\omega_2+v_0(k_1-k_2)=(\omega_1-\omega_2)(1-v_0^2)$, thus $\omega_1=\omega_2$ and finally $k_1=k_2$. Therefore, $\delta(\tilde k-\tilde k')$ implies $\delta( k- k')$ directly.  Similarly, $\delta(\tilde k+\tilde k')$ is proportional to $\delta(\tilde \omega-\tilde \omega')$, which is all what we need in this case. }

Equation (\ref{consistencyJ})  further implies $\Omega( A_R^{1,-}, A_R^{2,+}) = -\Omega( A_R^{2,+}, A_R^{1,-})=0$. As a result, we have $\Omega(J_{\rm out}A_R^1, J_{\rm out}A_R^2)=\Omega(A_R^1,A_R^2)$. Consequently, this complex structure endows our phase space  with  the (dimensionless) hermitian inner product:
\bea
\langle A_R^1,A_R^2\rangle_{\rm out}=\frac{1}{2\hbar}\left[\Omega(A_R^1,J_{\rm out}A_R^2)+i\Omega(A_R^1,A_R^2) \right]. \label{innerRout}
\eea
We can get an explicit form of this inner product by working out in detail each of the two terms above.   First note that $\Omega(A_R^1,A_R^2)=\Omega( A_R^{1,+},A_R^{2,+})+\Omega(A_R^{1,+}, A_R^{2,-})+\Omega(A_R^{1,-}, A_R^{2,+})+\Omega( A_R^{1.-}, A_R^{2,-})=\Omega( A_R^{1,+},A_R^{2,+})+\Omega( A_R^{1.-}, A_R^{2,-})$, where
\bea
\Omega( A_R^{1,+}, A_R^{2,+}) &=& \frac{1}{2}{\rm Re}\, \sum_{i\in H_1^{+>}}\sum_{j\in H_1^{+>}}\overline{z_{1,1i}^{R,{\rm out}}}{z^{R,{\rm out}}_{2,1j}}\int_\Sigma d\tilde\rho  d\tilde z d\tilde\phi \, \tilde\rho\,  i (-\tilde\omega_{j}-\tilde\omega_i) \overline{\tilde A_{a, 1i}^{R,{\rm out}}}{\tilde A^{R,{\rm out}}_{b,1j}}\tilde\eta^{ab}\nonumber\\
&+&\frac{1}{2}{\rm Re}\, \sum_{i\in H_1^{+>}}\sum_{j\in H_1^{-<}}\overline{z_{1,1i}^{R,{\rm out}}}\overline{z^{L,{\rm out}}_{2,1j}}\int_\Sigma d\tilde\rho  d\tilde z d\tilde\phi \, \tilde\rho\,  i (\tilde\omega_{j}-\tilde\omega_i) \overline{\tilde A_{a, 1i}^{R,{\rm out}}}\overline{\tilde A^{L,{\rm out}}_{b,1j}}\tilde\eta^{ab}\nonumber\\
&+& \frac{1}{2}{\rm Re}\, \sum_{i\in H_1^{-<}}\sum_{j\in H_1^{+>}}{z_{1,1i}^{L,{\rm out}}}{z^{R,{\rm out}}_{2,1j}}\int_\Sigma d\tilde\rho  d\tilde z d\tilde\phi \, \tilde\rho\,  i (-\tilde\omega_{j}+\tilde\omega_i)  {\tilde A_{a, 1i}^{L,{\rm out}}}{\tilde A^{R,{\rm out}}_{b,1j}}\tilde\eta^{ab}\nonumber\\
&+&\frac{1}{2}{\rm Re}\, \sum_{i\in H_1^{-<}}\sum_{j\in H_1^{-<}}{z_{1,1i}^{L,{\rm out}}}\overline{z^{L,{\rm out}}_{2,1j}}\int_\Sigma d\tilde\rho  d\tilde z d\tilde\phi \, \tilde\rho\,  i (\tilde\omega_{j}+\tilde\omega_i)  {\tilde A_{a, 1i}^{L,{\rm out}}}\overline{\tilde A^{L,{\rm out}}_{b,1j}}\tilde\eta^{ab}\nonumber\, ,
\eea
where again $i$ and $j$ are shorthands for $kmn$ and $k'm'n'$, respectively.
Using Proposition (\ref{circbasis2}) (with $h=h'=1$) the second and third lines are zero. Using again Proposition (\ref{circbasis2})  the first and fourth lines reduce to
\bea
\Omega( A_R^{1,+}, A_R^{2,+}) ={\rm Re}\,\left\{-\sum_{i\in H_1^{+>}}  i 2\pi^2 R^2   {J'}_m(j_{mn})^2  \overline{z_{1,1i}^{R,{\rm out}}}{z^{R,{\rm out}}_{2,1i}}\tilde\omega_i+\sum_{i\in H_1^{-<}}   i2\pi^2 R^2   {J'}_m(j_{mn})^2 \overline{z_{2,1i}^{L,{\rm out}}}{z^{L,{\rm out}}_{1,1i}}\tilde\omega_i\right\}\, .\nonumber
\eea
Similarly, we can get
\bea
\Omega( A_R^{1,-}, A_R^{2,-}) &=& \frac{1}{2}{\rm Re}\, \sum_{i\in H_1^{->}}\sum_{j\in H_1^{->}}{z_{1,1i}^{L,{\rm out}}}\overline{z^{L,{\rm out}}_{2,1j}}\int_\Sigma d\rho  dz d\phi \, \rho\,  i (\tilde\omega_{j} +\tilde\omega_i)  {\tilde A_{a, 1i}^{L,{\rm out}}}\overline{\tilde A^{L,{\rm out}}_{b,1j}}\tilde\eta^{ab}\nonumber\\
&+&\frac{1}{2}{\rm Re}\, \sum_{i\in H_1^{->}}\sum_{j\in H_1^{+<}}{z_{1,1i}^{L,{\rm out}}}{z^{R,{\rm out}}_{2,1j}}\int_\Sigma d\tilde\rho  d\tilde z d\tilde\phi \,\tilde \rho\,  i (-\tilde\omega_{j} +\tilde\omega_i)  {\tilde A_{a, 1i}^{L,{\rm out}}}{\tilde A^{R,{\rm out}}_{b,1j}}\tilde\eta^{ab}\nonumber\\
&+& \frac{1}{2}{\rm Re}\, \sum_{i\in H_1^{+<}}\sum_{j\in H_1^{->}}\overline{z_{1,1i}^{R,{\rm out}}}\overline{z^{L,{\rm out}}_{2,1j}}\int_\Sigma d\tilde\rho  d\tilde z d\tilde\phi \, \tilde\rho\,  i (\tilde\omega_{j}-\tilde\omega_i)  \overline{\tilde A_{a, 1i}^{R,{\rm out}}}\overline{\tilde A^{L,{\rm out}}_{b,1j}}\tilde\eta^{ab}\nonumber\\
&+&\frac{1}{2}{\rm Re}\, \sum_{i\in H_1^{+<}}\sum_{j\in H_1^{+<}}\overline{z_{1,1i}^{R,{\rm out}}}{z^{R,{\rm out}}_{2,1j}}\int_\Sigma d\tilde\rho  d\tilde z d\tilde\phi \,\tilde \rho\,  i (-\tilde\omega_{j} -\tilde\omega_i) \overline{\tilde A_{a, 1i}^{R,{\rm out}}}{\tilde A^{R,{\rm out}}_{b,1j}}\tilde\eta^{ab}\nonumber\, ,
\eea
and
\bea
\Omega( A_R^{1,-}, A_R^{2,-}) ={\rm Re}\left\{\sum_{i\in H_1^{->}}  i 2\pi^2 R^2   {J'}_m(j_{mn})^2\overline{z_{2,1i}^{L,{\rm out}}}{z^{L,{\rm out}}_{1,1i}}\tilde\omega_i-\sum_{i\in H_1^{+<}}  i 2\pi^2 R^2   {J'}_m(j_{mn})^2 \overline{z_{1,1i}^{R,{\rm out}}}{z^{R,{\rm out}}_{2,1i}}\tilde\omega_i\right\}\, .\nonumber
\eea
Thus,
\bea
\Omega( A_R^{1,+}, A_R^{2,+}) +\Omega( A_R^{1,-}, A_R^{2,-})&=&2\pi^2 R^2 {\rm Im}\,\left\{\sum_{i\in H_1^{+>}}  \tilde \omega_i    {J'}_m(j_{mn})^2  \overline{z_{1,1i}^{R,{\rm out}}}{z^{R,{\rm out}}_{2,1i}}+\sum_{i\in H_1^{->}}  \tilde \omega_i    {J'}_m(j_{mn})^2 \overline{z_{1,1i}^{L,{\rm out}}}{z^{L,{\rm out}}_{2,1i}} \right.\nonumber\\
&&\left. +\sum_{i\in H_1^{-<}}  \tilde\omega_i    {J'}_m(j_{mn})^2  \overline{z^{L,{\rm out}}_{1,1i}} {z_{2,1i}^{L,{\rm out}}}+\sum_{i\in H_1^{+<}}  \tilde\omega_i    {J'}_m(j_{mn})^2\overline{z^{R,{\rm out}}_{1,1i}} {z_{2,1i}^{R,{\rm out}}}\right\} \nonumber\\
&=& {\rm Im}\, \sum_{i\in H_{\rm in}}   \tilde\omega_i 2\pi^2 R^2   {J'}_m(j_{mn})^2\left[\overline{z_{1,1i}^{R,{\rm out}}}{z^{R,{\rm out}}_{2,1i}}+\overline{z_{1,1i}^{L,{\rm out}}}{z^{L,{\rm out}}_{2,1i}} \right]
\label{auxiliar1}\, .
\eea
where $H_{\rm in}\equiv \mathbb R\times \mathbb Z \times \mathbb N$.
On the other hand, $\Omega(A_R^1,JA_R^2)=\Omega( A_R^{1,+},i A_R^{2,+})+\Omega( A_R^{1,+},-i A_R^{2,-})+\Omega(A_R^{1,-},i A_R^{2,+})+\Omega( A_R^{1.-},-i A_R^{2,-})$. From (\ref{consistencyJ}) it is straightforward to see that $\Omega( A_R^{1,+},-i A_R^{2,-})=\Omega(A_R^{1,-},i A_R^{2,+})=0$. In addition, from (\ref{auxiliar1}) we find
\bea
\Omega( A_R^{1,+},i A_R^{2,+})+\Omega( A_R^{1.-},-i A_R^{2,-})&=& {\rm Re}\, \sum_{i\in H_{\rm in}}   \tilde\omega_i 2\pi^2 R^2   {J'}_m(j_{mn})^2  \left[\overline{z_{1,1i}^{R,{\rm out}}}{z^{R,{\rm out}}_{2,1i}}+\overline{z_{1,1i}^{L,{\rm out}}}{z^{L,{\rm out}}_{2,1i}} \right]\nonumber\, .
\eea
Summing up:
\bea
\langle A_R^1,A_R^2\rangle_{\rm out} &=& \frac{1}{2\hbar}\sum_{i\in H_{\rm in}}   \tilde\omega_i 2\pi^2 R^2   {J'}_m(j_{mn})^2 \left[\overline{z_{1,1i}^{R,{\rm out}}}{z^{R,{\rm out}}_{2,1i}}+\overline{z_{1,1i}^{L,{\rm out}}}{z^{L,{\rm out}}_{2,1i}} \right] \label{outinnerproductfinal}\, ,
\eea
which can be  reduced with a suitable normalization factor for $z_{hkmn}^{R,{\rm out}}$, $z_{hkmn}^{L,{\rm out}}$,
\bea
A_{kmn} &=&\frac{\sqrt{\hbar}}{\pi\sqrt{\tilde \omega_{kmn}R}} \frac{1}{{J'}_m(j_{mn})} \label{normalization2} \, ,
\eea
 as in (\ref{inner2}): 
\bea
\langle A_R^1,A_R^2\rangle_{\rm out}&=&R\int_{-\infty}^{\infty} dk\sum_{m=-\infty}^{\infty}\sum_{n=1}^{\infty}\,(\overline{z^{R,\rm out}_{1,1kmn}} z^{R,\rm out}_{2,1kmn}+ \overline{z^{L,\rm out}_{1,1kmn}} z^{L,\rm out}_{2,1kmn})\nonumber\\
&=&\sum_{kmn\in H_1^{+>}} \overline{z^{R,\rm out}_{1,kmn}} z^{R,\rm out}_{2,1kmn} +\sum_{kmn\in H_{-1}^{+>}} \overline{z^{R,\rm out}_{2,-1kmn}} z^{R,\rm out}_{1,-1kmn}\nonumber\\
&& + \sum_{kmn\in H_1^{->}} \overline{z^{L,\rm out}_{1,1kmn}} z^{L,\rm out}_{2,1kmn} + \sum_{kmn\in H_{-1}^{->}} \overline{z^{L,\rm out}_{2,-1kmn}} z^{L,\rm out}_{1,-1kmn} \label{outinnerproduct}\, .
\eea
which is manifestly positive definite and structurally identical to the corresponding ``in'' product.

\subsection{Quantization: construction of the out Fock space}

The completion of the complex vector space $(\Gamma_R, J_{\rm out})$ with respect to the inner product~(\ref{innerR}) defines the 1-particle Hilbert space~$\mathcal{H}$. Once $\mathcal{H}$ is specified, the bosonic Fock space~$\mathcal{F}$ is constructed in the standard way by taking the symmetric tensor algebra over~$\mathcal{H}$. Vectors of this Fock space are then interpreted physically as quantum states of photon excitations inside the waveguide.

The quantized electromagnetic field is represented by an operator-valued distribution acting on~$\mathcal{F}$. Using the late-time mode basis introduced in the previous subsection, the field operator can be expanded as (compare with Eq.~(\ref{ARb3})):
\bea
\hat{A}^R_a
 = \sum_{hkmn \in H_h^{+>}} a^{R,{\rm out}}_{hkmn}\, A^{R,{\rm out}}_{a,hkmn}
 + \sum_{hkmn \in H_h^{->}} a^{L,{\rm out}\dagger}_{hkmn}\,
   \overline{A^{L,{\rm out}}_{a,hkmn}} .
\label{OUTrep}
\eea
Here $a^{R,{\rm out}}_{hkmn}$ and $a^{L,{\rm out}}_{hkmn}$ are annihilation operators for right- and left-handed photons, while their Hermitian adjoints $a^{R,{\rm out}\dagger}_{hkmn}$ and $a^{L,{\rm out}\dagger}_{hkmn}$ create photons of the corresponding handedness. They obey the canonical commutation relations, $[a^{\lambda,{\rm out}}_{hkmn}, a^{\lambda',{\rm out}\dagger}_{h'k'm'n'}] = R^{-1}\,   \delta_{hh'}\,\delta(k-k')\,\delta_{mm'}\,\delta_{nn'}\,\delta_{\lambda\lambda'}$, while all other commutators vanish. These relations ensure the equal-time canonical commutation rules for the electromagnetic field~\footnote{The overall factor $R^{-1}$ arises from the normalization of the modes given in Eq.~(\ref{normalization2}).}.

The ``out'' vacuum state~$|{\rm out}\rangle$ is defined as the normalized vector annihilated by all operators $a^{R,{\rm out}}_{hkmn}$ and $a^{L,{\rm out}}_{hkmn}$, so that
\bea
a^{R,{\rm out}}_{hkmn}|{\rm out}\rangle = a^{L,{\rm out}}_{hkmn}|{\rm out}\rangle = 0
\quad \text{for all } (h,k,m,n).
\eea
This state represents the absence of photons as seen by the late-time comoving observers, whose natural notion of positive frequency is defined by the accelerating killing field $u^a$.

In the next subsection, we shall relate this late-time quantization to the early-time one through a Bogoliubov transformation, 
which encodes the particle creation effects induced by the acceleration of the waveguide.

\section{Classical duality symmetry and quantum anomaly}

Now that the in and out Fock spaces associated with the early- and late-time configurations of the waveguide have been constructed, we are  in a position to evaluate the vacuum expectation value of the operator that represents the classically conserved Noether charge in the quantum theory. We will compute its expectation value in the ``in'' vacuum state and analyze its behavior at both early and late times. This will allow us to identify and characterize the breaking of the classical electric-magnetic duality symmetry upon quantization.

\subsection{Physical interpretation of the classical Noether charge}

Before entering the quantum regime, let us first clarify the physical meaning of the classical charge $Q(A_R)$. 
The chiral Noether charge associated with electric-magnetic duality rotations~(\ref{freeq}) can be rewritten as
\bea
Q(A_R)=\int_{\Sigma} d\Sigma\,  {\rm Im} \,(A_{a,L} H^a_R) = \Omega(A_R,-i A_R) \, . \label{classicalQ} 
\eea
As discussed in Sec.~\ref{secII}, given any solution $A_R \in \Gamma_R$, if the duality transformed potential $-i A_R$ is also a solution (that is, if $-i A_R \in \Gamma_R$) then $Q(A_R)$ is conserved in time by virtue of the conservation of the symplectic structure (see also Appendix~\ref{appB}).

This quantity (\ref{classicalQ}) has a natural physical interpretation. Let us consider a generic decomposition of $A_R$ into positive- and negative-frequency modes, such as (\ref{ARb2}) or (\ref{ARb3}), with an associated complex structure~$J$.  Define two linear operators on phase space, $D: \Gamma_R \to \Gamma_R$ and $H: \Gamma_R \to \Gamma_R$, by $DA_R=-i A_R$ and  $HA_R^\pm=\pm \hbar A_R^\pm $, respectively, so that $HA_R=\hbar (A_R^+ - A_R^-)= -i\hbar J A_R$. Physically, $D$ generates electromagnetic duality transformations in phase space. The eigenvectors $A_R$ of $iD$ with eigenvalue $+1$ are therefore called ``self-dual'' or ``right-handed'', possessing positive handedness or chirality.\footnote{Similarly, the eigenvectors of $iD$ with eigenvalue $-1$, which are $A_a^L$, are called ``antiself-dual'' or ``left-handed'', corresponding to negative chirality.} 
On the other hand, the operator $H$ represents the usual helicity operator for massless fields: positive-frequency modes with positive chirality, $A_R^+$, have positive helicity $+\hbar$, while negative-frequency and right-handed modes, $A_R^-$, have negative helicity $-\hbar$.\footnote{By complex conjugation, $HA_L=\hbar(-A_L^+ + A_L^-)$, so that left-handed modes with positive (negative) frequency have negative (positive) helicity, contrary to right-handed modes.} Both operations $H$ and $D$ are closely related through $H=\hbar D J$~\cite{AA86}. 
While $D$ is vacuum-independent, the notion of helicity relies on the choice of  complex structure~$J$, and  can thus acquire different physical significance in the quantum theory, depending on the vacuum choice.  

The operator $D$ allows us to write the charge as $Q(A_R)=\Omega(A_R, D A_R)$, a purely classical identity showing that $Q(A_R)$ is the generator of duality transformations in phase space, as already known.  On the other hand, using the inner product property (\ref{antilinearity}),   the helicity operator produces $\langle A_R, H A_R\rangle=\langle A_R, -i \hbar J A_R\rangle=-\hbar i \langle A_R, i  A_R\rangle=-\frac{i}{2}\left[\Omega( A_R, i J A_R)+i \Omega( A_R, i A_R) \right]=-\frac{1}{2}Q(A_R)$\footnote{In this last equality we used the compatibility of $J$ and $\Omega$ to obtain $\Omega( A_R, i J A_R)=\Omega( JA_R, i J^2 A_R)=-\Omega( JA_R, i  A_R)=\Omega( iJA_R,   A_R)=-\Omega( A_R, i J A_R)$, thus $\Omega( A_R, i J A_R)=0$.}, which shows that $Q(A_R)$ is proportional to the expectation value of the helicity operator $H$ in the classical state $A_R$ (a ``first quantization'' interpretation).

To understand in more precise terms what this ``expectation value'' measures, let us rewrite the positive- and negative-frequency parts of the solution $A_R$ using the projectors $A_R^{\pm}=\frac{1\mp i J}{2}A_R$ (equivalent to $JA_R=iA_R^+-i A_R^-$). Acting with the helicity operator gives  $HA_R^{\pm}=\pm \hbar A_R^\pm$, i.e., for right-handed fields the positive-frequency part carries positive helicity while the negative-frequency part carries negative helicity. We can then compute
\bea
\langle A_R^\pm , A_R^\pm\rangle&=&\left< \frac{1\mp i J}{2}A_R,\frac{1\mp i J}{2}A_R\right> =\frac{1}{2\hbar} \Omega\left(\frac{1\mp i J}{2}A_R,\frac{J\pm i }{2}A_R \right)\nonumber\\
& =& \frac{1}{8\hbar} \left[ \Omega(A_R, JA_R)\mp \Omega(i J A_R, JA_R) \pm \Omega( A_R, i A_R) - \Omega(i J A_R, iA_R)\right] \label{auxmix}\, .
\eea
From this we obtain
\bea
||A_R^+||^2 + ||A_R^-||^2 = \frac{\Omega(A_R, JA_R) - \Omega(i J A_R, iA_R) }{4\hbar}= \frac{\Omega(A_R, JA_R) - \Omega( J A_R, A_R) }{4\hbar}=||A_R ||^2\, ,
\eea
 and, furthermore,
 \bea
 \langle  A_R^+,  A^-_R \rangle&=&\frac{1}{8\hbar} \left[ \Omega(A_R, JA_R)- \Omega(i J A_R, JA_R) - \Omega( A_R, i A_R) + \Omega(i J A_R, iA_R)\right]\nonumber\\
 &=&\frac{1}{8\hbar} \left[ \Omega(A_R, JA_R)- \Omega(i  A_R, A_R)  +\Omega( i A_R,  A_R) + \Omega( J A_R, A_R)\right]=0 \, , \label{orto}
 \eea
 which justifies the decomposition of the covariant phase space $\Gamma_R=\Gamma_R^+\oplus \Gamma_R^-$ into orthogonal subspaces of positive ($\Gamma_R^+$) and negative ($\Gamma_R^-$) frequency solutions.\footnote{Using (\ref{innerR}) and (\ref{antilinearity}) one can show more generally $\langle  A_R^{1,+},  A^{2,-}_R \rangle=0$. This  implies $\langle A_R^1, A_R^2\rangle= \langle  A_R^{1,+},  A^{2,+}_R \rangle+ \langle  A_R^{1,-},  A^{2,-}_R \rangle$. It is easy to find that $\langle  A_R^{1,+},  A^{2,+}_R \rangle$ is given by setting $z^L_{kmn}=0$ in (\ref{inner2}), while $\langle  A_R^{1,-},  A^{2,-}_R \rangle$ is given by setting instead $z^R_{kmn}=0$. In view of this, it is not surprising to find the RHS of (\ref{difhelicities}) when using (\ref{diffnorms}).}
More interestingly,
\bea
||A_R^+||^2 - ||A_R^-||^2 = \frac{\Omega( A_R, i A_R) - \Omega(i J A_R, JA_R) }{4\hbar}  = \frac{\Omega( A_R, i A_R) - \Omega(i  A_R, A_R) }{4\hbar} = \frac{\Omega( A_R, i A_R)}{2\hbar} \label{differentnorm} \, ,
\eea
where in the last equality we made use of the compatibility between the complex and symplectic structures.\footnote{Alternatively, $\Omega( A_R, i A_R) = 2\hbar \,{\rm Im}\, \langle  A_R, i A_R \rangle = 2\hbar \,{\rm Im}\, ( \langle  A_R, i A^+_R \rangle+ \langle  A_R, i A^-_R \rangle)= 2\hbar \,{\rm Im}\, ( \langle  A_R, JA^+_R \rangle - \langle  A_R, J A^-_R \rangle)=2\hbar \,{\rm Im}\, i ( \langle  A_R, A^+_R \rangle - \langle  A_R,  A^-_R \rangle)=2\hbar \,{\rm Im}\, i ( \langle  A^+_R, A^+_R \rangle - \langle  A^-_R,  A^-_R \rangle)=2\hbar (\langle  A^+_R, A^+_R \rangle - \langle  A^-_R,  A^-_R \rangle)$, where we used (\ref{orto}).} 
  From here we finally conclude 
\bea
Q(A_R)=   2\hbar \left( ||A_R^-||^2 - ||A_R^+||^2 \right)(=-2\langle A_R, H A_R\rangle) \label{diffnorms}\, .
\eea
Therefore, $Q(A_R)$ measures the net difference between positive and negative helicity mode amplitudes of the classical solution~$A_R$.  
Physically, this corresponds to the degree of circular polarization of the electromagnetic wave described by $A_R$. To see this more explicitly, consider a general  solution $A_R$ labeled by e.g. $\{z_{1kmn}^{R}, z_{1kmn}^{L}\}$ in (\ref{ARb2}). Multiplying by $-i$  produces another solution $-i A_R$  labeled by the pair $\{-iz_{1kmn}^{R}, i z_{1kmn}^{L}\}$. Using then (\ref{classicalQ}) and the identity $\Omega( A_R, -i A_R) = 2\hbar \,{\rm Im}\, \langle  A_R, -i A_R \rangle$, equation (\ref{inner2}) yields
\bea
Q(A_R) &=& 2 \hbar \, R \sum_{n=1}^{\infty} \sum_{m=-\infty}^{\infty}\int_{-\infty}^{\infty} dk ( |z^{L,\rm in}_{1kmn}|^2- |z^{R,\rm in}_{1kmn}|^2)\label{difhelicities}\\
&=&2 \hbar \, R\sum_{n=1}^{\infty} \sum_{m=-\infty}^{\infty}\int_{-\infty}^{\infty} dk \left|\frac{z_{1kmn}^{1}+i z_{1kmn}^{2}}{\sqrt{2}}\right|^2-\left|\frac{z_{1kmn}^{1}- i z_{1kmn}^{2}}{\sqrt{2}}\right|^2\, , \nonumber
\eea
Thus $Q(A_R)$ quantifies the net circular polarization. The result is just a constant, as expected from the duality symmetry, and it has the correct dimensions of angular momentum or helicity, $[Q]=\hbar$.

Finally, note that the classical Noether charge (\ref{classicalQ}) does not depend on the choice of complex structure $J$. Evaluating  $Q(A_R)=\Omega( A_R,- i A_R) = -2\hbar \,{\rm Im}\, \langle  A_R, i A_R \rangle_{\rm out}$   with the late-time inner product (\ref{outinnerproduct}) gives
\bea
Q(A_R)= 2 \hbar \, R  \sum_{n=1}^{\infty} \sum_{m=-\infty}^{\infty}\int_{-\infty}^{\infty} dk\, ( |z^{L, {\rm out}}_{1kmn}|^2- |z^{R, {\rm out}}_{1kmn}|^2) \label{classicresultforQ}\, .
\eea
This demonstrates explicitly that $Q(A_R)$ is conserved in time, as required by the duality symmetry. In the quantum theory, however, this  conservation will be modified by the anomaly discussed in the next subsections.

\subsection{Normal-ordering of the quantum operator}

Our goal in this subsection is to compute the  vacuum expectation value (VEV) of the chiral charge (\ref{classicalQ}) in the ``in'' vacuum. In the quantum theory this chiral charge is promoted to the operator
\bea
 \hat Q&=&\int_{\Sigma} d\Sigma\, \frac{ \hat A_{a}^{R \dagger} \hat H^{R,a}-\hat A_{a}^R \hat H^{R\dagger,a}  }{2i} \, . \label{quantumversion}
  \eea
This operator can be  expanded in terms of creation and annihilation operators using the mode expansion (\ref{OUTrep}) in the out representation. The spatial integrals can then be solved with Proposition (\ref{circbasis2}). Using the normalization factor (\ref{normalization2}) to simplify the resulting expression, a  computation yields:
\bea
\hat Q  &=&  \hbar \, R \sum_{n=1}^{\infty} \sum_{m=-\infty}^{\infty}\int_{-\infty}^{\infty} dk ( a^{L,\rm out}_{1kmn}a^{L,\rm out\dagger}_{1kmn} +  a^{L,\rm out\dagger}_{1kmn}a^{L,\rm out}_{1kmn}  - a^{R,\rm out\dagger}_{1kmn}a^{R,\rm out}_{1kmn} - a^{R,\rm out}_{1kmn}a^{R,\rm out \dagger}_{1kmn})\, . \label{qpure}
\eea

The formal VEV of this operator diverges, as expected for a quantity $\hat Q$ that is quadratic in the field operators. To obtain a finite and physically meaningful result, we have to renormalize  this observable. Since we are working in flat spacetime, it suffices to subtract the contribution of a global reference vacuum, while regularizing the  divergences with point-splitting \cite{Wald:1995yp}. We define the renormalized operator with respect to the instantanous vacuum as
\bea
\hat Q_{\rm ren}(t)=\frac{1}{2i}\int_{\Sigma_t} d\Sigma_t\,  \lim_{\vec y \to \vec x} \left[Q(t,\vec x,\vec y)-\mathbb I \langle t| Q(t,\vec x,\vec y) |t\rangle \right] \label{renresult}\, ,
\eea
where $Q(t,\vec x,\vec y):= \hat A_{a}^{R \dagger}(t,\vec x) \hat H^{R,a}(t,\vec y)-\hat A_{a}^R(t,\vec x) \hat H^{R\dagger,a}(t,\vec y)$ is the point-splitting version of the integrand in the charge operator (\ref{quantumversion}). The second term in brackets subtracts the vacuum expectation value in the chosen reference state, ensuring that all UV diverences vanish. 
At early times, the instantaneous vacuum $|t\rangle$ is simply the ``in'' vacuum, while at late times it represents the ``out'' vacuum. By construction, we have $\langle {\rm in}|\hat Q_{\rm ren}(t\to -\infty)|{\rm in}\rangle=0$. However, at late times we get $\langle {\rm in}|\hat Q_{\rm ren}(t\to \infty)|{\rm in}\rangle=\langle {\rm in}|:\hat Q:|{\rm in}\rangle$ where
\bea
:\hat Q: &=&  2\hbar \, R \sum_{n=0}^{\infty} \sum_{m=-\infty}^{\infty}\int_{-\infty}^{\infty} dk (a^{L,{\rm out}\dagger}_{1kmn}a^{L,{\rm out}}_{1kmn}  - a^{R,{\rm out}\dagger}_{1kmn}a^{R,{\rm out}}_{1kmn}) \, . \label{outnormalordering}
\eea

This normal-ordered operator is well-defined on the out Fock space and clearly shows that it measures the net difference between left- and right-handed circularly polarized photons. Its structure closely parallels the classical expression in Eq. (\ref{difhelicities}), written in terms of wave amplitudes. 

Finally, it is worth noting that the choice of normal ordering with respect to the ``in'' vacuum is the natural one for our purposes. When computing the vacuum expectation value at late times, the relation between the ``in'' and ``out'' operators through the Bogoliubov transformation will reveal that the anomaly arises from the mismatch between their respective positive-frequency splittings, which are affected by the acceleration and rotation of the waveguide.

\subsection{Bogoliubov transformations relating the in and out Fock representations}

To evaluate the expectation value of (\ref{outnormalordering}) in the  vacuum $|{\rm in}\rangle$, we need the unitary map that connects the in and out Fock Hilbert spaces. This can be obtained through a standard Bogoliubov analysis, which we develop in this subsection. While this analysis is widely known for scalar fields, the literature on Bogoliubov transformations for electromagnetic field modes is relatively sparse, and a careful treatment of each helicity sector is required. Consequently, we provide detailed derivations for completeness.

First, write the ``in'' representation as 
\bea
A_a^R&=&\sum_{h\omega_{kmn}>0} z^{R,{\rm in}}_{hkmn}A^{R,{\rm in}}_{a, hkmn}+  \overline{ z_{h k m n}^{L, {\rm in}}} \,\overline{ A^{L,{\rm in}}_{a, h k m n}}\\ 
&=&  \sum_{h\omega_{kmn}>0} z^{R,{\rm in}}_{hkmn}A^{R,{\rm in}}_{a, hkmn}+  z^{R,{\rm in}}_{-hkmn}A^{R,{\rm in}}_{a,-hkmn} \label{inrep2} \, ,
\eea
where  $h\in \{+1,-1\}$ and the sum runs for all  $(h,k,m,n)$  satisfying $h\omega_{kmn}>0$. The first term on the RHS contains the positive frequency modes of the in representation, while the second term involves the negative frequency modes. Since $\omega_{kmn}=\sqrt{k^2+\frac{j^2_{mn}}{R^2}}>0$,  only $h=1$ actually contributes in this sum; nevertheless, we keep $h$ explicit for  bookkeeping.  The second line follows from  (\ref{relationRL}) and the fact that $\omega_{kmn}=\omega_{-k-mn}$.

Next, write the ``out'' representation as
\bea
A_a^R&=&\sum_{(hkmn)\in H^{+>}_h} z^{R,{\rm out}}_{hkmn}A^{R,{\rm out}}_{a, hkmn}+ \sum_{(hkmn)\in H^{->}_h} \overline{ z_{h k m n}^{L, {\rm out}}} \,\overline{ A^{L,{\rm out}}_{a, h k m n}} \label{outrep}\\
&=&\sum_{(hkmn)\in H^{+>}_h} z^{R,{\rm out}}_{hkmn}A^{R,{\rm out}}_{a,hkmn}+ \sum_{(hkmn)\in H^{+<}_h}  z^{R,{\rm out}}_{hkmn}A^{R,{\rm out}}_{a,hkmn} \label{outrep2}\, .
\eea
We recall 
\bea
H^{\pm >}_h=\{(k,m,n)\in \mathbb R\times \mathbb Z \times \mathbb N   \, / \, h\sqrt{k^2+j_{mn}^2/R^2}-m\Omega_0-kv_0 \pm \Omega_0 > 0  \}\, ,\\
H^{\pm <}_h=\{(k,m,n)\in \mathbb R\times \mathbb Z \times \mathbb N   \, / \, h\sqrt{k^2+j_{mn}^2/R^2}-m\Omega_0-kv_0 \pm \Omega_0 < 0  \}\, ,
\eea
which satisfy the  relation: $H^{\pm >}_h=H^{\mp <}_{-h}$. The first term on the RHS of (\ref{outrep2}) (equivalent to (\ref{outpositive})) collect the positive-frequency modes with respect to the late-time  vector $u^a$,  while the second term (equivalent to (\ref{outnegative})) collects the negative-frequency modes.  The sums in (\ref{outrep}) run for all $(h,k,m,n)$  satisfying $h\omega_{kmn}-m\Omega_0-kv_0\pm\Omega_0>0$, so unlike the ``in'' case both $h=\pm1$ contribute.

Since the ``in'' and ``out'' representations are unitarily equivalent (they are related by time evolution), one may expand the modes of either basis in terms of the other.  More precisely, the positive (negative) frequency modes of the in representation, $A^{R,{\rm in}}_{1kmn}$ ($A^{R,{\rm in}}_{-1kmn}$), can be expressed as a mixed combination of positive ($A^{R,{\rm out}}_{h'k'm'n'}$ with $(h'k'm'n')\in H^{+>}_h$) and negative ($A^{R,{\rm out}}_{h'k'm'n'}$ with $(h'k'm'n')\in H^{+<}_h$)  frequency modes of the out representation:
\bea
A^{R,{\rm in}}_{1kmn}&=& \sum_{(h'k'm'n')\in H^{+>}_h} \alpha^R_{1kmn\atop h'k'm'n'} A^{R,{\rm out}}_{h'k'm'n'}+\sum_{(h'k'm'n')\in H^{+<}_h} \beta^R_{1kmn\atop h'k'm'n'} A^{R,{\rm out}}_{h'k'm'n'} \label{in1} \, ,\\
A^{R,{\rm in}}_{-1kmn}&=& \sum_{(h'k'm'n')\in H^{+<}_h} \alpha^R_{-1kmn\atop h'k'm'n'} A^{R,{\rm out}}_{h'k'm'n'}+\sum_{(h'k'm'n')\in H^{+>}_h} \beta^R_{-1kmn\atop h'k'm'n'} A^{R,{\rm out}}_{h'k'm'n'}  \label{in2}\, ,
\eea
and similarly with $A^L_a$. Substituting these equations into (\ref{inrep2}) and grouping positive- and negative-frequency pieces with respect to the ``out'' splitting,
\bea
A^R_a&=&  \sum_{(h'k'm'n')\in H^{+>}_h} \, \sum_{h\omega_{kmn}>0} \left( z^{R,{\rm in}}_{1kmn} \alpha^R_{1kmn\atop h'k'm'n'} +  z^{R,{\rm in}}_{-1kmn}\beta^R_{-1kmn\atop h'k'm'n'}\right) A^{R,{\rm out}}_{a, h'k'm'n'}\nonumber\\
&& +\sum_{(h'k'm'n')\in H^{+<}_h} \, \sum_{h\omega_{kmn}>0} \left( z^{R,{\rm in}}_{-1kmn} \alpha^R_{-1kmn\atop h'k'm'n'} +  z^{R,{\rm in}}_{1kmn}\beta^R_{1kmn\atop h'k'm'n'}\right) A^{R,{\rm out}}_{a, h'k'm'n'}\nonumber\\
&=& \sum_{(h'k'm'n')\in H^{+>}_h}\, \sum_{h\omega_{kmn}>0}  \left( z^{R,{\rm in}}_{1kmn} \alpha^R_{1kmn\atop h'k'm'n'} +  z^{R,{\rm in}}_{-1kmn}\beta^R_{-1kmn\atop h'k'm'n'}\right) A^{R,{\rm out}}_{a, h'k'm'n'}\nonumber\\
&& + \sum_{(h'k'm'n')\in H^{->}_h}\, \sum_{h\omega_{kmn}>0}  \left( z^{R,{\rm in}}_{-1kmn} \alpha^R_{-1kmn\atop -h'-k'-m'n'} +  z^{R,{\rm in}}_{1kmn}\beta^R_{1kmn\atop -h'-k'-m'n'}\right) \overline{A^{L,{\rm out}}_{a, h'k'm'n'}}\, .
\eea
Comparing with (\ref{outrep}) we can infer the Bogoliubov transformations:
\bea
z^{R,{\rm out}}_{h'k'm'n'} &=&\sum_{h\omega_{kmn}>0}  \alpha^R_{1kmn\atop h'k'm'n'}  z^{R,{\rm in}}_{1kmn} +  \beta^R_{-1kmn\atop h'k'm'n'} z^{R,{\rm in}}_{-1kmn}\, , \quad (h'k'm'n')\in H^{+>}_h\, , \label{B1}\\
\overline{z^{L,{\rm out}}_{h'k'm'n'}} &=&\sum_{h\omega_{kmn}>0}   \alpha^R_{-1kmn\atop -h'-k'-m'n'} \overline{z^{L,{\rm in}}_{1-k-mn}}+  \beta^R_{1kmn\atop -h'-k'-m'n'}z^{R,{\rm in}}_{1kmn}\, , \quad (h'k'm'n')\in H^{->}_h\, . \label{B2}
\eea

Conversely, the ``out'' modes can be expanded in the ``in'' basis.
\bea
A^{R,{\rm out}}_{a, hkmn}&=& \sum_{\omega_{k'm'n'}>0} \gamma^R_{hkmn\atop 1k'm'n'} A^{R,{\rm in}}_{a, 1k'm'n'}+ \delta^R_{hkmn\atop -1k'm'n'} A^{R,{\rm in}}_{a,-1k'm'n'}\, , \quad (hkmn)\in H^{+>}_h \label{out1}\\
A^{R,{\rm out}}_{a, hkmn}&=& \sum_{\omega_{k'm'n'}>0} \gamma^R_{hkmn\atop -1k'm'n'} A^{R,{\rm in}}_{a, -1k'm'n'}+ \delta^R_{hkmn\atop 1k'm'n'} A^{R,{\rm in}}_{a, 1k'm'n'}\, , \quad (hkmn)\in H^{+<}_h\, . \label{out2}
\eea
This is, positive frequency modes in the out representation ($A^{R,{\rm out}}_{h'k'm'n'}$ with $(h'k'm'n')\in H^{+>}_h$) can be written as a linear combination of positive ($A^{R,{\rm in}}_{a, 1k'm'n'}$) and negative ($A^{R,{\rm in}}_{a, -1k'm'n'}$) frequency modes of the in representation. Similarly with the negative frequency modes of the out representation  ($A^{R,{\rm out}}_{h'k'm'n'}$ with $(h'k'm'n')\in H^{+<}_h$). The coefficients $\gamma^R$ and $\delta^R$ are not independent but they are related to $\alpha^R$ and $\beta^R$ by the unitarity properties of the map. As shown in Appendix C, one has
\bea
\overline{\alpha^R_{1kmn\atop h'k'm'n'}} = {\gamma^R_{h'k'm'n'\atop 1kmn}}\, , \quad \beta^R_{1kmn\atop h'k'm'n'} = -{\delta^R_{h'k'm'n'\atop 1kmn}} \, , \quad \overline{\alpha^R_{-1kmn\atop h'k'm'n'}} = {\gamma^R_{h'k'm'n'\atop -1kmn}}\, ,\quad \beta^R_{-1kmn\atop h'k'm'n'} =-{\delta^R_{h'k'm'n'\atop -1kmn}}\, . \label{8coeficients}
\eea

Furthermore, the Bogoliubov coefficients  $\alpha^R$ and $\beta^R$ are not entirely arbitrary: they must  satisfy consistency (unitarity) constraints. More precisely (see Appendix C for computational details)
\bea
\delta_{kk'}\delta_{mm'}\delta_{nn'}&=&\sum_{(h''k''m''n'')\in H^{+>}_h}  \alpha^R_{1kmn\atop h''k''m''n''}  \overline{\alpha^R_{1k'm'n'\atop h''k''m''n''}}-\sum_{(h''k''m''n'')\in H^{+<}_h} \beta^R_{1kmn\atop h''k''m''n''}  \overline{\beta^R_{1k'm'n'\atop h''k''m''n''}}\, , \label{rel1}\\
\delta_{kk'}\delta_{mm'}\delta_{nn'}&=&\sum_{(h''k''m''n'')\in H^{+<}_h} \alpha^R_{-1kmn\atop h''k''m''n''}  \overline{\alpha^R_{-1k'm'n'\atop h''k''m''n''}}-\sum_{(h''k''m''n'')\in H^{+>}_h} \beta^R_{-1kmn\atop h''k''m''n''}  \overline{\beta^R_{-1k'm'n'\atop h''k''m''n''}}\, , \label{rel2}\\
\delta_{hh'}\delta_{kk'}\delta_{mm'}\delta_{nn'}&=&\sum_{\omega_{k''m''n''}>0} \overline{\alpha^R_{1k''m''n''\atop hkmn}}  \alpha^R_{1k''m''n''\atop h'k'm'n'}-{\beta^R_{-1k''m''n''\atop hkmn}}  \overline{\beta^R_{-1k''m''n''\atop h'k'm'n'}}\, , \quad(hkmn),(h'k'm'n')\in H^{+>}_h \label{ref3}\\
\delta_{hh'}\delta_{kk'}\delta_{mm'}\delta_{nn'}&=&\sum_{\omega_{k''m''n''}>0} \overline{\alpha^R_{-1k''m''n''\atop hkmn}}  \alpha^R_{-1k''m''n''\atop h'k'm'n'}-{\beta^R_{1k''m''n''\atop hkmn}}  \overline{\beta^R_{1k''m''n''\atop h'k'm'n'}}\, , \quad (hkmn),(h'k'm'n')\in H^{+<}_h\, . \label{rel4}
\eea
In particular, setting $h=h'$, $k=k'$, $m=m'$, $n=n'$ in these 4 expressions, we get
\bea
1&=&\sum_{(h'k'm'n')\in H^{+>}_h}   |\alpha^R_{1kmn\atop h'k'm'n'}|^2-\sum_{(h'k'm'n')\in H^{+<}_h} |\beta^R_{1kmn\atop h'k'm'n'}|^2\, ,\label{sum1}\\
1&=&\sum_{(h'k'm'n')\in H^{+<}_h}  |\alpha^R_{-1kmn\atop h'k'm'n'}|^2-\sum_{(h'k'm'n')\in H^{+>}_h}  |\beta^R_{-1kmn\atop h'k'm'n'}|^2\, ,\label{sum2}\\
1&=&\sum_{\omega_{k'm'n'}>0}  |\alpha^R_{1k'm'n'\atop hkmn}|^2 - |\beta^R_{-1k'm'n'\atop hkmn}|^2\, ,\quad  (hkmn)\in H^{+>}_h\, ,\label{sum3}\\
1&=&\sum_{\omega_{k'm'n'}>0}  |\alpha^R_{-1k'm'n'\atop hkmn}|^2 -  |\beta^R_{1k'm'n'\atop hkmn}|^2\, ,\quad  (hkmn)\in H^{+<}_h\, . \label{sum4}
\eea

These relations will play a central role in the next subsection, since they  allow us to derive a non-trivial identity that ultimately leads to the closed-form expression for the anomalous contribution to the Noether charge.   First, we combine  (\ref{sum1}) and (\ref{sum3}) to infer
\bea
-\sum_{\omega_{k'm'n'}>0} \left[\sum_{(hkmn)\in H^{+>}_h} |\beta^R_{-1k'm'n'\atop hkmn}|^2-\sum_{(hkmn)\in H^{+<}_h} |\beta^R_{1k'm'n'\atop hkmn}|^2\right]&=& \sum_{(hkmn)\in H^{+>}_h} 1 - \sum_{\omega_{k'm'n'}>0} 1\nonumber\\
&=& \sum_{(hkmn)\in H^{-<}_h} 1 - \sum_{\omega_{k'm'n'}>0} 1 \, ,
\eea
while if we combine (\ref{sum2}) and (\ref{sum4}) we similarly get 
\bea
-\sum_{\omega_{k'm'n'}>0} \left[\sum_{(hkmn)\in H^{+<}_h} |\beta^R_{1k'm'n'\atop hkmn}|^2-\sum_{(hkmn)\in H^{+>}_h} |\beta^R_{-1k'm'n'\atop hkmn}|^2\right]= \sum_{(hkmn)\in H^{+<}_h} 1 - \sum_{\omega_{k'm'n'}>0} 1 \, .
\eea
By subtracting these two last equations  one arrives at a key identity that isolates the net ``helicity mixing'' from positive to negative sectors:
\bea
-2\sum_{\omega_{k'm'n'}>0} \left[\sum_{(hkmn)\in H^{+<}_h} |\beta^R_{1k'm'n'\atop hkmn}|^2-\sum_{(hkmn)\in H^{+>}_h} |\beta^R_{-1k'm'n'\atop hkmn}|^2\right]= \sum_{(hkmn)\in H^{+<}_h} 1 - \sum_{(hkmn)\in H^{-<}_h} 1\, .
\eea
This expression admits a useful simplification. First, 
\bea
\sum_{(hkmn)\in H^{+<}_h} 1 - \sum_{(hkmn)\in H^{-<}_h} 1 &=& \sum_{(kmn)\in H^{+<}_1} 1 - \sum_{(kmn)\in H^{-<}_1} 1 + \sum_{(kmn)\in H^{+<}_{-1}} 1 - \sum_{(kmn)\in H^{-<}_{-1}} 1\nonumber\\
&=& \sum_{(kmn)\in H^{+<}_1} 1 - \sum_{(kmn)\in H^{-<}_1} 1 + \sum_{(kmn)\in H^{->}_{1}} 1 - \sum_{(kmn)\in H^{+>}_{1}} 1\nonumber\, ,
\eea
and since $H^{->}_1 = H_{\rm in}- H^{-<}_1$,   $H^{+>}_1 = H_{\rm in}- H^{+<}_1$, we have
\bea
\sum_{(kmn)\in H^{->}_{1}} 1 - \sum_{(kmn)\in H^{+>}_{1}} 1 = - \sum_{(kmn)\in H^{-<}_1} 1 +  \sum_{(kmn)\in H^{+<}_1} 1\, ,
\eea
thus
\bea
-\sum_{\omega_{k'm'n'}>0} \left[\sum_{(hkmn)\in H^{+<}_h} |\beta^R_{1k'm'n'\atop hkmn}|^2-\sum_{(hkmn)\in H^{+>}_h} |\beta^R_{-1k'm'n'\atop hkmn}|^2\right]&=&  \sum_{(kmn)\in H^{+<}_1} 1 - \sum_{(kmn)\in H^{-<}_1} 1\nonumber\\
&=&-\sum_{(kmn)\in H^{-<}_1-H^{+<}_1 }1 \, , \label{final}
\eea
where in the last equality we noticed that $H^{+<}_1\subset H^{-<}_1$ {\it provided that $\Omega_0>0$}. The last sum is convergent since
\bea
H^{-<}_1-H^{+<}_1 =\left\{(k,m,n)\in \mathbb R\times \mathbb Z\times \mathbb N\, /\, -\Omega_0 < \sqrt{k^2+\frac{j_{mn}^2}{R^2}}-m\Omega_0-kv_0< \Omega_0\right\} \, ,\label{set}
\eea
is a bounded set in $ \mathbb R\times \mathbb Z\times \mathbb N$. This is, for sufficiently high values of $k$, $m$, or $n$, $|\sqrt{k^2+\frac{j_{mn}^2}{R^2}}-m\Omega_0-kv_0|>\Omega_0$, as $j_{mn}\sim m$ for large $m$ and $j_{mn}\sim n\pi$ for large $n$. If both $k$ and $j_{mn}$ become large at equal rate, say $k\sim \alpha j_{mn}$, with $\alpha\neq 0$, then $|\sqrt{k^2+\frac{j_{mn}^2}{R^2}}-m\Omega_0-kv_0|\sim |\sqrt{\alpha^2+1}\frac{j_{mn}}{R}-m \Omega_0-\alpha \frac{j_{mn}}{R} v_0|> \frac{j_{mn}}{R} |\sqrt{\alpha^2+1}-R \Omega_0-\alpha  v_0|>\frac{j_{mn}}{R} |\sqrt{\alpha^2+1}-1-\alpha|$, where in the first inequality we used the property $j_{mn}>|m|>m$, and in the second inequality we used $R\Omega_0<1$ and $v_0<1$. Since $|\sqrt{\alpha^2+1}-1-\alpha|>0$ for any $\alpha\neq 0$, then $|\sqrt{k^2+\frac{j_{mn}^2}{R^2}}-m\Omega_0-kv_0|> \frac{j_{mn}}{R} |\sqrt{\alpha^2+1}-1-\alpha|$ grows without bound as $k\sim \alpha j_{mn}$ increases.  

Similarly, if  $\Omega_0<0$ then $H^{-<}_1\subset H^{+<}_1$ and
\bea
-\sum_{\omega_{k'm'n'}>0} \left[\sum_{(hkmn)\in H^{+<}_h} |\beta^R_{1k'm'n'\atop hkmn}|^2-\sum_{(hkmn)\in H^{+>}_h} |\beta^R_{-1k'm'n'\atop hkmn}|^2\right]=+\sum_{(kmn)\in H^{+<}_1-H^{-<}_1 }1  \, ,
\eea
where
\bea
H^{+<}_1-H^{-<}_1&=&\left\{(k,m,n)\in \mathbb R\times \mathbb Z\times \mathbb N\, /\, -|\Omega_0| < \sqrt{k^2+\frac{j_{mn}^2}{R^2}}+m|\Omega_0|-kv_0< |\Omega_0|\right\}\nonumber\\
&=&\left\{(k,m',n)\in \mathbb R\times \mathbb Z\times \mathbb N\, /\, -|\Omega_0| < \sqrt{k^2+\frac{j_{m'n}^2}{R^2}}-m'|\Omega_0|-kv_0< |\Omega_0|\right\} \, ,
\eea
is the same set as (\ref{set}) (recall that $j_{-mn}=j_{mn}$).

For $\Omega_0=0$ this set is empty, as one can trivially see. For $v_0=0$, given the lower bound $j_{mn}>|m|+j_{0n}$ valid for integer $|m|>0$ \cite{doi:10.1137/0514029}, we can find $ \sqrt{k^2+\frac{j_{mn}^2}{R^2}}-m\Omega_0\geq \frac{j_{mn}}{R}-m\Omega_0>\Omega_0(j_{mn}-m)> \Omega_0 j_{0n}>j_{01}\Omega_0>\Omega_0$ ($j_{01}=2.404...$), while for $m=0$ we simply have $\sqrt{k^2+\frac{j_{0n}^2}{R^2}}>\frac{j_{0n}}{R}>\Omega_0 j_{0n}>j_{01}\Omega_0>\Omega_0$. This is, for $v_0=0$ this set is also empty.

 For $\Omega_0\neq 0$ {\it and} $v_0\neq 0$, the set is not empty in general.  To see this,  consider the following example: $\{v_0=0.99, R\Omega_0=0.9\}$. Then, the mode $(k,m,n)=(10/R,0,1)$ satisfies $ R\sqrt{k^2+\frac{j_{mn}^2}{R^2}}-mR\Omega_0-Rk v_0=0.385095...$, which is inside the interval $(-0.9,0.9)$, so $H^{-<}_1-H^{+<}_1 \neq \emptyset$.

We can get more insights if we define $\tilde \omega=\gamma(\omega- kv_0)$ and $\tilde k=\gamma(k-v_0\omega)$. Because  $\tilde \omega^2-\tilde k^2=\omega^2-k^2=\frac{j_{mn}^2}{R^2}$, the set (\ref{set}) can be reduced to 
\bea
H^{+<}_1-H^{-<}_1&=&\left\{(\tilde k,m,n)\in \mathbb R\times \mathbb Z\times \mathbb N\, /\, -\gamma|\Omega_0| < \sqrt{\tilde k^2+\frac{j_{mn}^2}{R^2}}-m|\Omega_0|\gamma<\gamma |\Omega_0|\right\} \label{set2}\, ,
\eea
therefore the LHS of (\ref{final}) is a function of $\{R|\Omega_0| \gamma, {\rm sign}\, \Omega_0\}$ alone.

\subsection{Computation of the vacuum expectation value of $\hat Q$}

As indicated above, the normal-ordered operator takes the  form 
\bea
:Q:_{}=2 \hbar \, R  \sum_{n=1}^{\infty} \sum_{m=-\infty}^{\infty}\int_{-\infty}^{\infty} dk\, \left[a^{L {\rm out},\dagger}_{1kmn}a^{L {\rm out}}_{1kmn}-a^{R {\rm out},\dagger}_{1kmn}a^{R {\rm out}}_{1kmn} \right]\, . \label{outnormalordered}
\eea
 To evaluate its vacuum expectation value in the  state $|{\rm in}\rangle$, we need the Bogoliubov transformations that relate  the two sets of annihilation and creation operators. These relations follow from the classical mode transformations (\ref{B1}) and (\ref{B2}), and depend on the region of mode space to which each triplet $(k,m,n)$ belongs:
\bea
a^{R,{\rm out}}_{1k'm'n'} &=&\sum_{\omega_{kmn}>0}  \alpha^R_{1kmn\atop 1k'm'n'}  a^{R,{\rm in}}_{1kmn} +  \beta^R_{-1kmn\atop 1k'm'n'} a^{L,{\rm in}\dagger}_{1,-k-mn}\, , \quad (k'm'n')\in H^{+>}_1\, , \label{ar1}\\
{a^{R,{\rm out}}_{1,k'm'n'}} &=&\sum_{\omega_{kmn}>0}   \alpha^R_{-1kmn\atop 1k'm'n'} {a^{L,{\rm in}\dagger}_{1-k-mn}}+  \beta^R_{1kmn\atop 1k'm'n'}a^{R,{\rm in}}_{1kmn}\, , \quad (k'm'n')\in H^{->}_{-1}=H^{+<}_{1}\, , \label{ar2}\\
{a^{L,{\rm out}}_{1k'm'n'}} &=&\sum_{\omega_{kmn}>0}  \overline{ \alpha^R_{-1kmn\atop -1-k'-m'n'} }{a^{L,{\rm in}}_{1-k-mn}}+  \overline{\beta^R_{1kmn\atop -1-k'-m'n'}}a^{R,{\rm in}\dagger}_{1kmn}\, , \quad (k'm'n')\in H^{->}_1\, , \label{al1}\\
a^{L,{\rm out}}_{1k'm'n'} &=&\sum_{\omega_{kmn}>0}  \overline{\alpha^R_{1kmn\atop -1-k'-m'n'}}  a^{R,{\rm in}\dagger}_{1kmn} +  \overline{\beta^R_{-1kmn\atop -1-k'-m'n'} } a^{R,{\rm in}}_{1-k-mn}\, , \quad (k'm'n')\in H^{+>}_{-1}=H^{-<}_{1} \label{al2}\, .
\eea
It is convenient now to decompose the sums in Eq. (\ref{outnormalordered}) by noting that $H^{\pm >}_1\cap  H^{\pm <}_1=\emptyset$ and $H^{\pm >}_1\cup  H^{\pm <}_1=H_{\rm in}$. We may thus rewrite
\bea
:Q:_{}&=&2\hbar \, R\left[   \sum_{(kmn)\in H^{->}_1}a^{L {\rm out},\dagger}_{1kmn}a^{L {\rm out}}_{1kmn}+ \sum_{(kmn)\in H^{-<}_{1}}a^{L {\rm out},\dagger}_{1kmn}a^{L {\rm out}}_{1kmn} \right. \nonumber\\
&&\hspace{1cm} \left.  - \sum_{(kmn)\in H^{+>}_1}a^{R {\rm out},\dagger}_{1kmn}a^{R {\rm out}}_{1kmn}- \sum_{(kmn)\in H^{+<}_{1}}a^{R {\rm out},\dagger}_{1kmn}a^{R {\rm out}}_{1kmn} \right]\, .
\eea
Using all this and the standard commutation rules, we can evaluate the ``in''-vacuum expectation value:
\bea
\langle {\rm in}|:Q:_{}|{\rm in}\rangle=2\hbar \, R\sum_{\omega_{k'm'n'}>0}\left[   \sum_{(kmn)\in H^{->}_1}|\beta^R_{1k'm'n'\atop -1-k-mn}|^2 + \sum_{(kmn)\in H^{-<}_{1}}  |\alpha^R_{1k'm'n'\atop -1-k-mn}|^2 \right.\nonumber\\
\left. -\sum_{(kmn)\in H^{+>}_1} |\beta^R_{-1k'm'n'\atop 1kmn}|^2- \sum_{(kmn)\in H^{+<}_{1}} |\alpha^R_{-1k'm'n'\atop 1kmn}|^2\right]\,  .\nonumber\\
\eea
 Using the  identities (\ref{sum3})-(\ref{sum4}) derived previously, we can express this purely in terms of the $\beta$ coefficients, yielding
\bea
\langle {\rm in}|:Q:_{}|{\rm in}\rangle=2\hbar \, R\sum_{\omega_{k'm'n'}>0}\left[   \sum_{(kmn)\in H^{->}_1}|\beta^R_{1k'm'n'\atop -1-k-mn}|^2 - \sum_{(kmn)\in H^{-<}_{1}}  |\beta^R_{-1k'm'n'\atop -1-k-mn}|^2 \right.\nonumber\\
\left. -\sum_{(kmn)\in H^{+>}_1} |\beta^R_{-1k'm'n'\atop 1kmn}|^2+ \sum_{(kmn)\in H^{+<}_{1}} |\beta^R_{1k'm'n'\atop 1kmn}|^2\right]\,\nonumber\\
=2\hbar \, R\sum_{\omega_{k'm'n'}>0}\left[   \sum_{(kmn)\in H^{+<}_{-1}}|\beta^R_{1k'm'n'\atop -1-k-mn}|^2 - \sum_{(kmn)\in H^{+>}_{-1}}  |\beta^R_{-1k'm'n'\atop -1-k-mn}|^2 \right.\nonumber\\
\left. -\sum_{(kmn)\in H^{+>}_{1}} |\beta^R_{-1k'm'n'\atop 1kmn}|^2+ \sum_{(kmn)\in H^{+<}_{1}} |\beta^R_{1k'm'n'\atop 1kmn}|^2\right]\,\nonumber\\
=2\hbar \, R\sum_{\omega_{k'm'n'}>0}\left[   \sum_{(kmn)\in H^{+<}_{h}}|\beta^R_{1k'm'n'\atop hkmn}|^2 - \sum_{(kmn)\in H^{+>}_{h}}  |\beta^R_{-1k'm'n'\atop hkmn}|^2 \right]\nonumber\, .
\eea
This last result has a clear physical meaning: it measures the net imbalance in the number of created photons with opposite helicities between early and late times.

Interestingly, this particular combination of infinite sums can be computed in closed form even without direct knowledge of the acceleration dynamics of the waveguide, i.e. without direct knowledge of each $\beta^R_{h'k'm'n'\atop hkmn}$. As follows from the general identity (\ref{final}) derived in the previous subsection, and assuming $\Omega_0>0$ for simplicity, we obtain
\bea
\langle {\rm in}|:Q:_{}|{\rm in}\rangle=2\hbar \, R\left[ \sum_{(kmn)\in H^{-<}_1} 1 - \sum_{(kmn)\in H^{+<}_1} 1 \right]=2\hbar \, R\sum_{(kmn)\in H^{-<}_1-H^{+<}_1 }1 \, . \label{finalresult}
\eea
In other words, the expectation value of the normal-ordered charge operator in the ``in'' vacuum is determined by the difference in the number of modes belonging to the positive- and negative-helicity sectors that undergo mixing through the Bogoliubov transformation. This result turns out to be independent of the detailed background dynamics, and depends only on the spectral asymmetry at late times, $H^{-<}_1\neq H^{+<}_1$. This ``topological'' behavior is reminiscent of the Adler-Bell-Jackiw anomaly for spin-$\tfrac{1}{2}$ fermions \cite{PhysRevD.100.085014,PhysRevD.108.105025}.

When  the rotational parameter $\Omega_0$ vanishes, the sets $H^{-<}_1$ and $H^{-<}_1$ not only coincide, but vanish, and the expectation value (\ref{finalresult}) is zero, as required by classical duality invariance. However, for nonzero $\Omega_0$, $v_0\neq 0$
these sets are non-trivial and differ, giving rise to  $\langle {\rm in}|:Q:_{}|{\rm in}\rangle\neq 0$. This nonvanishing expectation value therefore encodes a quantum anomaly.

\section{Conclusions}

In this work we have analyzed the status of the classical electric-magnetic duality symmetry of Maxwell's theory when the electromagnetic field is quantized within an accelerating cylindrical waveguide. We have shown that the  Noether charge operator  fails to be conserved in time due to the background accelerated motion. The effect manifests as a time dependence in the vacuum expectation value,  $\langle {\rm in}|\hat Q_{\rm ren}|_{t\to \infty}|{\rm in}\rangle\neq \langle {\rm in}|\hat Q_{\rm ren}|_{t\to -\infty}|{\rm in}\rangle$, which signals the emergence of a \emph{quantum anomaly}. As in the case of the chiral anomaly for fermions, the breaking originates in a spectral asymmetry between right- and left-handed chiral sectors \cite{PhysRevD.108.105025}, induced by the helicity of the background at late times. The result is ``topological'' in nature and insensitive to the precise acceleration profile of the waveguide, depending only on the initial and final states of motion, just as for the Adler-Bell-Jackiw anomaly  \cite{PhysRevD.100.085014, PhysRevD.108.105025}.

Although our waveguide setup is idealized, it provides a concrete proof of concept: quantization of the electromagnetic field in a helically accelerated cavity leads to a net imbalance between right- and left-handed photons. The expected magnitude of the effect,  estimated at \(|\Delta N|\sim |R\Omega_0 v_0|/\sqrt{1-v_0^2}\), remains small for realistic parameters---of order a few photons even in the ultra-relativistic regime---but it might still be enhanced in more elaborate configurations. In particular, quantum-optical schemes employing squeezed states offer a promising route: by amplifying quantum correlations and suppressing vacuum noise, they could make the duality anomaly experimentally accessible, in close analogy with recent proposals in analogue gravity systems~\cite{PhysRevLett.128.091301, PhysRevD.106.105021}.

A key ingredient of our setup is confinement: the waveguide modifies the quantum vacuum fluctuations of the field and induces an effective dispersion relation for the allowed photon modes, \(\omega_{kmn} = \sqrt{k^2 + j_{mn}^2/R^2}\). This discrete spectrum effectively regularizes the sum over modes in (\ref{finalresult}), rendering the anomaly finite and physically meaningful. Extensions of the model to include dielectric media or more realistic boundary materials would naturally lead to alternative dispersion relations and hence distinct quantitative predictions for the photon-number imbalance at late times. Exploring these possibilities could help assess the robustness of the duality anomaly under different physical implementations.

For waveguide models with boundary conditions that are not duality-invariant---such as perfectly conducting plates supporting conventional TE and TM modes---the time evolution of $\langle {\rm in}|:Q:_{}|{\rm in}\rangle$ may already appear at the classical level, since, as explained in Sec. II and III,  electric-magnetic duality does not hold inside the cavity. In such cases, the total change of the expectation value of the Noether charge will generally contain two distinct contributions: a \emph{classical term} arising from explicit symmetry breaking by the boundary conditions, and a \emph{genuinely quantum contribution} associated with the anomaly itself. The framework developed here allows, in principle, to disentangle these two effects, by identifying the portion of $\langle {\rm in}|:Q:_{}|{\rm in}\rangle$ that remains when the classical contribution is eliminated.

The Noether charge operator \(\hat Q\) is a global quantity, integrated over the entire spatial section of the waveguide. In contrast, realistic detectors are inherently local. A natural question then arises: what would a local observer attached to the waveguide's surface actually measure? Understanding how the duality anomaly manifests in local observables (such as polarization-dependent energy fluxes, local helicity densities, or cross correlations between field components) remains an open and physically relevant problem. 

From a more mathematical viewpoint, the spectral asymmetry between right- and left-handed electromagnetic modes found at late times suggests a deep connection with the \(\eta\)-invariant term appearing in the Atiyah-Patodi-Singer (APS) index theorem. For the Dirac operator, Bar and Strohmaier have extended the APS theorem to encompass  Lorentzian geometries~\cite{Baer2016}. A similar analysis applied to the Maxwell operator could provide another   interpretation on the emergence of this duality anomaly, linking the variation of $\langle {\rm in}|:Q:_{}|{\rm in}\rangle$ to the spectral flow of self-dual and anti-self-dual modes in the rotating waveguide. This direction deserves further investigation.

In summary, our analysis shows that the electric-magnetic duality symmetry of Maxwell theory does not survive quantization in helically accelerated waveguides. The resulting duality anomaly represents the spin-1 analogue of the chiral anomaly for fermions, with the helicity of the waveguide acting as an effective ``axial background''. Physically, this quantum effect predicts that non-inertial observers could detect an imbalance in photon helicity, without requiring the presence of gravity. In a sense, this represents a chiral analogue of the Unruh effect \cite{PhysRevD.14.870,RevModPhys.80.787}. Beyond its conceptual significance, this effect opens new avenues for exploring fundamental aspects of quantum field theory in non-inertial and confined settings in the lab.

\appendix

\section{Orthonormality and completeness of the field modes}

In this appendix we collect several auxiliary results concerning Bessel functions that are used in the main text. Some of these results are not easily found in the literature, so we include short proofs for completeness.

\bel \label{ortobessel1}
For $m>-1$ and any pair of zeros $j_{mn}$, $j_{mn'}$ of  $J_m(x)$, we have
\bea
\int_0^1dx x J_m(j_{mn} x)J_m(j_{mn'} x)=  \delta_{nn'} \frac{J_m^{' 2}(j_{mn})}{2}\, .
\eea
\eel
\noindent See 6.521.1 in \cite{gradshteyn2007} or 10.22.37 in \cite{nist}.

\bel \label{ortoderivative}
For any pair of zeros $a$, $b$ of either $J_m(x)$ or $J'_m(x)$, we have
\bea
\int_0^1dx x J_m'(ax)J'_m(bx)= \int_0^1 dx x J_m(ax)J_m(bx)\left(1-\frac{m^2}{abx^2}\right)
\eea
\eel
\noindent{\bf Proof}. 
\bea
\int_0^1 dx x J'_m(ax)J'_m(bx)&=& \left. \frac{1}{a}x J_m(ax)J'_m(bx) \right |_{0}^1 -  \frac{1}{a}\int_0^1 dx  J_m(ax)J'_m(bx) -  \frac{b}{a}\int_0^1 dx x J_m(ax)J''_m(bx)\\
&=&  -  \frac{1}{a}\int_0^1 dx  J_m(ax)J'_m(bx) +\int_0^1 dx x J_m(ax)\left(\frac{1}{a x}J'_m(bx)+ \frac{b}{a }J_m(bx)\left(1-\frac{m^2}{b^2x^2}\right) \right) \nonumber\\
&=&  \int_0^1 dx x J_m(ax) \frac{b}{a }J_m(bx)\left(1-\frac{m^2}{b^2x^2}\right) = \int_0^1 dx x J_m(ax)J_m(bx)\left(1-\frac{m^2}{abx^2}\right) \nonumber
\eea
where in the second equality we used the ODE that defines the Bessel functions: $J_m''(bx)+\frac{1}{bx}J'_m(bx)+\left(1-\frac{m^2}{b^2x^2}\right)J_m(bx)=0$; and in the last equality we used  Lemma A.1.

\bel \label{completeness}
The Bessel functions satisfy the following completeness relation:
\bea
\sum_{n=1}^{\infty}\frac{J_m(j_{mn}x) J_m(j_{mn}x')}{{J'}^2_m(j_{mn})}=\frac{1}{2x}\delta(x-x')\, , \label{completenessid}
\eea
where $x\in [0,1]$.
\eel

\noindent{\bf Proof}. The Bessel functions solve a Sturm-Liouville problem with inner product
\bea
\langle J_m(j_{mn}x), J_m(j_{mn'}x)\rangle_{SL}:=\int_0^1 dx x J_m(j_{mn}x) J_m(j_{mn'}x)=  \delta_{nn'} \frac{J_m^{' 2}(j_{mn})}{2}\, ,
\eea
where the last equality follows from Lemma \ref{ortobessel1}. Because the Bessel functions form a complete orthogonal set with respect to this product, we can define an orthonormal basis of the corresponding space of solutions by 
$u_{mn}(x):=\frac{\sqrt{2}}{J_m'(j_{mn})}J_m(j_{mn}x)$. This is a complete basis in the solution space of the Bessel equation, therefore any solution can be expanded in this basis as $F=\sum_{n=1}^{\infty}c_{mn} u_{mn}$, where $c_{mn}=\langle F,u_{mn} \rangle_{SL}$. This implies
\bea
F(x)=\sum_{n=1}^{\infty} \left(\int_0^1 dx' x' F(x')u_{mn}(x') \right)u_{mn}(x)=\int_0^1 dx' x' F(x') \left(\sum_{n=1}^{\infty} u_{mn}(x')u_{mn}(x)\right)\, .
\eea
Since this equation holds for {\it all} $F(x)$ defined over $[0,1]$,   this equation implies the equality  in (\ref{completenessid}).

\bel \label{ladderoperators}

The spin-weighted vector fields $\Bm^a$, $\bar \Bm^a$ as ladder operators raising and lowering the values of $m$ when acting on the Bessel functios:
\bea
\Bm^a \nabla_a (J_{m}(\sqrt{\omega^2-k^2}\rho)e^{-i m \phi}) \equiv \eth  (J_{m}(\sqrt{\omega^2-k^2}\rho)e^{-i m \phi})=- \frac{\sqrt{\omega^2-k^2}}{\sqrt{2}}J_{m+1}(\sqrt{\omega^2-k^2}\rho)e^{-i m \phi}\, ,\\
\bar \Bm^a \nabla_a (J_{m}(\sqrt{\omega^2-k^2}\rho)e^{-i m \phi}) \equiv \bar \eth  (J_{m}(\sqrt{\omega^2-k^2}\rho)e^{-i m \phi})= \frac{\sqrt{\omega^2-k^2}}{\sqrt{2}}J_{m-1}(\sqrt{\omega^2-k^2}\rho)e^{-i m \phi}\, .
\eea
\noindent These relations follow directly from the standard recurrence identities Eqs.~10.6.1-10.6.3 in \cite{nist}.

\eel

\bel \label{ortobessel3}
For $m>-1$ and any pair of zeros $j_{mn}$, $j_{mn'}$ of  $J_m(x)$, we have
\bea
\int_0^1dx x J_{m\pm 1}(j_{mn} x)J_{m\pm 1}(j_{mn'} x)=  \delta_{nn'} \frac{J_m^{' 2}(j_{mn})}{2}\, .
\eea

\eel

\noindent{\bf Proof}.
Using the identities: $\pm\partial_x J_m(j_{mn}x)+\frac{m}{x}J_m(j_{mn}x)=j_{mn} J_{m\pm 1}(j_{mn}x)$  (see 8.471.1, 8.471.2 in \cite{gradshteyn2007}), we can obtain
\bea
\int_0^1dx x J_{m\pm 1}(j_{mn} x)J_{m\pm 1}(j_{mn'} x)&=&\frac{1}{j_{mn}j_{mn'}} \int_0^1dx x \left[ \partial_x J_m(j_{mn}x)\partial_x J_m(j_{mn'}x)+\frac{m^2}{x^2}J_m(j_{mn}x)J_m(j_{mn'}x)\right. \nonumber\\ 
&&\left. \pm \frac{m}{x}\left(J_m(j_{mn'}x)\partial_x J_m(j_{mn}x)+J_m(j_{mn}x)\partial_x J_m(j_{mn'}x) \right)\right]\nonumber\\
&=&\frac{1}{j_{mn}j_{mn'}} \int_0^1dx x \left[ J_m(j_{mn}x)J_m(j_{mn'}x)\left[j_{mn}j_{mn'}-\frac{m^2}{x^2}\right]+\frac{m^2}{x^2}J_m(j_{mn}x)J_m(j_{mn'}x)\right. \nonumber\\ 
&&\left. \pm \frac{m}{j_{mn}j_{mn'}} \left. J_m(j_{mn'}x) J_m(j_{mn}x)\right|_0^1 \right]\nonumber\\
&=& \int_0^1dx x  J_m(j_{mn}x)J_m(j_{mn'}x)=  \delta_{nn'} \frac{J_m^{' 2}(j_{mn})}{2}\, ,
\eea
where in the second equality we used Lemma \ref{ortoderivative}, and in the fourth one we used  Lemma \ref{ortobessel1}.

\bec   \label{completeness2}
The Bessel functions satisfy the following completeness relation:
\bea
\sum_{n=1}^{\infty}\frac{J_{m\pm 1}(j_{mn}x) J_{m\pm 1}(j_{mn}x')}{{J'}^2_m(j_{mn})}=\frac{1}{2x}\delta(x-x')\, , \label{completenessid2}
\eea
where $x\in [0,1]$.
\eec

\noindent{\bf Proof.} Using the orthogonality condition in Lemma \ref{ortobessel3} we can obtain a complete basis of the solution space of Bessel functions in terms of $J_{m\pm 1}$, similarly to Lemma \ref{completeness}. The proof in Lemma \ref{completeness2} then follows straightforwardly.

\bep \label{circbasis2}
The two independent circular polarization vectors obtained in  (\ref{ar})-(\ref{al}) satisfy 
\bea
\int_{\mathbb R}dz\int_0^{2\pi} d\phi\int_0^1 dx x\,  \eta^{ab}\, A^\lambda_{a,h k m n}A^{\lambda'}_{b, h'k'm'n'}&=&  2\pi^2  (-1)^{m+1}   {J'}_m(j_{mn})^2  \delta_{n,n'}\delta_{m,-m'}\delta(k+k')\delta_{\lambda,-\lambda'}e^{-i(h+h')\omega_{kmn} t} \, , \nonumber\\
 \int_{\mathbb R}dz\int_0^{2\pi} d\phi\int_0^1 dx x\,  \eta^{ab}\overline{A^{\lambda}_{a,h k m n}}A^{\lambda'}_{b, h'k'm'n'}&=& 2 \pi^2   {J'}_m(j_{mn})^2  \delta_{n,n'}\delta_{m,m'}\delta(k-k') \delta_{\lambda,\lambda'}e^{i(h-h')\omega_{kmn} t} \, ,\nonumber
\eea
for any $\lambda,\lambda'\in\{R,L\}$
\eep

\noindent {\bf Proof. } Immediate from the definitions and Lemma \ref{ortobessel3} above.\\

\section{Explicit verification of the time-independence of the symplectic structure}\label{appB}

Although the time-independence of the symplectic structure is guaranteed by construction in the covariant phase-space formalism, it is instructive to verify this property explicitly for our setup, for which the spacetime boundary is non-trivial. In this appendix, we will show that the symplectic form (\ref{sp0}), defined on the spacelike hypersurface $\Sigma_0=\{t=t_0\}$ and subject to the boundary conditions (\ref{boundaryconditions0}), is conserved in time. The proof is purely geometric and independent of any coordinate choice.

Let $n^a=-g^{ab}\nabla_b t$ be the future-directed unit normal to $\Sigma$.  
The Lie derivative of $\Omega_0$ along $n^a$ reads
\[
\mathcal{L}_n\Omega_0(A_1,A_2)
=\mathcal{L}_n\!\int_\Sigma\!\!(A_1\wedge{^*F_2}-A_2\wedge{^*F_1})
=\int_\Sigma\!\big(\mathcal{L}_n A_1\wedge{^*F_2}-\mathcal{L}_n A_2\wedge{^*F_1}
+A_1\wedge\mathcal{L}_n{^*F_2}-A_2\wedge\mathcal{L}_n{^*F_1}\big).
\]
Using Cartan's identity, $\mathcal{L}_n=i_n\!\circ\! d+d\!\circ\! i_n$,  the source-free Maxwell equations $dF=d\, {^*F}=0$,  integrating by parts and then using the Leibniz rule for $i_n$, we obtain
\bea
\mathcal{L}_n\Omega_0(A_1,A_2)&=& \int_\Sigma  (di_nA_1+i_n F_1) \wedge ^*F_2 - (di_n A_2+i_n F_2)\wedge ^*F_1 + A_1 \wedge  di_n{^*F}_2 - A_2\wedge di_n{^*F}_1  \nonumber\\
&=& \int_\Sigma  d(i_nA_1 \wedge ^*F_2 -i_nA_2 \wedge ^*F_1)+i_n F_1 \wedge ^*F_2-i_n F_2 \wedge ^*F_1\nonumber\\&&+ F_1 \wedge  i_n{^*F}_2 - F_2\wedge i_n{^*F}_1-d( A_1 \wedge  i_n{^*F}_2 - A_2\wedge i_n{^*F}_1)\nonumber\\
&=& \int_\Sigma  di_n(A_1 \wedge ^*F_2 -A_2 \wedge ^*F_1)+i_n (F_1 \wedge ^*F_2- F_2 \wedge ^*F_1)\nonumber\, .
\eea
For any two-forms $F_1$ and $F_2$, one has $F_1\wedge{^*F_2}=F_2\wedge{^*F_1}$\footnote{Namely, $F_1 \wedge ^*F_2=F_{1,ab} \frac{1}{2}\epsilon_{cdmn}F_{2}^{mn}\epsilon^{abcd}d^4x=F_{2,mn} \frac{1}{2}\epsilon_{cdab}F_{1}^{ab}\epsilon^{mncd}d^4x=F_2 \wedge ^*F_1$.}, so the bulk terms cancel, leaving only a boundary contribution:
\[
\mathcal{L}_n\Omega_0(A_1,A_2)
=\int_{\partial\Sigma}\!i_n(A_1\wedge{^*F_2}-A_2\wedge{^*F_1}).
\]
Applying the Leibniz rule, and using the identity $\epsilon^{abcd}\epsilon_{ebfg}=2( \delta^{a}_{[e}\delta^c_{f]}\delta^d_g+\delta^{a}_{[f}\delta^c_{|g|}\delta^d_{e]}+\delta^{a}_{g}\delta^c_{[e}\delta^d_{f]})$, we find
\bea
\mathcal{L}_n\Omega_0(A_1,A_2)&=&\int_{\partial\Sigma} (i_n A_1)  ^*F_2 -(i_n A_2)  ^*F_1 -(A_1 \wedge i_n\, ^*F_2 -A_2 \wedge i_n\, ^*F_1)\nonumber\\
&=&\int_{\partial\Sigma} \frac{1}{2}[(i_n A_1)  ^*F_{2,ab} -(i_n A_2)  ^*F_{1,ab}]\epsilon^{abcd} n_c \nabla_d\rho- (A_{1,a}   \,^*F_{2,nb} -A_{2,a} \,  ^*F_{1,nb})\epsilon^{abcd} n_c \nabla_d\rho\nonumber\\
&= & \int_{\partial\Sigma}  -[(i_n A_1)  F_{2}^{cd} -(i_n A_2)  F_{1}^{cd}] n_c \rho_d+ (A_{1,a}   F_{2}^{ef} -A_{2,a} F_{1}^{ef})(\delta^{a}_{e}\delta^c_f\delta^d_g+\delta^{a}_f\delta^c_g\delta^d_e+\delta^{a}_{g}\delta^c_e\delta^d_f)n^g n_c \rho_d\nonumber\\
&= & \int_{\partial\Sigma}   (A_{1,a}   F_{2}^{ef} -A_{2,a} F_{1}^{ef})(\delta^{a}_{e}\delta^c_f\delta^d_g+\delta^{a}_f\delta^c_g\delta^d_e)n^g n_c \rho_d\nonumber\\
&= & \int_{\partial\Sigma}  (A_{1,a}   F_{2}^{ad} -A_{2,a} F_{1}^{ad}) \rho_d\nonumber\, ,
\eea
where $\rho_a=\nabla_a\rho$ is the unit outward normal to the cylindrical boundary, which is tangent to $\Sigma$, thus $n^a\rho_a=0$. Finally, in the Newman--Penrose basis $\{m^a,\bar m^a\}$, this result becomes
\bea
\mathcal{L}_n\Omega_0
\propto\!\int_{\partial\Sigma}
(A_{1,m}F_{2,\rho\bar m}+A_{1,\bar m}F_{2,\rho m}
-A_{2,m}F_{1,\rho\bar m}-A_{2,\bar m}F_{1,\rho m})\nonumber\\
\propto 
\int_{\partial\Sigma}
(A_{1,m}F_{2,\rho\phi}-A_{1,\bar m}F_{2,\rho\phi}
-A_{2,m}F_{1,\rho\phi}+A_{2,\bar m}F_{1,\rho\phi}).\nonumber
\eea
The last integral vanishes under the boundary condition $F_{ab}\rho^a\phi^b|_{\rho=R}=0$. Hence, $\mathcal{L}_n\Omega_0=0$, and the symplectic form is constant in time.

This confirms, through an explicit calculation, that the symplectic structure of the Maxwell field remains invariant within our waveguide, provided the normal vector $n^a$ is transverse to the boundary and satisfies $n^a\rho_a=0$. The result is  fully covariant and independent of the particular $3+1$ decomposition.

\section{Useful identities of the Bogoliubov coefficients}

In this subsection we provide computational details concerning all the identities of the Bogoliubov coefficients used along the main text. 

We first establish the connection between the Bogoliubov coefficients in the expansion (\ref{out1})-(\ref{out2}) and those in (\ref{in1})-(\ref{in2}), specified in (\ref{8coeficients}) in the main text.  First, let us remark that $A^{R,{\rm in}}_{a, h_0k_0m_0n_0}$ is given by (\ref{ARb2}) for the choice $z^{R,{\rm in}}_{hkmn}=\delta_{hh_0}\delta_{kk_0}\delta_{mm_0}\delta_{nn_0}$, $z^{L,{\rm in}}_{hkmn}=0$ for any $(h,k,m,n)\in H_{\rm in}$, and so $A^{R,{\rm in}}_{a, h_0k_0m_0n_0}\in \Gamma_R$ (note the shorthand $\delta_{kk_0}\equiv R^{-1}\delta(k-k_0)$). Similarly, $\overline{A^{L,{\rm in}}_{a, h_0k_0m_0n_0}}\in \Gamma_R$. Using the result (\ref{inner2}) we infer
\bea
\langle A^{R,{\rm in}}_{h_0k_0m_0n_0},A^{R,{\rm in}}_{h_1k_1m_1n_1}\rangle_{\rm in}& =& \delta_{h_1h_0}\delta_{k_1k_0}\delta_{m_1m_0}\delta_{n_1n_0}\, ,\label{11}\\
\langle \overline{A^{L,{\rm in}}_{h_0k_0m_0n_0}},\overline{A^{L,{\rm in}}_{h_1k_1m_1n_1}}\rangle_{\rm in}& =& \delta_{h_1h_0}\delta_{k_1k_0}\delta_{m_1m_0}\delta_{n_1n_0}\, ,\label{-1-1}\\
\langle A^{R,{\rm in}}_{h_0k_0m_0n_0},\overline{A^{L,{\rm in}}_{h_1k_1m_1n_1}}\rangle_{\rm in}& =& 0\, . \label{1-1}
\eea
We can apply a similar reasoning using (\ref{outrep}). Together with (\ref{outinnerproduct}), this leads us to
\bea
\langle A^{R,{\rm out}}_{h_0k_0m_0n_0},A^{R,{\rm out}}_{h_1k_1m_1n_1}\rangle_{\rm out}& =& \delta_{h_1h_0}\delta_{k_1k_0}\delta_{m_1m_0}\delta_{n_1n_0}\, ,\label{out11}\\
\langle \overline{A^{L,{\rm out}}_{h_0k_0m_0n_0}},\overline{A^{L,{\rm out}}_{h_1k_1m_1n_1}}\rangle_{\rm out}& =& \delta_{h_1h_0}\delta_{k_1k_0}\delta_{m_1m_0}\delta_{n_1n_0}\, ,\\
\langle A^{R,{\rm out}}_{h_0k_0m_0n_0},\overline{A^{L,{\rm out}}_{h_1k_1m_1n_1}}\rangle_{\rm out}& =& 0\, .
\eea
Using these orthonormal properties, from (\ref{out1}) and (\ref{out2}) we can get, respectively\footnote{From (\ref{antilinearity}) one infers that $\langle \alpha A^1_R,  A^2_R \rangle_{\rm in}=\bar \alpha \langle A^1_R,  A^2_R \rangle_{\rm in}$ if $A^2_R$ belongs to the subspace of positive-frequency modes, so that $J A^2_R=i A^2_R$, while $\langle \alpha A^1_R,  A^2_R \rangle_{\rm in}=\alpha \langle A^1_R,  A^2_R \rangle_{\rm in}$ if $A^2_R$ belongs to the subspace of negative-frequency modes, i.e. $J A^2_R=-i A^2_R$.}
\bea
\overline{ \gamma^R_{hkmn\atop 1k'm'n'} }= \langle A^{R,{\rm out}}_{hkmn}, A^{R,{\rm in}}_{1k'm'n'}\rangle_{\rm in}\, ,\quad  \forall k'm'n', \, (hkmn)\in H^{+>}_h\, ,\\
 \delta^R_{hkmn\atop -1k'm'n'} = \langle A^{R,{\rm out}}_{hkmn}, A^{R,{\rm in}}_{-1k'm'n'}\rangle_{\rm in}\, ,\quad   \forall k'm'n', \, (hkmn)\in H^{+>}_h \, ,\\
\overline{ \gamma^R_{hkmn\atop -1k'm'n'}} = \langle A^{R,{\rm out}}_{hkmn}, A^{R,{\rm in}}_{-1k'm'n'}\rangle_{\rm in}\, ,\quad  \forall k'm'n', \, (hkmn)\in H^{+<}_h\, ,\\
 \delta^R_{hkmn\atop 1k'm'n'} = \langle A^{R,{\rm out}}_{hkmn}, A^{R,{\rm in}}_{1k'm'n'}\rangle_{\rm in}\, ,\quad  \forall k'm'n', \, (hkmn)\in H^{+<}_h \, .
\eea
Similarly, from (\ref{in1}) and (\ref{in2}) we can get, respectively
\bea
\overline{ \alpha^R_{1kmn\atop h'k'm'n'} }= \langle A^{R,{\rm in}}_{1kmn}, A^{R,{\rm out}}_{h'k'm'n'}\rangle_{\rm out} \, ,\quad \forall kmn, \, (h'k'm'n')\in H^{+>}_h\, ,\\
  \beta^R_{1kmn\atop h'k'm'n'} = \langle A^{R,{\rm in}}_{1kmn}, A^{R,{\rm out}}_{h'k'm'n'}\rangle_{\rm out}  \, ,\quad \forall kmn, \, (h'k'm'n')\in H^{+<}_h\, ,\\
\overline{   \alpha^R_{-1kmn\atop h'k'm'n'} }= \langle A^{R,{\rm in}}_{-1kmn}, A^{R,{\rm out}}_{h'k'm'n'}\rangle_{\rm out}   \, ,\quad \forall kmn, \, (h'k'm'n')\in H^{+<}_h\, ,\\
  \beta^R_{-1kmn\atop h'k'm'n'} = \langle A^{R,{\rm in}}_{-1kmn}, A^{R,{\rm out}}_{h'k'm'n'}\rangle_{\rm out}  \, ,\quad \forall kmn, \, (h'k'm'n')\in H^{+>}_h\, .
\eea

Now,  these 8 coefficients are related to each other. 
For all $kmn, \, (h'k'm'n')\in H^{+>}_h$, we have
\bea
\overline{\alpha^R_{1kmn\atop h'k'm'n'}} &=& \langle A^{R,{\rm in}}_{1kmn}, A^{R,{\rm out}}_{h'k'm'n'}\rangle_{\rm out}=\frac{1}{2\hbar}\left[\Omega(A^{R,{\rm in}}_{1kmn},J_{\rm out}A^{R,{\rm out}}_{h'k'm'n'})+i\Omega(A^{R,{\rm in}}_{1kmn},A^{R,{\rm out}}_{h'k'm'n'}) \right]\nonumber\\
&=&\frac{1}{2\hbar}\left[\Omega(A^{R,{\rm in}}_{1kmn},iA^{R,{\rm out}}_{h'k'm'n'})+i\Omega(A^{R,{\rm in}}_{1kmn},A^{R,{\rm out}}_{h'k'm'n'}) \right]\nonumber\\
&=&\frac{1}{2\hbar}\left[\Omega(-iA^{R,{\rm in}}_{1kmn},A^{R,{\rm out}}_{h'k'm'n'})-i\Omega(A^{R,{\rm out}}_{h'k'm'n'}, A^{R,{\rm in}}_{1kmn}) \right]\nonumber\\
&=&\frac{1}{2\hbar}\left[\Omega(-J_{\rm in}A^{R,{\rm in}}_{1kmn},A^{R,{\rm out}}_{h'k'm'n'})-i\Omega(A^{R,{\rm out}}_{h'k'm'n'}, A^{R,{\rm in}}_{1kmn}) \right]\nonumber\\
&=&\frac{1}{2\hbar}\left[\Omega(A^{R,{\rm out}}_{h'k'm'n'}, J_{\rm in}A^{R,{\rm in}}_{1kmn})-i\Omega(A^{R,{\rm out}}_{h'k'm'n'}, A^{R,{\rm in}}_{1kmn}) \right]\nonumber\\
&=&\overline{\langle A^{R,{\rm out}}_{h'k'm'n'}, A^{R,{\rm in}}_{1kmn}\rangle_{\rm in}}= {\gamma^R_{h'k'm'n'\atop 1kmn}}\, . \label{alphagamma1}
\eea
For all $kmn, \, (h'k'm'n')\in H^{+<}_h$, we have
\bea
\beta^R_{1kmn\atop h'k'm'n'} &=& \langle A^{R,{\rm in}}_{1kmn}, A^{R,{\rm out}}_{h'k'm'n'}\rangle_{\rm out}=\frac{1}{2\hbar}\left[\Omega(A^{R,{\rm in}}_{1kmn},J_{\rm out}A^{R,{\rm out}}_{h'k'm'n'})+i\Omega(A^{R,{\rm in}}_{1kmn},A^{R,{\rm out}}_{h'k'm'n'}) \right]\nonumber\\
&=&\frac{1}{2\hbar}\left[\Omega(A^{R,{\rm in}}_{1kmn},-iA^{R,{\rm out}}_{h'k'm'n'})+i\Omega(A^{R,{\rm in}}_{1kmn},A^{R,{\rm out}}_{h'k'm'n'}) \right]\nonumber\\
&=&\frac{1}{2\hbar}\left[\Omega(iA^{R,{\rm in}}_{1kmn},A^{R,{\rm out}}_{h'k'm'n'})-i\Omega(A^{R,{\rm out}}_{h'k'm'n'}, A^{R,{\rm in}}_{1kmn}) \right]\nonumber\\
&=&\frac{1}{2\hbar}\left[\Omega(J_{\rm in}A^{R,{\rm in}}_{1kmn},A^{R,{\rm out}}_{h'k'm'n'})-i\Omega(A^{R,{\rm out}}_{h'k'm'n'}, A^{R,{\rm in}}_{1kmn}) \right]\nonumber\\
&=&\frac{-1}{2\hbar}\left[\Omega(A^{R,{\rm out}}_{h'k'm'n'}, J_{\rm in}A^{R,{\rm in}}_{1kmn})+i\Omega(A^{R,{\rm out}}_{h'k'm'n'}, A^{R,{\rm in}}_{1kmn}) \right]\nonumber\\
&=&-{\langle A^{R,{\rm out}}_{h'k'm'n'}, A^{R,{\rm in}}_{1kmn}\rangle_{\rm in}}= -{\delta^R_{h'k'm'n'\atop 1kmn}}\, . \label{betadelta1}
\eea
For all $kmn, \, (h'k'm'n')\in H^{+<}_h$, we have
\bea
\overline{\alpha^R_{-1kmn\atop h'k'm'n'}} &=& \langle A^{R,{\rm in}}_{-1kmn}, A^{R,{\rm out}}_{h'k'm'n'}\rangle_{\rm out}=\frac{1}{2\hbar}\left[\Omega(A^{R,{\rm in}}_{-1kmn},J_{\rm out}A^{R,{\rm out}}_{h'k'm'n'})+i\Omega(A^{R,{\rm in}}_{-1kmn},A^{R,{\rm out}}_{h'k'm'n'}) \right]\nonumber\\
&=&\frac{1}{2\hbar}\left[\Omega(A^{R,{\rm in}}_{-1kmn},-iA^{R,{\rm out}}_{h'k'm'n'})+i\Omega(A^{R,{\rm in}}_{-1kmn},A^{R,{\rm out}}_{h'k'm'n'}) \right]\nonumber\\
&=&\frac{1}{2\hbar}\left[\Omega(iA^{R,{\rm in}}_{-1kmn},A^{R,{\rm out}}_{h'k'm'n'})-i\Omega(A^{R,{\rm out}}_{h'k'm'n'}, A^{R,{\rm in}}_{-1kmn}) \right]\nonumber\\
&=&\frac{1}{2\hbar}\left[\Omega(-J_{\rm in}A^{R,{\rm in}}_{-1kmn},A^{R,{\rm out}}_{h'k'm'n'})-i\Omega(A^{R,{\rm out}}_{h'k'm'n'}, A^{R,{\rm in}}_{-1kmn}) \right]\nonumber\\
&=&\frac{1}{2\hbar}\left[\Omega(A^{R,{\rm out}}_{h'k'm'n'}, J_{\rm in}A^{R,{\rm in}}_{-1kmn})-i\Omega(A^{R,{\rm out}}_{h'k'm'n'}, A^{R,{\rm in}}_{-1kmn}) \right]\nonumber\\
&=&\overline{\langle A^{R,{\rm out}}_{h'k'm'n'}, A^{R,{\rm in}}_{-1kmn}\rangle_{\rm in}}= {\gamma^R_{h'k'm'n'\atop -1kmn}}\, . \label{alphagamma2}
\eea
Finally, for all $kmn, \, (h'k'm'n')\in H^{+>}_h$, we have
\bea
\beta^R_{-1kmn\atop h'k'm'n'} &=& \langle A^{R,{\rm in}}_{-1kmn}, A^{R,{\rm out}}_{h'k'm'n'}\rangle_{\rm out}=\frac{1}{2\hbar}\left[\Omega(A^{R,{\rm in}}_{-1kmn},J_{\rm out}A^{R,{\rm out}}_{h'k'm'n'})+i\Omega(A^{R,{\rm in}}_{-1kmn},A^{R,{\rm out}}_{h'k'm'n'}) \right]\nonumber\\
&=&\frac{1}{2\hbar}\left[\Omega(A^{R,{\rm in}}_{-1kmn},iA^{R,{\rm out}}_{h'k'm'n'})+i\Omega(A^{R,{\rm in}}_{-1kmn},A^{R,{\rm out}}_{h'k'm'n'}) \right]\nonumber\\
&=&\frac{1}{2\hbar}\left[\Omega(-iA^{R,{\rm in}}_{-1kmn},A^{R,{\rm out}}_{h'k'm'n'})-i\Omega(A^{R,{\rm out}}_{h'k'm'n'}, A^{R,{\rm in}}_{-1kmn}) \right]\nonumber\\
&=&\frac{1}{2\hbar}\left[\Omega(J_{\rm in}A^{R,{\rm in}}_{-1kmn},A^{R,{\rm out}}_{h'k'm'n'})-i\Omega(A^{R,{\rm out}}_{h'k'm'n'}, A^{R,{\rm in}}_{-1kmn}) \right]\nonumber\\
&=&\frac{-1}{2\hbar}\left[\Omega(A^{R,{\rm out}}_{h'k'm'n'}, J_{\rm in}A^{R,{\rm in}}_{-1kmn})+i\Omega(A^{R,{\rm out}}_{h'k'm'n'}, A^{R,{\rm in}}_{-1kmn}) \right]\nonumber\\
&=&-{\langle A^{R,{\rm out}}_{h'k'm'n'}, A^{R,{\rm in}}_{-1kmn}\rangle_{\rm in}}= -{\delta^R_{h'k'm'n'\atop -1kmn}}\, . \label{betadelta2}
\eea

We now proceed to establish the unitary identities of the Bogoliubov coefficients, given in (\ref{rel1})-(\ref{rel4}) of the main text.  To see this, let us first notice  that equation (\ref{11}) implies
\bea
0={\langle  A^{R,{\rm in}}_{1kmn}, A^{R,{\rm in}}_{-1k'm'n'}\rangle_{\rm in}}=\frac{1}{2\hbar}\left[\Omega( A^{R,{\rm in}}_{1kmn}, J_{\rm in}A^{R,{\rm in}}_{-1k'm'n'})+i\Omega( A^{R,{\rm in}}_{1kmn}, A^{R,{\rm in}}_{-1k'm'n'}) \right]\nonumber\, .
\eea
Since the symplectic structure is real-valued, taking the real and imaginary  parts yield 
\bea
\Omega( A^{R,{\rm in}}_{1kmn}, A^{R,{\rm in}}_{-1k'm'n'})=0\, , \quad  \Omega( A^{R,{\rm in}}_{1kmn}, J_{\rm in}A^{R,{\rm in}}_{-1k'm'n'})=\Omega( A^{R,{\rm in}}_{1kmn},-iA^{R,{\rm in}}_{-1k'm'n'})=0
\eea
 (this last equation can  also be inferred from the equality ${\langle  A^{R,{\rm in}}_{1kmn}, A^{R,{\rm in}}_{-1k'm'n'}\rangle_{\rm in}}=\overline{\langle  A^{R,{\rm in}}_{-1kmn}, A^{R,{\rm in}}_{1k'm'n'}\rangle_{\rm in}}$). 
 Similarly, for all $(hkmn)\in H^{+>}_h$ and $(h'k'm'n')\in H^{+<}_h$ we have
\bea
0={\langle  A^{R,{\rm out}}_{hkmn}, A^{R,{\rm out}}_{h'k'm'n'}\rangle_{\rm out}}=\frac{1}{2\hbar}\left[\Omega( A^{R,{\rm out}}_{hkmn}, J_{\rm out}A^{R,{\rm out}}_{h'k'm'n'})+i\Omega( A^{R,{\rm out}}_{hkmn}, A^{R,{\rm out}}_{h'k'm'n'}) \right]\, ,
\eea
so the real and imaginary parts yield, respectively, $\Omega( A^{R,{\rm out}}_{hkmn}, A^{R,{\rm out}}_{h'k'm'n'})=0$ and 
\bea
\Omega( A^{R,{\rm out}}_{hkmn}, J_{\rm out}A^{R,{\rm out}}_{h'k'm'n'})=\Omega( A^{R,{\rm out}}_{hkmn},-iA^{R,{\rm out}}_{h'k'm'n'})=0\, , \label{outorto}\quad (hkmn)\in H^{+>}_h, (h'k'm'n')\in H^{+<}_h\, ,
\eea
(this last equation can  also be inferred from the equality ${\langle  A^{R,{\rm out}}_{hkmn}, A^{R,{\rm out}}_{h'k'm'n'}\rangle_{\rm out}}=\overline{\langle  A^{R,{\rm out}}_{h'k'm'n'}, A^{R,{\rm out}}_{hkmn}\rangle_{\rm out}}$). On the other hand, equation (\ref{11}) with $h'=h$  implies
\bea
\delta_{kk'}\delta_{mm'}\delta_{nn'}={\langle  A^{R,{\rm in}}_{hkmn}, A^{R,{\rm in}}_{hk'm'n'}\rangle_{\rm in}}=\frac{1}{2\hbar}\left[\Omega( A^{R,{\rm in}}_{hkmn}, J_{\rm in}A^{R,{\rm in}}_{hk'm'n'})+i\Omega( A^{R,{\rm in}}_{hkmn}, A^{R,{\rm in}}_{hk'm'n'}) \right]\nonumber\, .
\eea
Again, since the symplectic structure is real-valued, taking the imaginary part yields $\Omega( A^{R,{\rm in}}_{hkmn}, A^{R,{\rm in}}_{hk'm'n'})=0$ for both $h=\pm 1$, while taking the real part produces $\Omega( A^{R,{\rm in}}_{hkmn}, J_{\rm in}A^{R,{\rm in}}_{hk'm'n'})=\Omega( A^{R,{\rm in}}_{hkmn},{\rm sign}\, h\, i A^{R,{\rm in}}_{hk'm'n'})=2\hbar \delta_{kk'}\delta_{mm'}\delta_{nn'}$, which implies 
\bea
\Omega( A^{R,{\rm in}}_{1kmn}, i A^{R,{\rm in}}_{1k'm'n'})=-\Omega( A^{R,{\rm in}}_{-1kmn}, i A^{R,{\rm in}}_{-1k'm'n'})=2\hbar \delta_{kk'}\delta_{mm'}\delta_{nn'}\, .  \label{in12}
\eea
Finally, using (\ref{out11}), if either $(hkmn), (h'k'm'n')\in H^{+>}_h$ or $(hkmn), (h'k'm'n')\in H^{+<}_h$, we have
\bea
\delta_{hh'}\delta_{kk'}\delta_{mm'}\delta_{nn'}={\langle  A^{R,{\rm out}}_{hkmn}, A^{R,{\rm out}}_{h'k'm'n'}\rangle_{\rm out}}=\frac{1}{2\hbar}\left[\Omega( A^{R,{\rm out}}_{hkmn}, J_{\rm out}A^{R,{\rm out}}_{h'k'm'n'})+i\Omega( A^{R,{\rm out}}_{hkmn}, A^{R,{\rm out}}_{h'k'm'n'}) \right]\, .
\eea
 Taking the imaginary part yields 
 \bea
 \Omega( A^{R,{\rm out}}_{hkmn}, A^{R,{\rm out}}_{h'k'm'n'})=0\, , \label{always0}
 \eea
 while taking the real part produces
 \bea
 \Omega( A^{R,{\rm out}}_{hkmn}, J_{\rm out}A^{R,{\rm out}}_{h'k'm'n'})&=&\Omega( A^{R,{\rm out}}_{hkmn}, i A^{R,{\rm out}}_{h'k'm'n'}) =2\hbar\, \delta_{hh'}\delta_{kk'}\delta_{mm'}\delta_{nn'}\, ,\quad (hkmn), (h'k'm'n')\in H^{+>}_h\, ,\label{outid1}\\
\Omega( A^{R,{\rm out}}_{hkmn}, J_{\rm out}A^{R,{\rm out}}_{h'k'm'n'})&=&- \Omega( A^{R,{\rm out}}_{hkmn}, iA^{R,{\rm out}}_{h'k'm'n'})=  2\hbar\, \delta_{hh'}\delta_{kk'}\delta_{mm'}\delta_{nn'}\, , \quad (hkmn), (h'k'm'n')\in H^{+<}_h\, . \label{outid2}
 \eea

If we now evaluate $\Omega( A^{R,{\rm in}}_{\pm 1kmn}, i A^{R,{\rm in}}_{\pm 1k'm'n'})$ by expanding as in (\ref{in1}) and (\ref{in2}), and we use the partial results (\ref{outorto}), (\ref{outid1}), (\ref{always0}), (\ref{outid2}) and (\ref{in12}), we can obtain, respectively:
\bea
\delta_{kk'}\delta_{mm'}\delta_{nn'}=\sum_{(h''k''m''n'')\in H^{+>}_h}  {\rm Re}\left(\alpha^R_{1kmn\atop h''k''m''n''}  \overline{\alpha^R_{1k'm'n'\atop h''k''m''n''}}\right)-\sum_{(h''k''m''n'')\in H^{+<}_h} {\rm Re}\left(\beta^R_{1kmn\atop h''k''m''n''}  \overline{\beta^R_{1k'm'n'\atop h''k''m''n''}}\right)\, ,\\
\delta_{kk'}\delta_{mm'}\delta_{nn'}=\sum_{(h''k''m''n'')\in H^{+<}_h} {\rm Re}\left(\alpha^R_{-1kmn\atop h''k''m''n''}  \overline{\alpha^R_{-1k'm'n'\atop h''k''m''n''}}\right)-\sum_{(h''k''m''n'')\in H^{+>}_h} {\rm Re}\left(\beta^R_{-1kmn\atop h''k''m''n''}  \overline{\beta^R_{-1k'm'n'\atop h''k''m''n''}}\right)\, .
\eea
A similar evaluation with $\Omega( A^{R,{\rm in}}_{\pm 1kmn},  A^{R,{\rm in}}_{\pm 1k'm'n'})$ using the same equations produces
\bea
0=-\sum_{(h''k''m''n'')\in H^{+>}_h}  {\rm Im}\left(\alpha^R_{1kmn\atop h''k''m''n''}  \overline{\alpha^R_{1k'm'n'\atop h''k''m''n''}}\right)+\sum_{(h''k''m''n'')\in H^{+<}_h} {\rm Im}\left(\beta^R_{1kmn\atop h''k''m''n''}  \overline{\beta^R_{1k'm'n'\atop h''k''m''n''}}\right)\, ,\\
0=-\sum_{(h''k''m''n'')\in H^{+<}_h} {\rm Im}\left(\alpha^R_{-1kmn\atop h''k''m''n''}  \overline{\alpha^R_{-1k'm'n'\atop h''k''m''n''}}\right)+\sum_{(h''k''m''n'')\in H^{+>}_h} {\rm Im}\left(\beta^R_{-1kmn\atop h''k''m''n''}  \overline{\beta^R_{-1k'm'n'\atop h''k''m''n''}}\right)\, .
\eea
Combining both results, we get
\bea
\delta_{kk'}\delta_{mm'}\delta_{nn'}=\sum_{(h''k''m''n'')\in H^{+>}_h}  \alpha^R_{1kmn\atop h''k''m''n''}  \overline{\alpha^R_{1k'm'n'\atop h''k''m''n''}}-\sum_{(h''k''m''n'')\in H^{+<}_h} \beta^R_{1kmn\atop h''k''m''n''}  \overline{\beta^R_{1k'm'n'\atop h''k''m''n''}}\, ,\\
\delta_{kk'}\delta_{mm'}\delta_{nn'}=\sum_{(h''k''m''n'')\in H^{+<}_h} \alpha^R_{-1kmn\atop h''k''m''n''}  \overline{\alpha^R_{-1k'm'n'\atop h''k''m''n''}}-\sum_{(h''k''m''n'')\in H^{+>}_h} \beta^R_{-1kmn\atop h''k''m''n''}  \overline{\beta^R_{-1k'm'n'\atop h''k''m''n''}}\, .
\eea
Doing a similar computation for $\Omega( A^{R,{\rm out}}_{hkmn}, iA^{R,{\rm out}}_{h'k'm'n'})$ and $\Omega( A^{R,{\rm out}}_{hkmn}, A^{R,{\rm out}}_{h'k'm'n'})$, using both expansions (\ref{out1}) and (\ref{out2}), we obtain, respectively: 
\bea
\delta_{hh'}\delta_{kk'}\delta_{mm'}\delta_{nn'}=\sum_{\omega_{k''m''n''}>0} \overline{\alpha^R_{1k''m''n''\atop hkmn}}  \alpha^R_{1k''m''n''\atop h'k'm'n'}-{\beta^R_{-1k''m''n''\atop hkmn}}  \overline{\beta^R_{-1k''m''n''\atop h'k'm'n'}}\, , \quad(hkmn),(h'k'm'n')\in H^{+>}_h\\
\delta_{hh'}\delta_{kk'}\delta_{mm'}\delta_{nn'}=\sum_{\omega_{k''m''n''}>0} \overline{\alpha^R_{-1k''m''n''\atop hkmn}}  \alpha^R_{-1k''m''n''\atop h'k'm'n'}-{\beta^R_{1k''m''n''\atop hkmn}}  \overline{\beta^R_{1k''m''n''\atop h'k'm'n'}}\, , \quad (hkmn),(h'k'm'n')\in H^{+<}_h
\eea
where we have already substituted the coefficients $\delta$ and $\gamma$ in terms of $\alpha$ and $\beta$ according to (\ref{alphagamma1}), (\ref{betadelta1}), (\ref{alphagamma2}), (\ref{betadelta2}).

%%%%%%%%%%%%%%%%%%%%%%%%%%%%
{\bf \em Acknowledgments.} 
%%%%%%%%%%%%%%%%%%%%%%%%%%%%
I thank Ivan Agullo,  Albert Ferrando, Dimitrios Kranas, Jose Navarro-Salas and Juan Margalef for  useful discussions and feedback. 
I acknowledge support through {\it Atracci\'on de Talento Cesar Nombela} grant No 2023-T1/TEC-29023, funded by Comunidad de Madrid (Spain); as well as   financial support  via the Spanish Grant  PID2023-149560NB-C21, funded by MCIU/AEI/10.13039/501100011033/FEDER, UE.

\appendix

\bibliography{references}

\end{document}